\documentclass[a4paper,aps,prd,10pt,preprintnumbers,showpacs,twocolumn,superscriptaddress,nofootinbib,amsmath,amssymb]{revtex4-1}
\usepackage{graphicx}
\usepackage{cmap}
\usepackage[utf8]{inputenc}
\usepackage[T1]{fontenc}
\usepackage{color}

\DeclareMathAlphabet{\pazocal}{OMS}{zplm}{m}{n}

\begin{document}

\title{Review of  analytic results on quasinormal modes of black holes}
\author{S. V. Bolokhov}\email{bolokhov-sv@rudn.ru}
\affiliation{Peoples' Friendship University of Russia (RUDN University), 6 Miklukho-Maklaya Street, Moscow, 117198, Russia}
\author{Milena Skvortsova}\email{milenas577@mail.ru }
\affiliation{Peoples' Friendship University of Russia (RUDN University), 6 Miklukho-Maklaya Street, Moscow, 117198, Russia}

\begin{abstract}
We present a concise review of known analytic results for quasinormal modes of black holes and related spacetimes. Our emphasis is on those regimes where the perturbation equations admit exact or perturbative solutions, providing insights complementary to numerical or semi-analytic approaches. We discuss solvable cases in lower-dimensional spacetimes, algebraically special modes, and exact results in higher-curvature gravity theories. Particular attention is given to the eikonal regime and its correspondence with null geodesics, as well as to beyond-eikonal approximations based on inverse multipole expansions in parametrized metrics. We review analytic solutions obtained in the near-extremal limit of Schwarzschild–de Sitter black holes, in the regime of large field mass, and in pure de Sitter and anti–de Sitter spacetimes, where boundary conditions play a crucial role. While not exhaustive, this overview highlights the diversity of techniques and physical insights made possible by analytic treatments of quasinormal spectra.
\end{abstract}

\maketitle

\section{Introduction}

The advent of gravitational wave astronomy and observation in electromagnetic spectra has opened an unprecedented observational window into the strong-field regime of gravity \cite{LIGOScientific:2017vwq,EventHorizonTelescope:2019dse,EventHorizonTelescope:2019ggy,Goddi:2016qax,NANOGrav:2023gor,NANOGrav:2023hvm,NANOGrav:2023pdq,LISA:2022yao,LISA:2022kgy}. The detection of gravitational waves from binary black hole mergers by the LIGO and Virgo collaborations \cite{TheLIGOScientific:2016src, TheLIGOScientific:2016pea, TheLIGOScientific:2016qqj} has confirmed one of the most striking predictions of general relativity and enabled the study of black holes as astrophysical objects. A particularly important phase of the gravitational wave signal is the \textit{ringdown}, during which the merged object settles into a stationary black hole. This stage is governed by quasinormal modes (QNMs), which encode the characteristic oscillations of the perturbed black hole spacetime \cite{Kokkotas:1999bd,Nollert:1999ji,Konoplya:2011qq}. Since QNMs depend only on the geometry of the final black hole, their detection provides a direct probe of the near-horizon structure of spacetime and the first few overtones of quasinormal spectrum, essential for the early ringdown phase \cite{Giesler:2019uxc,Giesler:2024hcr}, are especially sensitive to the geometry extremely close to the event horizon \cite{Konoplya:2023kem, Konoplya:2023ahd}.

Future gravitational wave observatories, such as the space-based LISA mission \cite{LISA:2017pwj}, will dramatically increase the sensitivity to the ringdown phase, allowing for precision tests of gravity in the strong-field regime. Moreover, QNMs offer a powerful tool for testing the no-hair theorem and for distinguishing black holes from exotic compact objects. The goal of the present work is to review analytic results on QNMs in black hole spacetimes, highlighting the mathematical techniques and physical insights obtained in those special cases where exact or approximate solutions are available.

Despite their significance, obtaining closed-form expressions for quasinormal modes remains a formidable challenge. In most cases, the differential equations governing linear perturbations are too complex to admit exact analytic solutions. Even for the Schwarzschild spacetime—the simplest vacuum black hole solution in general relativity—the Regge–Wheeler equation describing axial gravitational perturbations \cite{Regge:1957td} does not yield analytic expressions for the QNM spectrum. As a result, one must rely on a variety of numerical and semi-analytic methods, such as the continued fraction method \cite{Leaver:1985ax}, the WKB approximation \cite{Schutz:1985km,Iyer:1986np,Konoplya:2019hlu,Konoplya:2003ii,Konoplya:2004ip,Matyjasek:2017psv}, time-domain integration \cite{Gundlach:1993tp}, and spectral methods \cite{Konoplya:2022zav,Fortuna:2020obg,Konoplya:2025hgp}, to compute the complex QNM frequencies.

The situation becomes even more intricate for rotating or charged black holes, where variable separation may be less straightforward, and the effective potentials often lack the frequency-independent, single-peak barrier structure required for many approximation techniques (see, for instance, \cite{Leaver:1990zz,Leaver:1985ax,Konoplya:2006br,Kanti:2006ua,Konoplya:2007zx,Kokkotas:2010zd,Onozawa:1996ux,Andersson:1996xw}). As a result, exact analytic solutions for quasinormal modes are exceedingly rare and typically exist only in highly symmetric or fine-tuned scenarios. Nevertheless, analytic progress has been made in certain regimes—such as near-extremality, asymptotic limits, or high multipole numbers—where simplifications allow closed-form or perturbative analytic results. This review aims to systematize such cases and illustrate the physical insight they offer.

One of the most fruitful analytic regimes is the so-called eikonal limit, which corresponds to high-frequency perturbations with large multipole numbers $\ell$. In this limit, the behavior of quasinormal modes can be intuitively understood using a geometric-optics analogy, where perturbations follow unstable circular null geodesics around the black hole. Specifically, the real part of the QNM frequency corresponds to the angular frequency of the orbit, while the imaginary part is related to the Lyapunov exponent that governs its instability \cite{Cardoso:2008bp}. This elegant correspondence allows for the derivation of compact analytic expressions in the eikonal regime and has been applied successfully in a wide variety of spacetimes \cite{Cardoso:2008bp,Konoplya:2017wot,Konoplya:2022gjp,Bolokhov:2023dxq}. Moreover, null geodesics play a critical role in critical collapse and black hole formation \cite{Ianniccari:2024eza,Hod:2024ihh}.

The eikonal approximation is also the regime where the WKB method reaches its highest accuracy, enabling semi-analytic computations of QNMs. In this context, QNM spectra often become nearly independent of the spin of the perturbing field and converge rapidly to their asymptotic behavior even at moderate $\ell$ values. Analytic formulas in this regime have been extensively developed across a broad range of black hole and compact object models,  \cite{Glampedakis:2019dqh,Silva:2019scu,Bryant:2021xdh,Zhidenko:2008fp,Konoplya:2005sy,Konoplya:2001ji,Chen:2022nlw,Allahyari:2018cmg,Bolokhov:2023bwm,Dubinsky:2024aeu,Dubinsky:2024gwo}.

Beyond the eikonal limit, analytic techniques have continued to evolve. Recent work has demonstrated that controlled expansions in inverse powers of $\ell$ allow one to compute subleading corrections to the eikonal frequency, producing analytic expressions that remain accurate at intermediate values of $\ell$ \cite{Konoplya:2023moy}. These beyond-eikonal approximations have been successfully applied in numerous recent studies to analyze black hole perturbations in various alternative gravity theories and exotic scenarios \cite{Bolokhov:2023dxq,Dubinsky:2024rvf,Malik:2024sxv,Malik:2024tuf,Malik:2024voy,2753764,Bolokhov:2023dxq,Malik:2023bxc}. Such approaches offer a compelling balance between analytical clarity and quantitative accuracy.

Analytic results for quasinormal modes, although confined to special cases or regimes, are of high value both theoretically and practically. Firstly, they make explicit the dependence of QNM frequencies on the physical parameters of the black hole or the underlying theory of gravity, thereby providing direct physical insight. Secondly, they serve as benchmarks to validate numerical computations and test the convergence and accuracy of semi-analytical methods. Thirdly, they are indispensable in parameter scans across large theory spaces or families of compact objects, where numerical methods may be computationally expensive or unstable. In this way, analytic approaches enhance our understanding of black hole spectroscopy.


Our paper is organized as follows. In Section II, we review cases in which analytic expressions for quasinormal modes have been obtained in lower-dimensional spacetimes, where the perturbation equations are typically simpler and exactly solvable. Section III is devoted to algebraically special modes, which, while not comprising the full QNM spectrum, admit exact solutions in special settings and reveal important mathematical structure.

In Section IV, we discuss exactly solvable models arising in modified gravity theories and higher-curvature corrections, including cases where the perturbation equations reduce to hypergeometric or P\"oschl–Teller forms. Section V focuses on the eikonal regime and its correspondence with unstable null geodesics, while Section VI details the derivation of analytic QNMs in terms of Lyapunov exponents and photon sphere characteristics. In Section VII, we explore beyond-eikonal analytic approximations based on inverse multipole expansions in parametrized spherically symmetric spacetimes. Section VIII presents analytic results in the near-extremal limit of de Sitter black holes, where the wave equation reduces to exactly solvable forms such as the P\"oschl–Teller potential. In Section IX, we analyze the regime of large field mass \( \mu M \gg 1 \), where the WKB approach provides highly accurate expressions through a \( 1/\mu \) expansion. Section X reviews quasinormal modes in pure de Sitter and anti–de Sitter spacetimes, highlighting the role of boundary conditions and causal structure. Finally, in Section XI, we summarize our conclusions and comment on prospects for future work.

\section{Analytic Quasinormal Modes in the BTZ Black Hole Background}\label{sec:lowerdim}

One of the most important exactly solvable models for quasinormal modes is provided by the $(2+1)$-dimensional Banados-Teitelboim-Zanelli (BTZ) black hole \cite{Banados:1992wn}. Unlike in higher-dimensional cases, the wave equations governing perturbations of this spacetime are analytically solvable for scalar, electromagnetic, and even spinor (Weyl/Dirac) fields. This makes the BTZ background a valuable testing ground for exploring the properties of quasinormal modes (QNMs) and their role in holography, especially in light of the AdS/CFT correspondence.

The BTZ metric for a non-rotating black hole in asymptotically AdS$_3$ spacetime is given by
\begin{equation}
ds^2 = \left(-M + \frac{r^2}{\ell^2} \right) dt^2 - \left(-M + \frac{r^2}{\ell^2} \right)^{-1} dr^2 - r^2 d\phi^2, \label{eq:btz_metric}
\end{equation}
where $M$ is the black hole mass and $\ell$ is the AdS radius, related to the cosmological constant via $\Lambda = -1/\ell^2$. The event horizon is located at $r = r_+ = \ell \sqrt{M}$.

The scalar field $\Phi$ obeys the Klein-Gordon equation
\begin{equation}
\nabla^\mu \nabla_\mu \Phi = 0. \label{eq:scalar_eom}
\end{equation}
Using the separation of variables ansatz
\begin{equation}
\Phi(t,r,\phi) = \frac{1}{r^{1/2}} f(r) e^{-i\omega t + im\phi},
\end{equation}
and transforming to the tortoise coordinate $r^*$ via
\begin{equation}
\frac{dr^*}{dr} = \left( \frac{r^2}{\ell^2} - M \right)^{-1},
\end{equation}
the equation reduces to a Schr\"odinger-like wave equation:
\begin{equation}
\frac{d^2 f}{dr_*^2} + \left[\omega^2 - V(r) \right] f = 0, \label{eq:master_eq}
\end{equation}
where the effective potential is
\begin{equation}
V(r) = \frac{3r^2}{4\ell^4} - \frac{M}{2\ell^2} - \frac{M^2}{4r^2} + \frac{m^2}{\ell^2} - \frac{Mm^2}{r^2}. \label{eq:btz_potential}
\end{equation}

At the horizon ($r^* \to -\infty$), the physical boundary condition is purely ingoing waves:
\begin{equation}
\Psi(r^*) \sim e^{-i \omega r^*}, \quad r^* \to -\infty.
\end{equation}
At spatial infinity ($r^* = 0$), the potential diverges, implying that the solution must vanish:
\begin{equation}
\Psi(r^*) \to 0, \quad r^* \to 0.
\end{equation}
These boundary conditions define the quasinormal spectrum.

Introducing the new variable $x = 1/\cosh^2(\sqrt{M} r^*) \in [0,1]$, the wave equation transforms into a hypergeometric equation. By suitable rescaling of the wavefunction,
\begin{equation}
f(r^*) = (1 - x)^{3/4} x^{i\omega/2\sqrt{M}} y(x),
\end{equation}
the function $y(x)$ satisfies the standard hypergeometric form:
\begin{equation}
x(1 - x) y'' + [c - (a + b + 1)x] y' - ab y = 0,
\end{equation}
with the parameters
\begin{align}
a &= 1 + \frac{im}{2\sqrt{M}} - \frac{i\omega}{2\sqrt{M}}, \\
b &= 1 - \frac{im}{2\sqrt{M}} - \frac{i\omega}{2\sqrt{M}}, \\
c &= 1 - \frac{i\omega}{\sqrt{M}}.
\end{align}

The ingoing condition at the horizon ($x=0$) and vanishing at infinity ($x=1$) impose quantization:
\begin{equation}
a = -n \quad \text{or} \quad b = -n, \quad n = 0,1,2,\dots
\end{equation}
leading to exact quasinormal frequencies:
\begin{equation}
\omega = \pm m - 2i \sqrt{M}(n + 1). \label{eq:btz_qnm}
\end{equation}
These results hold for both scalar and electromagnetic fields due to the equivalence of their wave equations in three dimensions.

The exact solvability of the scalar and electromagnetic QNMs in the BTZ background provides a valuable benchmark for testing numerical methods and understanding the physics of AdS black holes. The QNM frequencies scale with the black hole radius, and hence with its temperature, consistent with expectations from the AdS/CFT correspondence \cite{Kovtun:2004de,Policastro:2001yc,Horowitz:1999jd}. In the dual CFT, these frequencies govern the thermalization timescales of perturbations, linking gravitational dynamics in the bulk with relaxation in the boundary theory \cite{Birmingham:2001pj}. The BTZ black hole is thus an important toy model for exploring black hole spectroscopy and holography.

In~\cite{Konoplya:2004ik}, the authors analyzed the quasinormal modes (QNMs) of a \((2+1)\)-dimensional BTZ black hole, taking into account the quantum backreaction from a conformally coupled scalar field. The backreaction arises from the expectation value of the stress-energy tensor \(\langle T_{\mu\nu} \rangle\) due to Hawking radiation and leads to a modified spacetime geometry. The corrected semiclassical Einstein equations take the form
\begin{equation}
G_{\mu \nu} + \Lambda g_{\mu \nu} = \kappa \langle T_{\mu \nu} \rangle.
\end{equation}

The modified BTZ metric is given by
\begin{equation}
ds^2 = -f(r)\,dt^2 + \frac{dr^2}{f(r)} + r^2\,d\phi^2,
\end{equation}
with the corrected lapse function
\begin{equation}
f(r) = \frac{r^2}{L^2} - M - \frac{2\ell_p F(M)}{r},
\end{equation}
where \(L\) is the AdS radius, \(M\) the black hole mass, \(\ell_p = \hbar / 8\) is the Planck length in \((2+1)\) dimensions, and \(F(M)\) encodes the one-loop backreaction correction due to the scalar field:
\begin{equation}
F(M) = \frac{M^{3/2}}{2\sqrt{2}} \sum_{n=1}^{\infty} \frac{e^{-in\delta} \left[\cosh(2\pi n \sqrt{M}) + 3 \right]}{\left[ \cosh(2\pi n \sqrt{M}) - 1 \right]^{3/2}},
\end{equation}
where $\delta$ is an arbitrary phase~\cite{Konoplya:2004ik}.

The evolution of a conformally coupled scalar field \(\Phi(t, r, \phi)\) is governed by the wave equation:
\begin{equation}
\square \Phi = \frac{1}{8} R \Phi.
\end{equation}
Using the ansatz \(\Phi(t, r, \phi) = e^{-i\omega t} e^{i L \phi} \frac{\psi(r)}{\sqrt{r}}\) and the tortoise coordinate \(dr_*/dr = 1/f(r)\), the wave equation reduces to
\begin{equation}
\left( \frac{d^2}{dr_*^2} + \omega^2 - V(r) \right) \psi(r_*) = 0,
\end{equation}
with the effective potential
\begin{equation}
V(r) = \left( \frac{M + 4 L^2}{4 r^2} - \frac{3 \ell_p F(M)}{2 r^3} \right) f(r).
\end{equation}

The appropriate boundary conditions for quasinormal modes in this asymptotically AdS spacetime are:
\begin{itemize}
    \item purely ingoing waves at the horizon (\(r_* \to -\infty\)),
    \item Dirichlet boundary condition at spatial infinity (\(r_* \to 0\)).
\end{itemize}

By expanding the potential near the boundary, it was shown that the leading effect of the backreaction is equivalent to a constant energy shift:
\begin{equation}
\omega^2 \to \tilde{\omega}^2 = \omega^2 - \left(4L^2 + M\right) \left( \frac{L}{M} \right)^{3/2} F(M)\, \ell_p.
\end{equation}
The wave equation in this form admits an exact solution in terms of modified Bessel functions:
$$G(r_*, \xi; \tilde{\omega}) = $$
\begin{equation}
\frac{I_n(z(\xi)) \left[ I_n(Z_0) K_n(z(r_*)) - K_n(Z_0) I_n(z(r_*)) \right]}{ \sqrt{LM} I_n(Z_0) },
\end{equation}
where \(n = -i \tilde{\omega} / \sqrt{LM}\) and \(Z_0 = \sqrt{V_0 / (LM)}\). 
Here, \( I_\nu(z) \) and \( K_\nu(z) \) denote the modified Bessel functions of the first and second kind, respectively. These special functions appear as solutions to the differential equation arising in the Green function method and are chosen to satisfy the appropriate boundary conditions at the horizon and at spatial infinity.

The quasinormal frequencies are determined from the poles of the Green function, that is, the condition
\begin{equation}
I_n(Z_0) = 0.
\end{equation}

The physical implications of these corrections are significant: backreaction due to Hawking radiation leads to QNMs with higher real frequencies and reduced damping rates. This makes small quantum-corrected black holes better oscillators (with higher quality factors \(Q = |\omega_{\text{Re}}| / |\omega_{\text{Im}}|\)) compared to classical macroscopic black holes.

The exact solutions were generalized for other $2+1$ dimensional black holes \cite{Becar:2013qba,Gonzalez:2014voa}.  However, adding higher curvature corrections in the Gauss-Bonnet form \cite{Konoplya:2020ibi} already leads to the spectral problem which cannot be solved analytically \cite{Skvortsova:2023zca}. 
The two dimensional black holes sometimes also admit exact solutions as shown in \cite{Becar:2014jia,Becar:2014aka}.

\section{Algebraically special modes}\label{sec:algebraic}

\subsection{Algebraically special modes for Schwarzschild black hole and Schwarzschild singularity}

An intriguing class of analytic solutions in black hole perturbation theory arises from so-called \textit{algebraically special modes}. These modes, originally identified in the context of the Teukolsky formalism and characterized by vanishing of specific invariants (e.g., the Starobinsky constant), often admit exact analytical expressions and are closely associated with unique features in the structure of the perturbation equations.

In~\cite{Cardoso:2006yj}, Cardoso and Cavaglià explored the role of algebraically special modes in the stability analysis of four-dimensional negative-mass Schwarzschild and (anti-)de Sitter spacetimes. These spacetimes exhibit naked singularities and, though unphysical from the standpoint of the cosmic censorship conjecture, they serve as testbeds for studying perturbative instabilities.

The background metric considered has the form
\begin{equation}
ds^2 = f(r) dt^2 - \frac{dr^2}{f(r)} - r^2(d\theta^2 + \sin^2\theta d\phi^2), 
\end{equation}
\begin{equation}
f(r) = \alpha \frac{r^2}{R^2} + 1 - \frac{2M}{r},    
\end{equation}
where $\alpha = +1$ corresponds to AdS, $\alpha = -1$ to dS, and $R$ is the AdS/dS curvature radius. The mass parameter $M$ is allowed to take negative values, and the perturbations are classified into odd (Regge-Wheeler) and even (Zerilli) sectors.

The master wave equations for both parity sectors reduce to Schr\"odinger-like forms:
\begin{equation}
\frac{d^2 \Psi_\pm}{dr_*^2} + \left[ \omega^2 - V_\pm(r) \right] \Psi_\pm = 0,
\end{equation}
where $r_*$ is the tortoise coordinate defined by $dr/dr_* = f(r)$. The effective potentials $V_\pm$ can be written in a supersymmetric form as
\begin{equation}
V_\pm = W^2 \pm \frac{dW}{dr_*} + \beta,
\end{equation}
with $W(r)$ and $\beta$ being functions determined by the background geometry and multipole number $\ell$.

Algebraically special modes are defined as those for which the quasinormal frequency satisfies $\omega^2 = \beta$, leading to the solution
\begin{equation}
\Psi_\pm(r_*) = \chi_\pm(r_*) \left( C_1 + C_2 \int^{r_*} \frac{dr'_*}{\chi_\pm^2(r'_*)} \right), \end{equation}
\begin{equation}
\chi_\pm(r_*) = \exp\left( \pm \int W(r'_*) dr'_* \right).
\end{equation}

Remarkably, these modes can be constructed exactly, allowing a detailed investigation of their behavior near singularities and boundaries. The authors demonstrate that for the negative-mass Schwarzschild and Schwarzschild-de Sitter spacetimes, the even-parity (Zerilli) algebraically special modes are regular and grow exponentially in time, implying instability:
\begin{equation}
\omega_s = i \frac{(\ell - 1) \ell (\ell + 1)(\ell + 2)}{12 |M|}.
\end{equation}
In contrast, the odd-parity modes are found to be stable under the same conditions.

The analysis further reveals that these instabilities persist even in the presence of a cosmological constant, although the precise behavior depends on the boundary conditions imposed at spatial infinity. In asymptotically AdS spacetimes, the choice of boundary conditions (e.g., reflective vs. transparent) can determine the presence or absence of instability. Overall, this work shows that algebraically special modes are not merely mathematical curiosities but can be physically significant in diagnosing instabilities in geometries beyond the usual positive-mass black holes.

\subsection{Algebraically Special Modes and Darboux Transformations}

Algebraically special modes form a distinguished class of solutions to black hole perturbation equations, characterized by their reflectionless nature: waves purely ingoing at the horizon and outgoing at infinity. Historically, such solutions were first studied in the context of Schwarzschild perturbations, and later extended to Kerr spacetime and other settings. A seminal analysis of these modes for Schwarzschild black holes was presented in~\cite{Chandrasekhar:1983book}, while further developments appeared in~\cite{MaassenvandenBrink:2000zh}.

A central observation is that algebraically special solutions are often tied to exact, closed-form expressions for the wavefunction at specific (imaginary) frequencies. For instance, for the Regge-Wheeler equation governing axial perturbations of Schwarzschild black holes,
\begin{equation}
\frac{d^2 X}{dx^2} + \left[\omega^2 - V_{\text{RW}}(r) \right] X = 0,
\end{equation}
the algebraically special frequency is given by~\cite{Chandrasekhar:1984zz}
\begin{equation}
\omega_* = -i \frac{n(n+1)}{3M}, \qquad n = \frac{1}{2}(\ell - 1)(\ell + 2),
\end{equation}
and the exact solution is
\begin{equation}
X_*(r) = \left(n + \frac{3M}{r} \right) e^{-i \omega_* x},
\end{equation}
describing a purely ingoing wave with no reflection. A corresponding Zerilli solution can be obtained via the Chandrasekhar transformation.

These special solutions play a pivotal role in the theory of Darboux transformations, which relate pairs of second-order linear differential equations while preserving the spectrum under suitable conditions. This connection has been fully explored in~\cite{Glampedakis:2017rar}, where the authors identified Chandrasekhar's transformation between the Regge-Wheeler and Zerilli equations as a classical Darboux transformation.

Let us briefly outline the mechanism. Consider two Schr\"odinger-type wave equations
\begin{align}
\frac{d^2 y}{dx^2} + q(x) y &= 0, \\
\frac{d^2 Y}{dx^2} + Q(x) Y &= 0,
\end{align}
connected by the Darboux transformation
\begin{equation}
Y = y' + f(x) y,
\end{equation}
where $f(x)$ satisfies the Riccati equation
\begin{equation}
f' - f^2 - q(x) = -c,
\end{equation}
and $Q(x) = q(x) - 2f'(x)$. The generator function $f(x)$ is typically constructed from a known solution $u(x)$ of the original equation via
\begin{equation}
f(x) = -\frac{d}{dx} \ln u(x).
\end{equation}

In the Schwarzschild case, using the algebraically special solution $X_*(x)$ as the generating function yields the Zerilli potential from the Regge-Wheeler one, confirming their isospectrality. Importantly, this approach shows that the reflection and transmission coefficients, and hence the quasinormal spectra, are preserved under the Darboux map — provided the potentials are short-ranged and regular~\cite{Glampedakis:2017rar}.

The analysis is further enriched in the Kerr case. Although the Teukolsky equation with its long-range potential does not allow a standard Darboux mapping to a short-range equation, a generalized Darboux transformation (GDT) can be invoked. The GDT takes the form
\begin{equation}
Y = \beta(x) y' + f(x) y,
\end{equation}
and allows, under proper constraints, the transformation of long-range potentials into short-range ones. This technique underlies the derivation of the Sasaki-Nakamura equation from the Teukolsky equation~\cite{Sasaki:1981sx}.

Moreover, the existence of algebraically special solutions simplifies the construction of such transformations. For example, the Kerr algebraically special modes are known in closed form and can be used to generate GDT mappings between Teukolsky-derived equations~\cite{Chandrasekhar:1984zz, Glampedakis:2017rar}.

Summarizing, the connection between algebraically special modes and Darboux transformations has been explored  in~\cite{Glampedakis:2017rar} and shown that certain special solutions can be mapped between different types of perturbation equations (e.g., axial and polar) via Darboux or supersymmetric quantum mechanical transformations, further underscoring the algebraic structure behind these modes. Notably, an explicit analysis of this mechanism shows how algebraically special solutions relate to reflectionless scattering.

It is worth mentioning that algebraically special modes should not be confused with the purely imaginary modes which appear frequently in the spectra of various black hole models \cite{Konoplya:2008ix,Cuyubamba:2016cug,Konoplya:2017ymp,Gonzalez:2017gwa,Konoplya:2017zwo,Cuyubamba:2018jdl,Konoplya:2020bxa,Starinets:2002br,Konoplya:2020juj,Cardoso:2003cj,Konoplya:2013sba,Nunez:2003eq,Konoplya:2022zav,Konoplya:2024ptj,Konoplya:2008au,Konoplya:2025mvj,Konoplya:2018qov}.  The latter may indicate the hydrodynamics mode in the gauge/gravity duality, purely de Sitter branch of modes in asymptotically de Sitter spacetimes or simply the onset of instability \cite{Konoplya:2008yy,Konoplya:2008ix}. The algebraically special modes for Schwarzschild stars have been found in \cite{Konoplya:2019nzp}, while for four and higher dimensional Schwarzschild-de Sitter black hole in \cite{Cardoso:2003cj,Konoplya:2022xid,Konoplya:2003dd}.

\subsection{Are algebraically special modes special?}

Algebraically special modes occupy a unique place in the study of black hole perturbations. While they share many features with quasinormal modes (QNMs), they do not fully qualify as QNMs in the conventional sense. These modes correspond to special discrete frequencies at which the black hole potential becomes reflectionless. That is, a wave sent from infinity toward the black hole at this special frequency will be entirely absorbed, with no reflection, and vice versa. 

Mathematically, both QNMs and algebraically special modes obey similar-looking boundary conditions: they are purely ingoing at the black hole horizon and purely outgoing at spatial infinity. However, the crucial distinction is that QNMs describe damped oscillations with complex frequencies and non-zero reflection, whereas algebraically special modes are typically associated with purely imaginary frequencies and zero reflection. In fact, at the algebraically special frequency, the reflection coefficient vanishes, which implies that the effective potential becomes transparent for waves at this specific energy.

One of the earliest and most influential studies of such modes is due to Chandrasekhar, who demonstrated their existence in the Schwarzschild geometry through a remarkable factorization of the Regge-Wheeler and Zerilli equations~\cite{Chandrasekhar:1984}. These results were further refined in later works such as~\cite{MaassenvandenBrink:2000iwh}, which examined the analytic structure of the Green's function and showed that these modes, while satisfying the same asymptotic boundary conditions as QNMs, do not appear as poles in the Green’s function and therefore do not dominate the time evolution of perturbations.

Thus, algebraically special modes can be viewed as a mathematical curiosity that provides insights into the symmetry structure of black hole perturbation theory. However, they do not generally correspond to physically excited modes in dynamical processes such as black hole mergers. This distinction is summarized in Table I.

\begin{widetext}
\begin{table*}
\centering
\renewcommand{\arraystretch}{1.3}
\begin{tabular}{|l|c|c|}
\hline
\textbf{Feature} & \textbf{QNMs} & \textbf{Algebraically Special Modes} \\
\hline
Horizon behavior & Ingoing & Ingoing \\
Infinity behavior & Outgoing & Outgoing \\
Frequency type & Complex (damped) & Usually purely imaginary \\
Reflection coefficient & Non-zero & Zero (reflectionless) \\
Excitation in dynamics & Yes (dominates late-time signal) & No (typically not excited) \\
Appears in Green’s function & Yes & No \\
Physical interpretation & Observable in ringdown & Mathematical symmetry/structure \\
\hline
\end{tabular}
\caption{Comparison between quasinormal modes and algebraically special modes.}
\label{tab:QNM_vs_AS}
\end{table*}
\end{widetext}

In summary, while algebraically special modes formally satisfy the same boundary conditions as quasinormal modes and share certain spectral properties, they do not appear in the physical spectrum that governs the time evolution of perturbations. Nevertheless, their role in the mathematical structure of black hole perturbation theory—especially their connection to Darboux transformations and potential isospectrality—makes them a valuable topic of theoretical interest.

\section{Particular cases in theories with higher curvature corrections and modified theories of gravity}\label{sec:exact}

An intriguing exactly solvable case for quasinormal modes (QNMs) appears in the context of Einstein-Gauss-Bonnet (EGB) gravity with a negative cosmological constant. In particular, for a special value of the Gauss-Bonnet coupling constant $\alpha = R^2/2$, where $R$ is the AdS radius, the metric of a $D$-dimensional EGB-AdS black hole simplifies considerably and allows for an analytical treatment of perturbations~\cite{Gonzalez:2017gwa}.

The background metric in this case is given by
\begin{equation}
ds^2 = -f(r)dt^2 + \frac{dr^2}{f(r)} + r^2 d\Omega_{D-2}^2,
\end{equation}
where
\begin{equation}
f(r) = \frac{r^2}{R^2} \left( 1 - \left(\frac{r_H}{r} \right)^{D-1} \right),
\end{equation}
and $r_H$ denotes the event horizon. This is structurally equivalent to the BTZ-like form, with the effective potential taking a form that allows for exact solutions.

The scalar field perturbations in this geometry are governed by the standard Klein-Gordon equation:
\begin{equation}
\frac{1}{\sqrt{-g}} \partial_\mu \left( \sqrt{-g} g^{\mu\nu} \partial_\nu \phi \right) = m^2 \phi,
\end{equation}
which, after separation of variables and transforming to the tortoise coordinate $r_*$, reduces to a Schr\"odinger-like equation.

In this specific background, the radial part of the perturbation equation transforms into a hypergeometric equation, enabling exact determination of the QNMs. The solution satisfying the boundary conditions of ingoing waves at the horizon and vanishing at the AdS boundary leads to the discrete spectrum of quasinormal frequencies~\cite{Gonzalez:2017gwa}:
\[
\omega = -\frac{i}{R^2} \left( 2 r_H (2 + n) \pm \sqrt{R^2 (2 + \ell) + 4 r_H^2} \right),
\]
where \( r_H \) is the event horizon radius, \( n \) the overtone number, and \( \ell \) the multipole number. This expression demonstrates how the imaginary part increases linearly with \( n \), reflecting the increasing damping of higher overtones. The presence of the square root reflects the influence of both the spacetime curvature and the angular momentum on the oscillatory part of the spectrum.

These analytic results not only offer a rare opportunity to probe the spectrum of higher-curvature black holes in exact form but also reveal the onset of eikonal instability at large \( \ell \) \cite{Takahashi:2011du,Takahashi:2011qda,Takahashi:2010gz,Dotti:2004sh,Dotti:2005sq,Gleiser:2005ra} \cite{Cuyubamba:2016cug,Konoplya:2017zwo,Konoplya:2017ymp,Konoplya:2008ix,Konoplya:2008ix}, as the square root term can become imaginary depending on the parameters.  It is worth mentioning that similar simplification, allowing for exact solutions, takes place in 4-dimensional Lifshitz Black Hole and topological BHs \cite{Gonzalez:2012de,Becar:2012bj,Catalan:2013eza}.

\section{Eikonal Regime and First-Order WKB Approximation}\label{sec:eikonal}

In the study of black hole perturbations, one particularly useful approximation arises in the eikonal regime, where the angular momentum number $\ell$ is large. In this limit, the perturbation equations simplify significantly, and analytical expressions for quasinormal frequencies can often be obtained using the Wentzel–Kramers–Brillouin (WKB) method. The WKB formula for finding quasinormal modes of black holes was first found in \cite{Schutz:1985km} and later developed to higher orders in \cite{Iyer:1986np, Konoplya:2003ii,Konoplya:2004ip,Matyjasek:2017psv}. It has been used in  great number of publications as a quick and relatively accurate method for finding quasinormal modes and grey-body factors  (see, for example, \cite{Zinhailo:2019rwd,Becar:2023zbl,Konoplya:2006ar,Dubinsky:2024aeu,Chen:2023akf,Konoplya:2006rv,Dubinsky:2024hmn,Dubinsky:2024fvi,Al-Badawi:2023lvx,Konoplya:2020jgt,Guo:2023nkd,Skvortsova:2024atk,Kodama:2009bf,Skvortsova:2023zmj,Skvortsova:2024eqi,Churilova:2021tgn,Lutfuoglu:2025hjy,Hamil:2025cms,Gong:2023ghh} and references therein). It should be noted that the WKB method is applicable only in spacetimes that are asymptotically flat or de Sitter, where the boundary condition corresponds to a purely outgoing wave—either at spatial infinity or at the cosmological horizon. In asymptotically anti--de Sitter (AdS) spacetimes, the boundary at infinity is timelike and requires different boundary conditions (typically Dirichlet), which are incompatible with the assumptions of the WKB approximation.

The general form of the master wave equation for linear perturbations of black holes is
\begin{equation}
\label{eq:wave_eq}
\frac{d^2 \Psi}{dr_*^2} + \left[ \omega^2 - V(r) \right] \Psi = 0,
\end{equation}
where $r_*$ is the tortoise coordinate defined by $dr_*/dr = 1/f(r)$ and $V(r)$ is the effective potential which depends on the spin of the field and the black hole background. In the eikonal limit ($\ell \gg 1$), the potential typically forms a single peak, making it amenable to a WKB-type treatment.

At first order, the WKB method gives the following quantization condition for quasinormal modes \cite{Schutz:1985km}:
\begin{equation}
\label{eq:wkb1}
\frac{i(\omega^2 - V_0)}{\sqrt{-2V_0''}} = n + \frac{1}{2}, \quad n = 0, 1, 2, \dots,
\end{equation}
where $V_0$ is the maximum of the potential and $V_0''$ is the second derivative with respect to $r_*$ evaluated at the peak. Solving for $\omega$ yields:
\begin{equation}
\label{eq:wkb1sol}
\omega = \sqrt{V_0} - i \left( n + \frac{1}{2} \right) \sqrt{ -\frac{V_0''}{2 V_0} }.
\end{equation}

This formula provides analytic access to the quasinormal spectrum at high $\ell$ and gives reasonably accurate results even at moderate $\ell$, especially for low overtones.

\textbf{Schwarzschild Black Hole.} For the Schwarzschild black hole, the effective potential for scalar and gravitational perturbations in the eikonal limit is:
\begin{equation}
V(r) = f(r) \frac{\ell(\ell+1)}{r^2}, \qquad f(r) = 1 - \frac{2M}{r}.
\end{equation}
This potential attains its maximum at
\begin{equation}
r_0 = 3M,
\end{equation}
which corresponds to the location of the unstable circular photon orbit (though we postpone the geometrical interpretation to the next section).

Evaluating $V_0$ and $V_0''$ at $r_0 = 3M$, one obtains the well-known analytic expression for quasinormal modes in the eikonal regime \cite{Mashhoon:1982im,Ferrari:1984zz}:
\begin{equation}
\omega = \frac{1}{3\sqrt{3}M} \left( \ell + \frac{1}{2} - i \left( n + \frac{1}{2} \right) \right) + \mathcal{O}\left( \frac{1}{\ell} \right).
\end{equation}

\textbf{Kerr Black Hole (Slow Rotation).} For the slowly rotating Kerr black hole, the perturbation equations are more involved but can still be treated analytically in the eikonal limit. In this case, the quasinormal frequencies for a perturbation with azimuthal number $m$ and multipole number $\ell$ are approximated as \cite{Mashhoon:1982im, Ferrari:1984zz}:
\begin{equation}
\omega \approx \omega_{\text{Sch}} + a \, m \, \delta\omega + \mathcal{O}(a^2),
\end{equation}
where $\omega_{\text{Sch}}$ is the Schwarzschild eikonal frequency, and the correction $\delta\omega$ depends on the field's spin and parity. The linear term in $a$ reflects the Zeeman-like splitting of modes due to the black hole's rotation.

Eikonal formulas derived for various black hole models, fields' configurations and gravitational theories are summarized in table \ref{tab:gravity_theories}.

\vspace{1em}

The derivation and physical interpretation of the eikonal formula in terms of the photon sphere and its instability properties will be discussed in detail in the next section, where we explore the correspondence between quasinormal modes and null geodesics.

\begin{table}[h!]
\centering
\renewcommand{\arraystretch}{1.3} 
\begin{tabular}{|l|l|}
\hline
\hline
\textbf{Gravitational Theory} & \textbf{Publication} \\
\hline
Schwarzschild, Kerr, R-N & \cite{Mashhoon:1982im, Ferrari:1984zz, Andersson:1996xw, Malik:2024voy} \\
Einstein-Gauss-Bonnet & \cite{Konoplya:2004xx,Abdalla:2005hu,Konoplya:2017wot,Konoplya:2020bxa,Konoplya:2020der}  \\
Einstein-scalar-Gauss-Bonnet & \cite{Konoplya:2019hml,Paul:2023eep}\\
Einstein-Weyl Gravity & \cite{Kokkotas:2017zwt,Zinhailo:2018ska} \\
Dilaton Gravity & \cite{Konoplya:2001ji,Kokkotas:2015uma,Konoplya:2022zav,Malybayev:2021lfq} \\
LQG-inspired metrics & \cite{Konoplya:2025hgp,Konoplya:2024lch,Bolokhov:2023bwm,Skvortsova:2024atk,Malik:2024elk} \\
Starobinsky-Bel-Robinson gravity & \cite{Bolokhov:2023dxq} \\
Tangherlini metric & \cite{Konoplya:2003ii} \\
BH in the Goedel Universe & \cite{Konoplya:2005sy} \\
brane-world models & \cite{Kanti:2005xa,Malik:2024itg}  \\
Preston-Poisson BH & \cite{Konoplya:2012vh} \\
SdS & \cite{Zhidenko:2003wq,Konoplya:2023moy} \\
Kazakov-Solodukhin BH & \cite{Konoplya:2019xmn} \\
general parametrized BHs & \cite{Konoplya:2020hyk,Churilova:2019jqx,Dubinsky:2024rvf,Konoplya:2023owh} \\
BH in galactic halo & \cite{Konoplya:2021ube,Konoplya:2022hbl} \\
Effective Field Theory & \cite{Konoplya:2023ppx} \\
T-duality inspired BHs & \cite{Konoplya:2023ahd} \\
Regular BHs in higher curvature gravity & \cite{Konoplya:2024hfg} \\
pure Weyl gravity & \cite{Konoplya:2025mvj} \\
BHs with anisotropic fluid & \cite{Bolokhov:2022rqv} \\
scalar-tensor theories & \cite{Glampedakis:2019dqh,Silva:2019scu,Bryant:2021xdh} \\
Regular BHs in ASG & \cite{Konoplya:2022hll} \\
Regular BHs in electrodynamics & \cite{Skvortsova:2024wly,Toshmatov:2015wga} \\
M-T wormholes & \cite{Konoplya:2018ala,Malik:2024wvs} \\
scalar coupled to Einstein tensor & \cite{Konoplya:2018qov} \\
\hline
\hline
\end{tabular}
\caption{Examples of gravitational theories and corresponding key publications on the eikonal limit of quasinormal modes.}
\label{tab:gravity_theories}
\end{table}

\section{Eikonal Quasinormal Modes and Null Geodesic Correspondence}\label{sec:eikonal2}

\begin{table*}
\centering
\begin{tabular}{|c|c|c|}
\hline
\textbf{Quantity} & \textbf{Expression} & \textbf{Depends on} \\
\hline
Photon sphere radius \( r_c \) & \( \displaystyle \frac{d}{dr} \left( \frac{f(r)}{r^2} \right) = 0 \) & Geometry \( f(r) \) \\
\hline
Angular frequency \( \Omega_c \) & \( \displaystyle \left. \frac{\sqrt{f(r)}}{r} \right|_{r = r_c} \) & \( f(r) \) at \( r_c \) \\
\hline
Lyapunov exponent \( \lambda \) & \( \displaystyle \left. \sqrt{ \frac{g(r)}{2} \left( \frac{2f}{r^2} - \frac{f''}{f} + \left( \frac{f'}{f} \right)^2 \right) } \right|_{r = r_c} \) & \( f(r), g(r) \) at \( r_c \) \\
\hline
Eikonal QNM frequency \( \omega \) & \( \displaystyle \omega = \Omega_c \ell - i\left(n + \frac{1}{2} \right)\lambda \) & Photon orbit properties \\
\hline
Shadow impact parameter \( b_c \) & \( \displaystyle \left. \frac{r}{\sqrt{f(r)}} \right|_{r = r_c} \) & \( f(r) \) at \( r_c \) \\
\hline
\end{tabular}
\caption{Quantities governing eikonal QNMs and the black hole shadow, all derived from the photon sphere.}
\end{table*}

In the eikonal (large-\( \ell \)) limit, the quasinormal mode (QNM) spectrum and the black hole shadow are both governed by the geometry of unstable circular null geodesics (the photon sphere). For a general static, spherically symmetric metric of the form
\[
ds^2 = -f(r)\,dt^2 + \frac{dr^2}{g(r)} + r^2 d\Omega^2,
\]
perturbations of massless fields are governed by a wave equation that reduces to a Schr\"odinger-like form for the radial part:
\[
\frac{d^2 \Psi}{dr_*^2} + \left[ \omega^2 - V_{\text{eff}}(r) \right] \Psi = 0,
\]
where the tortoise coordinate is defined by
\[
\frac{dr_*}{dr} = \frac{1}{\sqrt{f(r) g(r)}}.
\]
In the eikonal limit \( \ell \gg 1 \), the effective potential is approximated as
\[
V_{\text{eff}}(r) \approx f(r) \frac{\ell(\ell+1)}{r^2}.
\]
This potential has a peak near the radius of the unstable circular photon orbit \( r = r_c \), and the QNMs can be computed using the first-order WKB formula (e.g., Schutz and Will, 1985):
\[
\frac{i(\omega^2 - V_0)}{\sqrt{-2V_0''}} = n + \frac{1}{2},
\]
where \( V_0 = V_{\text{eff}}(r_c) \) and \( V_0'' \) is the second derivative of the potential with respect to the tortoise coordinate, evaluated at \( r = r_c \).

Solving for \( \omega \), one obtains:
\[
\omega = \sqrt{V_0} - i \left( n + \frac{1}{2} \right) \sqrt{ -\frac{V_0''}{2V_0} }.
\]
Using the eikonal approximation \( \ell \gg 1 \), we have:
\[
V_0 = f(r_c) \frac{\ell^2}{r_c^2}, \qquad \sqrt{V_0} = \ell \left. \frac{\sqrt{f(r)}}{r} \right|_{r = r_c} \equiv \Omega_c \ell.
\]
Similarly, one can show (see ~\cite{Cardoso:2008bp}) that the second derivative of the potential in tortoise coordinates yields the Lyapunov exponent:
\[
\lambda = \left. \sqrt{ \frac{g(r)}{2} \left( \frac{2f(r)}{r^2} - \frac{f''(r)}{f(r)} + \left( \frac{f'(r)}{f(r)} \right)^2 \right) } \right|_{r = r_c}.
\]
Substituting into the WKB result, the QNM frequencies take the form \cite{Cardoso:2008bp}:
\[
\omega_{\text{QNM}} = \Omega_c \ell - i \left( n + \frac{1}{2} \right) \lambda,
\]
where:
\[
\Omega_c = \left. \frac{\sqrt{f(r)}}{r} \right|_{r = r_c}.
\]
The same photon orbit determines the black hole shadow observed at infinity. The critical impact parameter for a massless particle approaching the black hole is:
\[
b_c = \left. \frac{r}{\sqrt{f(r)}} \right|_{r = r_c}.
\]
This corresponds to the apparent radius of the shadow \( R_{\text{sh}} = b_c \) for an observer at asymptotic infinity. Hence, both the real part of the QNM frequency (via \( \Omega_c \)) and the size of the shadow (via \( b_c \)) are determined by the geometry of the photon sphere  \cite{Jusufi:2019ltj,Jusufi:2020dhz}.

This geometric interpretation provides a powerful link between QNMs and the geodesic structure of spacetime, and has been verified in many cases including Schwarzschild and Kerr black holes. However, this correspondence is not universally valid.

In particular, the correspondence relies on the form of the effective potential being dominated by a standard centrifugal term $\propto \ell^2/r^2$ in the eikonal regime. If the potential is modified—for example, by higher curvature corrections or matter couplings—the centrifugal term may acquire $\ell$-independent contributions or deviate from the usual form, breaking the assumptions of the WKB expansion and thus invalidating the geodesic correspondence.

In~\cite{Konoplya:2017wot,Konoplya:2019hml,Konoplya:2020bxa}, it was shown that for Einstein-Gauss-Bonnet black holes, the eikonal QNM frequencies of gravitational perturbations do not coincide with the predictions from the geodesic correspondence, due to the modification of the effective potential by the higher curvature terms. This effect is usually absent for test fields in asymptotically flat black hole background, where the correspondence still holds, as was shown in various theories of gravity \cite{Zinhailo:2019rwd,Malik:2024elk,Malik:2024tuf,Malik:2024qsz,Konoplya:2025hgp,Bolokhov:2023bwm}.

Further refinements were given in~\cite{Konoplya:2022gjp}, where it was demonstrated that in asymptotically de Sitter spacetimes, an additional “de Sitter branch” of QNMs exists, for which the correspondence with photon spheres fails. The standard correspondence applies only to the Schwarzschild-like branch. A more recent work~\cite{Bolokhov:2023dxq} emphasized that when the effective potential lacks a well-defined peak or if the Taylor expansion near the peak fails to match the full WKB series, the correspondence becomes unreliable.

The eikonal regime may also signal the onset of a catastrophic dynamical instability, rendering the linear approximation invalid~\cite{Dotti:2004sh,Dotti:2005sq,Takahashi:2011du,Konoplya:2017lhs}, and consequently breaking the correspondence with null geodesics. This so-called \emph{eikonal instability} arises at high multipole numbers \( \ell \), since in certain modified gravity theories, increasing \( \ell \) not only raises the potential barrier but also deepens a negative gap in the effective potential. This structure allows for the formation of bound states with negative energy, triggering the instability~\cite{Konoplya:2017lhs}. 

Thus, while the geodesic-QNM correspondence is elegant and holds in many cases, its validity is ultimately contingent upon the applicability of the WKB method and the dominance of a standard centrifugal barrier in the effective potential.

\section{Beyond the eikonal limit}\label{sec:beyondeikonal}

A significant step toward obtaining analytic results for quasinormal modes (QNMs) and grey-body factors of black holes was proposed in~\cite{Konoplya:2023moy}. There, the authors extended the WKB approach beyond the eikonal limit to construct compact analytic expressions that remain remarkably accurate even for low multipole numbers $\ell$. The general strategy is to match the Taylor expansion of the effective potential near its peak with the asymptotic boundary conditions, order by order in inverse powers of $\kappa = \ell + 1/2$.

The starting point is the wave-like equation
\begin{equation}
\frac{d^2\Psi}{dr_*^2} + \left( \omega^2 - V(r_*) \right)\Psi = 0,
\end{equation}
with QNM boundary conditions:
\begin{equation}\label{BCs1}
\Psi(r_* \to -\infty) \propto e^{-i\omega r_*}, \quad \Psi(r_* \to +\infty) \propto e^{+i\omega r_*}.
\end{equation}
In this framework, the first-order WKB approximation gives the leading-order eikonal result. However, the authors of~\cite{Konoplya:2023moy} systematically expanded the frequency $\omega$ in powers of $\kappa^{-1}$:
\begin{equation}
\omega = \Omega \kappa - i \lambda K + \sum_{n=1}^{\infty} \frac{C_n}{\kappa^n},
\end{equation}
where $\Omega$ and $\lambda$ are respectively related to the real and imaginary parts of the eikonal QNM frequency, and $K = n + 1/2$ is the overtone index.

Applying this expansion to Schwarzschild-de Sitter spacetime, they obtained analytic formulas for scalar, electromagnetic, Dirac, and gravitational perturbations up to sixth order in $\kappa^{-1}$. For example, for gravitational perturbations, the analytic expression reads:
$$\omega = \frac{\varsigma}{3\sqrt{3}M}\left[ \kappa - \frac{7\varsigma^2}{432\kappa} - \frac{5K^2 \varsigma^2}{36\kappa} \right] -$$
\begin{equation}
 iK \frac{\varsigma}{3\sqrt{3}M} \left[ 1 - \frac{7\varsigma^2}{216\kappa^2} + \frac{385\varsigma^4}{15552\kappa^2} + \frac{235K^2\varsigma^4}{3888\kappa^2} \right] + \mathcal{O}\left( \kappa^{-3} \right),
\end{equation}
where $\varsigma = \sqrt{1 - 9\Lambda M^2}$. Similar expressions were obtained for Dirac and Maxwell fields. Despite being derived through an asymptotic expansion, these formulas yield excellent agreement with exact numerical values even for low $\ell$ (as demonstrated in detailed tables comparing with Page's results).

An important extension of this method was its application to black holes in effective field theories  \cite{Konoplya:2023ppx}, where an analogous expansion in $\kappa^{-1}$ was constructed for the modified effective potential. Furthermore, the method was also adapted to compute analytic grey-body factors using similar expansions. In particular, the authors proposed a technique to extrapolate the WKB-derived transmission coefficients into the infrared and ultraviolet regimes, improving the accuracy of Hawking radiation spectra computed from analytic approximations.

This analytic formalism is not only efficient but also flexible, as evidenced by its implementation in a publicly available Mathematica notebook \cite{Konoplya:2023moy}. The method has since been applied to various modified gravity models, including scalar-tensor, Einstein-dilaton-Gauss-Bonnet, and regular black hole spacetimes~\cite{Konoplya:2025hgp,Konoplya:2024lch,Malik:2024wvs,Malik:2024tuf,Malik:2024bmp,Malik:2024sxv,Bolokhov:2024ixe,Malik:2024voy,2753764,Malik:2024elk,Malik:2024qsz,Malik:2024zoo}.
The same approach can be applied to finding the analytic expressions for grey-body factors \cite{Dubinsky:2024nzo}. The same boundary conditions~\eqref{BCs1} apply to asymptotically flat or de Sitter wormholes, where the purely incoming wave at minus infinity now corresponds to a wave traversing the wormhole throat and propagating into the other universe~\cite{Konoplya:2005et}. Analytic quasinormal modes beyond the eikonal limit for various traversable wormhole geometries~\cite{Bronnikov:1973fh,Morris:1988cz} have been recently obtained in~\cite{Malik:2024wvs}.

\section{Near-extreme black holes}

In this section, we review quasinormal modes of black holes that are not asymptotically flat but instead asymptotically de Sitter. The spectrum of such black holes exhibits several distinctive features, and a substantial body of literature is devoted to the study of their quasinormal frequencies~\cite{Konoplya:2004uk,Dyatlov:2011jd,Dyatlov:2011zz,Hintz:2016gwb,Konoplya:2007zx,Jansen:2017oag,Jing:2003wq,Konoplya:2022xid,Aragon:2020qdc,Dubinsky:2024gwo,Konoplya:2025mvj}, including their implications for the Strong Cosmic Censorship conjecture~\cite{Cardoso:2017soq,Dias:2018ynt,Dias:2018etb,Mo:2018nnu,Konoplya:2022kld}. In the near-extremal limit, where the event horizon approaches the cosmological (de Sitter) horizon, the spectral problem simplifies dramatically and admits exact analytic solutions.

\subsection{Bosonic fields in the background of near-extremal Schwarzschild--de Sitter Black Holes}

A remarkable regime in which analytic expressions for quasinormal modes can be obtained is the near-extremal limit of the Schwarzschild--de Sitter (SdS) black hole. In this limit, the black hole horizon \( r_b \) approaches the cosmological horizon \( r_c \), i.e.,
\[
\frac{r_c - r_b}{r_b} \ll 1.
\]
In such cases, the effective potential for scalar, electromagnetic, and gravitational perturbations simplifies drastically and becomes exactly solvable, as first shown by Cardoso and Lemos~\cite{Cardoso:2003sw}.

The SdS metric is given by the following metric function:
\[
f(r) = 1 - \frac{2M}{r} - \frac{r^2}{a^2}, \quad a^2 = \frac{3}{\Lambda}.
\]
In the near-extremal limit, \( f(r) \) can be approximated in the region between the two nearly coinciding horizons as
\[
f(r) \approx \frac{(r - r_b)(r_c - r)}{r_b^2},
\]
which leads to a remarkably simple form for the wave equation:
\[
\frac{d^2 \phi}{dr_*^2} + \left[\omega^2 - \frac{V_0}{\cosh^2(\kappa_b r_*)} \right] \phi = 0,
\]
where \( r_* \) is the tortoise coordinate, \( \kappa_b \) is the surface gravity of the black hole horizon, and \( V_0 \) is a constant depending on the type of perturbation:
\[
V_0 =
\begin{cases}
\kappa_b^2 \ell(\ell+1), & \text{scalar and electromagnetic}, \\
\kappa_b^2 (l+2)(l-1), & \text{gravitational}.
\end{cases}
\]
The potential above is the well-known \emph{P\"oschl--Teller potential}, whose quasinormal spectrum is known exactly~\cite{Ferrari:1984zz}:
\[
\omega = \kappa_b \left[ -i \left( n + \frac{1}{2} \right) + \sqrt{\frac{V_0}{\kappa_b^2} - \frac{1}{4}} \right],
\quad n = 0, 1, 2, \dots.
\]

Explicitly, this yields:
\begin{align*}
\omega &= \kappa_b \left[ -i\left(n + \frac{1}{2}\right) + \sqrt{\ell(\ell + 1) - \frac{1}{4}}\, \right] \quad \text{(scal., el-mag.)}, \\
\omega &= \kappa_b \left[ -i\left(n + \frac{1}{2}\right) + \sqrt{(l+2)(l-1) - \frac{1}{4}}\, \right] \quad \text{(grav.)}.
\end{align*}

These expressions are \emph{exact} in the near-extremal limit and provide a rare analytic handle on the entire QNM spectrum, including highly damped modes. The exact solvability also explains the numerical success of earlier studies where the SdS potential was fitted to the P\"oschl--Teller form~\cite{Moss:2001ga}.

The appearance of this solvable potential is geometrically tied to the fact that, in the near-extremal limit, the region between the two horizons becomes a narrow throat, akin to a nearly symmetric potential well in the tortoise coordinate. The wave equation thus mimics that of quantum mechanical scattering in a hyperbolic potential.

This approach was extended to the D-dimensional charged black hole in \cite{Molina:2003ff} and a generalized P\"oschl--Teller potential was introduced in \cite{Cardona:2017scd}.

\subsection{Bosonic and Fermionic Fields in the Kerr-Newman-de Sitter background: Exact Solutions Beyond the P\"oschl--Teller Potential}

While scalar, electromagnetic, and gravitational perturbations of the near-extremal Schwarzschild--de Sitter black hole reduce to a wave equation with the P\"oschl--Teller potential, fermionic fields do not. As shown in~\cite{Churilova:2021nnc}, the effective potentials for Dirac and Rarita--Schwinger fields take a more complicated form which, nevertheless, still admits an exact analytic treatment.

The Dirac equation in this background yields the effective potential
\begin{widetext}
\[
V_{\pm 1/2}(r_*) = \kappa_e^2 \left( \ell + \frac{1}{2} \right) \left[ \ell + \frac{1}{2} \mp \sinh(\kappa_e r_*) \right] \frac{1}{\cosh^2(\kappa_e r_*)} + \mathcal{O}(\kappa_e^3),
\]
which is manifestly \emph{not} of the P\"oschl--Teller form due to the asymmetry introduced by the hyperbolic sine. Similarly, the Rarita--Schwinger field exhibits an analogous potential structure (see Appendix A of~\cite{Churilova:2021nnc}).

Despite this, using a Frobenius expansion and analytic continuation methods, it was shown that the QNM spectrum can still be expressed analytically:
\[
\frac{\omega}{\kappa_e} = \pm \sqrt{(\ell + s)(\ell + 1 - s) + \frac{(2s - 1)(2s - 5)}{12}} - i \left(n + \frac{1}{2} \right) + \mathcal{O}(\kappa_e),
\quad n = 0, 1, 2, \dots,
\]
\end{widetext}
where \( s = \pm 1/2 \) or \( \pm 3/2 \) is the spin of the field. This expression reduces to the P\"oschl--Teller result for \( s = 1 \) and \( s = 2 \), and provides new exact QNM frequencies for half-integer spin fields.

Interestingly, the deviation from the P\"oschl--Teller form does not obstruct solvability, because in the near-extremal limit \( \kappa_e \ll 1 \), the wave equation simplifies sufficiently for a closed-form Frobenius solution to truncate. Thus, exact results can still be obtained, even though the effective potential lacks full symmetry.

This case highlights that analytic solvability in the near-extremal regime is not exclusive to P\"oschl--Teller-type potentials, but can emerge from other structural simplifications, as is the case for fermions. Furthermore, this analysis has been extended to include charged and rotating de Sitter black holes, allowing analytic QNM formulas for arbitrary spin fields in the near-extremal Kerr--Newman--de Sitter geometry~\cite{Churilova:2021nnc}.

\section{Quasinormal Modes in the Large Field Mass Regime}

Massive fields play an important role in the study of quasinormal modes for several theoretical and phenomenological reasons. First, many well-motivated extensions of the Standard Model predict the existence of massive bosonic fields, including scalar (e.g., dilaton, axion-like), vector (Proca), and tensor fields. In the context of black hole perturbation theory, massive fields introduce qualitatively new features into the spectrum, such as the possibility of quasi-bound states and long-lived modes, often referred to as ``quasiresonances''  \cite{Ohashi:2004wr,Konoplya:2004wg,Konoplya:2006br,Zinhailo:2024jzt,Bolokhov:2023bwm,Konoplya:2017tvu,Konoplya:2005hr,Lutfuoglu:2025hjy}.  These modes can dominate the late-time response of the black hole and may have observable astrophysical consequences \cite{Konoplya:2023fmh}. Moreover, the interaction between black holes and massive fields is relevant for models of dark matter, especially ultralight bosons, which could form clouds around rotating black holes through superradiance \cite{Brito:2015oca,Annulli:2020lyc,Chung:2021roh}. Understanding the quasinormal spectrum of massive fields in realistic spacetimes is therefore essential for interpreting gravitational wave signals and probing fundamental physics in the strong-field regime. In addition, even originally massless fields can behave as effectively massive in the presence of strong magnetic fields \cite{Kokkotas:2010zd,Konoplya:2008hj,Konoplya:2007yy,Davlataliev:2024mjl} or in certain extra-dimensional scenarios  \cite{Seahra:2004fg} where tidal forces modify the dispersion relation. Finally, the late-time behavior of massive fields differs significantly from the massless case: instead of pure power-law tails, the field exhibits slowly decaying oscillatory tails that can dominate the signal at late times \cite{Koyama:2001ee,Koyama:2000hj,Konoplya:2006gq,Rogatko:2007zz,Gibbons:2008gg,Konoplya:2024wds,Konoplya:2013rxa,Moderski:2001tk,Jing:2004zb,Konoplya:2006gq,Churilova:2019qph}.

In addition to the eikonal limit and near-extremal regimes, analytic expressions for quasinormal modes (QNMs) can also be obtained when the mass \( \mu \) of the perturbing field is large compared to the black hole mass \( M \), i.e., \( \mu M \gg 1 \). In this limit, the effective potential for a massive scalar field in the Schwarzschild--de Sitter background develops a single sharp peak, making it amenable to a WKB treatment similar to that used in the large-\( \ell \) expansion.

The Klein--Gordon equation for a massive scalar field in this background leads to the standard wave-like equation where  the effective potential reads
\[
V(r) = f(r) \left( \frac{\ell(\ell+1)}{r^2} + \mu^2 + \frac{f'(r)}{r} \right).
\]

For \( \mu M \gg 1 \), the potential becomes dominated by the mass term and possesses a single maximum between the event and cosmological horizons. This justifies the application of the WKB method and an expansion in powers of \( 1/\mu \). Following~\cite{Konoplya:2024ptj}, one can express the QNM frequencies as
\begin{widetext}
\[
\omega_n = \mu \sqrt{1 - \sigma^2}
- i \frac{K \sigma^3 \sqrt{1 - \sigma^2}}{3M}
+ \mathcal{O}\left( \frac{1}{\mu} \right),
\quad
K = n + \frac{1}{2},
\]
where
\[
\sigma = (9M^2 \Lambda)^{1/6}.
\]
Higher-order terms in \( 1/\mu \) can also be systematically included. Up to second order, the expansion reads
\begin{align*}
\omega_n =\; & \mu \sqrt{1 - \sigma^2}
- i \frac{K \sigma^3 \sqrt{1 - \sigma^2}}{3M} \\
& - \frac{\sigma^4 \sqrt{1 - \sigma^2}}{1296 M^2 \mu}
\left( -72\ell^2 - 72\ell + 12K^2(\sigma^2 - 1) + 29\sigma^2 - 11 \right) \\
& + i \frac{K \sigma^5 (1 - \sigma^2)^{3/2}}{46656 M^3 \mu^2}
\left( 864\ell^2 + 864\ell + 76K^2(\sigma^2 - 1) + 865\sigma^2 + 167 \right)
+ \mathcal{O}\left( \frac{1}{\mu^3} \right).
\end{align*}

\end{widetext}
These expressions are in excellent agreement with both numerical WKB results and time-domain simulations.

This large-mass expansion is conceptually similar to the eikonal expansion in \( 1/\ell \), but more suited for fields with high Compton frequency \( \mu \gg 1/M \). It also demonstrates the emergence of oscillatory exponential decay at late times for massive fields, in contrast to the purely exponential tails in the massless case. Such analytic approximations hold promise for modeling black hole perturbations from realistic, massive fields (e.g., ultralight scalars or massive standard model fields), particularly in astrophysical or cosmological contexts where \( \Lambda \neq 0 \). This approach of expansion in terms of $1/\mu$ was extended to the charged black holes in \cite{Konoplya:2024ptj,Bolokhov:2024ixe}.

\section{Beyond-Eikonal Analytic Quasinormal Modes of Parametrized Spherically Symmetric Black Holes}

In many cases, quasinormal modes (QNMs) in the eikonal limit \( \ell \gg 1 \) can be related to properties of null geodesics. However, to obtain more accurate analytic expressions for lower multipoles, one can systematically go beyond the leading order in \( 1/\ell \). This is especially useful when the black hole spacetime is not known in closed form, but is given in a general parametrized way. One powerful and systematic framework is the Rezzolla--Zhidenko parametrization~\cite{Rezzolla:2014mua}, which provides a convergent expansion for any spherically symmetric and asymptotically flat black hole.

This parametrization combines two types of expansion: a \( 1/r \)-expansion at large distances and a continued-fraction expansion near the event horizon. As such, it provides a general and systematic framework for describing asymptotically flat, spherically symmetric black hole spacetimes in arbitrary metric theories of gravity. Due to its flexibility and convergence properties, the parametrization has been extended to include axially symmetric and higher-dimensional black holes~\cite{Konoplya:2016jvv,Younsi:2016azx,Konoplya:2020kqb}, as well as other compact objects~\cite{Bronnikov:2021liv}. It has been employed in a wide range of recent studies to analyze various physical phenomena around black holes and to construct analytic approximations to numerically obtained black hole solutions~\cite{Kocherlakota:2020kyu,Zhang:2024rvk,Cassing:2023bpt,Li:2021mnx,Ma:2024kbu,Shashank:2021giy,Konoplya:2021slg,Kokkotas:2017ymc,Yu:2021xen,Konoplya:2021qll,Toshmatov:2023anz,Konoplya:2019fpy,Nampalliwar:2019iti,Ni:2016uik,Konoplya:2018arm,Konoplya:2022iyn,Konoplya:2022tvv,Konoplya:2019ppy,Konoplya:2019goy}. Remarkably, in many cases, only the first few terms in the expansion are sufficient to approximate the black hole metric with high accuracy~\cite{Konoplya:2020hyk}, which is why truncated forms of the parametrized metric have been widely studied in the works cited above.

The line element is written as
\[
ds^2 = -f(r)\,dt^2 + \frac{dr^2}{g(r)} + r^2 d\Omega^2,
\]
with two independent functions \( f(r) \) and \( g(r) \). These are re-expressed via functions \( N(r) \) and \( B(r) \) as
\[
f(r) = N^2(r), \qquad g(r) = \frac{N^2(r)}{B^2(r)}.
\]
A compactified radial coordinate is introduced:
\[
x = 1 - \frac{r_0}{r}, \quad x = 0 \text{ at the horizon}, \quad x = 1 \text{ at infinity}.
\]
The functions \( N(x) \) and \( B(x) \) are expanded as:
\begin{widetext}
\begin{align*}
A(x) &\equiv N^2(x) = x \left[ 1 - \epsilon(1 - x) + (a_0 - \epsilon)(1 - x)^2 + \tilde{A}(x)(1 - x)^3 \right], \\
B(x) &= 1 + b_0(1 - x) + \tilde{B}(x)(1 - x)^2,
\end{align*}
where \( \tilde{A}(x) \) and \( \tilde{B}(x) \) are expressed via continued fractions:
\[
\tilde{A}(x) = \frac{a_1}{1 + \frac{a_2 x}{1 + \frac{a_3 x}{1 + \dots}}}, \quad
\tilde{B}(x) = \frac{b_1}{1 + \frac{b_2 x}{1 + \frac{b_3 x}{1 + \dots}}}.
\]
The parameter \( \epsilon = \frac{2M - r_0}{r_0} \) quantifies deviation from the Schwarzschild horizon radius. The coefficients \( a_0, b_0 \) are related to post-Newtonian parameters, while \( a_i, b_i \) describe higher-order deformations determining the near-horizon behavior.

The radial equation for scalar and electromagnetic and gravitational  fields usually reduces to a wave equation of the form:
\[
\frac{d^2 \Psi}{dr_*^2} + \left[ \omega^2 - V_{s}(r) \right] \Psi = 0, \qquad \frac{dr_*}{dr} = \frac{1}{\sqrt{f(r)g(r)}}.
\]
The effective potential takes the general form:
\[
V_s(r) = \frac{\ell(\ell + 1)f(r)}{r^2} + (1 - s)\left( \frac{g(r)f'(r) + f(r)g'(r)}{2r} - \frac{s(f(r)g(r) - f(r))}{r^2} \right),
\]
\end{widetext}
where \( s = 0, 1, 2 \) corresponds to scalar, electromagnetic, and gravitational perturbations, respectively. Note that in this context we consider only axial gravitational perturbations within Einstein's theory, coupled to a general anisotropic fluid background~\cite{Ashtekar:2018lag,Bouhmadi-Lopez:2020oia,Konoplya:2024lch}. The inclusion of genuinely non-Einsteinian gravity sectors typically leads to a qualitatively different effective potential structure.

To go beyond the eikonal limit, one expands the effective potential near its peak in powers of \( 1/\kappa \), where \( \kappa = \ell + \tfrac{1}{2} \). Using the WKB method, the QNM frequency is expressed as a power series:
\[
\omega = \sum_{n = -1}^\infty \frac{\omega_n}{\kappa^n}.
\]
Retaining corrections up to \( \mathcal{O}(\kappa^{-1}) \), the fundamental mode \( (n = 0) \) frequency for a general parametrized black hole is given analytically as~\cite{Dubinsky:2024rvf}:
\[
\omega = \frac{2\kappa}{3\sqrt{3}} - \frac{2iK}{3\sqrt{3}} + \mathcal{O}(\kappa^{-1}),
\]
with spin corrections and deviations from Schwarzschild encoded via \( \epsilon, a_1, b_1 \), yielding:
\begin{align*}
\omega =\; & \frac{2\kappa}{3\sqrt{3}} - \frac{2iK}{3\sqrt{3}} 
+ \epsilon \left( \cdots \right) + a_1 \left( \cdots \right) \\
& + b_1 \left( \cdots \right) + \mathcal{O}(\kappa^{-2}),
\end{align*}
where \( K = n + \tfrac{1}{2} \), and the full expression appears in Eq.~(15) of~\cite{Dubinsky:2024rvf}.

This formula is remarkably accurate even for low multipoles, especially for scalar and electromagnetic fields. Comparison with numerical WKB-Padé results shows sub-percent deviations for \( \ell = 2 \). For gravitational perturbations, accuracy is slightly reduced, suggesting the benefit of extending to second order in \( 1/\kappa \), which remains tractable and compact.

This method allows for model-independent constraints on QNMs, enabling checks of phenomenological bounds (such as Hod’s inequality \cite{Hod:2006jw,Hod:2006jw}) across a wide class of black holes with parametrically small deviations from Schwarzschild geometry.

\section{Quasinormal Modes in Pure de Sitter and Anti--de Sitter Spacetimes}

While quasinormal modes are typically associated with perturbations of black holes, they can also arise in maximally symmetric spacetimes such as pure de Sitter and anti--de Sitter space. In these cases, the causal boundaries (cosmological or timelike) replace the role of black hole horizons in defining the appropriate boundary conditions for quasinormal ringing.

\subsection*{de Sitter Space}

In the static patch of \(D\)-dimensional de Sitter space, the metric is
\[
ds^2 = \left(1 - \frac{r^2}{L^2}\right) dt^2 - \left(1 - \frac{r^2}{L^2}\right)^{-1} dr^2 - r^2 d\Omega_{D-2}^2,
\]
where \( L = \sqrt{3/\Lambda} \) is the de Sitter radius. Despite the absence of a black hole, this spacetime possesses a cosmological horizon at \( r = L \), which acts as a causal boundary. As a result, one can meaningfully define QNMs as solutions to the wave equations that are regular at the origin and purely outgoing at the cosmological horizon~\cite{Lopez-Ortega:2006aal,Lopez-Ortega:2012xvr,Du:2004jt}.

In this setup, the Klein--Gordon or Maxwell equations reduce to hypergeometric-type equations, leading to the exact QNM spectrum. For a massless field of spin \(s\), the quasinormal frequencies in \(D=4\) take the simple form:
\[
\omega_n^{(\text{dS})} = -\frac{i}{L} \left( \ell + n + 1 - \delta_{s0}\delta_{n0} \right),
\quad n = 0, 1, 2, \dots,
\]
as derived in~\cite{Lopez-Ortega:2006aal} and reaffirmed numerically in~\cite{Konoplya:2022kld}. These frequencies are purely imaginary, reflecting non-oscillatory, exponentially decaying behavior of perturbations.

Furthermore, in the presence of a small black hole (\( r_0 \ll r_c \)), the QNM frequencies approach their pure de Sitter values according to the universal law~\cite{Konoplya:2022kld}:
\[
\omega_n = \omega_n^{(\text{dS})} \left( 1 - \frac{M}{r_c} + \mathcal{O}\left(\frac{M^2}{r_c^2}\right) \right),
\]
indicating that the dominant part of the spectrum depends only on the cosmological scale \( r_c \), and not on the near-horizon geometry. This regime ensures that the damping rates satisfy both Hod's proposal and the Strong Cosmic Censorship bound:
\[
|\text{Im}(\omega)| \ll \frac{\kappa_0}{2} \leq \frac{\kappa_i}{2},
\]
where $\kappa_0$ and $\kappa_i$ are surface gravity at the event and Cuachy horizons respectively.

\subsection*{Anti--de Sitter Space}

In contrast, pure anti--de Sitter (AdS) space is not globally hyperbolic, and its boundary at spatial infinity requires additional conditions to render the wave dynamics well-posed. The metric in static coordinates reads:
\[
ds^2 = \left(1 + \frac{r^2}{L^2} \right) dt^2 - \left(1 + \frac{r^2}{L^2} \right)^{-1} dr^2 - r^2 d\Omega_{D-2}^2.
\]
Here, quasinormal modes are defined as solutions regular at the origin and satisfying Dirichlet, Neumann, or more generally Robin-type boundary conditions at spatial infinity~\cite{Burgess:1984ti}. These modes are especially significant in the context of the AdS/CFT correspondence, where they describe poles in correlation functions of the dual conformal field theory~\cite{Horowitz:1999jd,Kovtun:2004de,Son:2007vk}. 

In global anti--de Sitter (AdS) spacetime, the Klein--Gordon equation for a scalar field admits an exact spectrum of quasinormal (or, more precisely, normal) modes when suitable boundary conditions are imposed at spatial infinity. For Dirichlet boundary conditions, the mode frequencies are discrete and take the form
\[
\omega_n = \frac{2n + \Delta}{L}, \quad n = 0, 1, 2, \dots,
\]
where \( L \) is the AdS curvature radius, and \( \Delta \) is the conformal dimension of the dual operator in the boundary CFT. The value of \( \Delta \) is related to the mass of the bulk scalar field via
\[
\Delta = \frac{D - 1}{2} + \sqrt{\frac{(D - 1)^2}{4} + m^2 L^2}.
\]
This result is derived by solving the Klein--Gordon equation in AdS and identifying the quantization condition that ensures regularity at the origin and vanishing (or appropriate falloff) at the timelike boundary. The general expression, including mass dependence and dimensionality, is presented in Eq.~(51) of~\cite{Lopez-Ortega:2006aal}. For a massless scalar field in four dimensions, this simplifies to \( \Delta = 3 \), recovering the spectrum
\[
\omega_n = \frac{2n + 3}{L}, \quad n = 0, 1, 2, \dots.
\]
When the black hole is immersed in AdS spacetime, the quasinormal modes of black holes transition to the modes of pure AdS spacetime, when the radius of the event horizon is much smaller than the AdS radius \cite{Konoplya:2002zu}.

Overall, the existence of well-defined QNMs in pure de Sitter and AdS spacetimes emphasizes that quasinormal ringing is not exclusive to black holes, but rather a general feature of spacetimes with causal boundaries. These spectra also serve as important benchmarks for testing numerical methods, analytical techniques, and quantum gravity conjectures such as dS/CFT and AdS/CFT.

\section{Conclusions}

In this review, we have surveyed a wide range of analytic results for quasinormal modes of black holes, highlighting the special regimes and symmetries that allow exact or perturbative expressions to be obtained. These include exactly solvable cases in lower dimensions and higher-curvature theories, algebraically special modes, the eikonal regime and its geometric interpretation, near-extremal and large-field-mass limits, and systematic beyond-eikonal expansions in parametrized backgrounds. We also examined the role of boundary conditions in defining QNMs in pure de Sitter and anti–de Sitter spacetimes, where the presence of cosmological or timelike boundaries leads to discrete spectra even in the absence of event horizons.

While our emphasis has been on recent developments and analytically tractable regimes, we did not attempt to cover all known analytic results. In particular, the high-overtone asymptotics of quasinormal modes, including their connection to the area spectrum and quantum gravity proposals \cite{Hod:1998vk,Motl:2003cd,Motl:2002hd,Cardoso:2004up}, have already been comprehensively reviewed in earlier works. 

Analytic methods, though often limited in scope, provide indispensable insights into the dynamics of black hole perturbations and serve as powerful tools for theory development and observational interpretation. Future progress may come from extending these techniques to rotating or dynamical backgrounds, exploring new symmetry reductions, or further connecting analytic QNMs to fundamental aspects of quantum gravity and holography.

\textbf{Acknowledgments.}
The authors are very grateful to R. A. Konoplya and K. A. Bronnikov for the fruitful discussions and helpful advice that provided invaluable assistance in the preparation of this review. This work was supported by RUDN University Project FSSF-2023-0003.
\bibliography{Bibliography}

\begin{thebibliography}{249}%
\makeatletter
\providecommand \@ifxundefined [1]{%
 \@ifx{#1\undefined}
}%
\providecommand \@ifnum [1]{%
 \ifnum #1\expandafter \@firstoftwo
 \else \expandafter \@secondoftwo
 \fi
}%
\providecommand \@ifx [1]{%
 \ifx #1\expandafter \@firstoftwo
 \else \expandafter \@secondoftwo
 \fi
}%
\providecommand \natexlab [1]{#1}%
\providecommand \enquote  [1]{``#1''}%
\providecommand \bibnamefont  [1]{#1}%
\providecommand \bibfnamefont [1]{#1}%
\providecommand \citenamefont [1]{#1}%
\providecommand \href@noop [0]{\@secondoftwo}%
\providecommand \href [0]{\begingroup \@sanitize@url \@href}%
\providecommand \@href[1]{\@@startlink{#1}\@@href}%
\providecommand \@@href[1]{\endgroup#1\@@endlink}%
\providecommand \@sanitize@url [0]{\catcode `\\12\catcode `\$12\catcode `\&12\catcode `\#12\catcode `\^12\catcode `\_12\catcode `\%12\relax}%
\providecommand \@@startlink[1]{}%
\providecommand \@@endlink[0]{}%
\providecommand \url  [0]{\begingroup\@sanitize@url \@url }%
\providecommand \@url [1]{\endgroup\@href {#1}{\urlprefix }}%
\providecommand \urlprefix  [0]{URL }%
\providecommand \Eprint [0]{\href }%
\providecommand \doibase [0]{http://dx.doi.org/}%
\providecommand \selectlanguage [0]{\@gobble}%
\providecommand \bibinfo  [0]{\@secondoftwo}%
\providecommand \bibfield  [0]{\@secondoftwo}%
\providecommand \translation [1]{[#1]}%
\providecommand \BibitemOpen [0]{}%
\providecommand \bibitemStop [0]{}%
\providecommand \bibitemNoStop [0]{.\EOS\space}%
\providecommand \EOS [0]{\spacefactor3000\relax}%
\providecommand \BibitemShut  [1]{\csname bibitem#1\endcsname}%
\let\auto@bib@innerbib\@empty
\bibitem [{\citenamefont {Abbott}\ \emph {et~al.}(2017{\natexlab{a}})\citenamefont {Abbott} \emph {et~al.}}]{LIGOScientific:2017vwq}%
  \BibitemOpen
  \bibfield  {author} {\bibinfo {author} {\bibfnamefont {B.~P.}\ \bibnamefont {Abbott}} \emph {et~al.} (\bibinfo {collaboration} {LIGO Scientific, Virgo}),\ }\href {\doibase 10.1103/PhysRevLett.119.161101} {\bibfield  {journal} {\bibinfo  {journal} {Phys. Rev. Lett.}\ }\textbf {\bibinfo {volume} {119}},\ \bibinfo {pages} {161101} (\bibinfo {year} {2017}{\natexlab{a}})},\ \Eprint {http://arxiv.org/abs/1710.05832} {arXiv:1710.05832 [gr-qc]} \BibitemShut {NoStop}%
\bibitem [{\citenamefont {Akiyama}\ \emph {et~al.}(2019{\natexlab{a}})\citenamefont {Akiyama} \emph {et~al.}}]{EventHorizonTelescope:2019dse}%
  \BibitemOpen
  \bibfield  {author} {\bibinfo {author} {\bibfnamefont {K.}~\bibnamefont {Akiyama}} \emph {et~al.} (\bibinfo {collaboration} {Event Horizon Telescope}),\ }\href {\doibase 10.3847/2041-8213/ab0ec7} {\bibfield  {journal} {\bibinfo  {journal} {Astrophys. J. Lett.}\ }\textbf {\bibinfo {volume} {875}},\ \bibinfo {pages} {L1} (\bibinfo {year} {2019}{\natexlab{a}})},\ \Eprint {http://arxiv.org/abs/1906.11238} {arXiv:1906.11238 [astro-ph.GA]} \BibitemShut {NoStop}%
\bibitem [{\citenamefont {Akiyama}\ \emph {et~al.}(2019{\natexlab{b}})\citenamefont {Akiyama} \emph {et~al.}}]{EventHorizonTelescope:2019ggy}%
  \BibitemOpen
  \bibfield  {author} {\bibinfo {author} {\bibfnamefont {K.}~\bibnamefont {Akiyama}} \emph {et~al.} (\bibinfo {collaboration} {Event Horizon Telescope}),\ }\href {\doibase 10.3847/2041-8213/ab1141} {\bibfield  {journal} {\bibinfo  {journal} {Astrophys. J. Lett.}\ }\textbf {\bibinfo {volume} {875}},\ \bibinfo {pages} {L6} (\bibinfo {year} {2019}{\natexlab{b}})},\ \Eprint {http://arxiv.org/abs/1906.11243} {arXiv:1906.11243 [astro-ph.GA]} \BibitemShut {NoStop}%
\bibitem [{\citenamefont {Goddi}\ \emph {et~al.}(2016)\citenamefont {Goddi} \emph {et~al.}}]{Goddi:2016qax}%
  \BibitemOpen
  \bibfield  {author} {\bibinfo {author} {\bibfnamefont {C.}~\bibnamefont {Goddi}} \emph {et~al.},\ }\href {\doibase 10.1142/9789813226609_0046} {\bibfield  {journal} {\bibinfo  {journal} {Int. J. Mod. Phys. D}\ }\textbf {\bibinfo {volume} {26}},\ \bibinfo {pages} {1730001} (\bibinfo {year} {2016})},\ \Eprint {http://arxiv.org/abs/1606.08879} {arXiv:1606.08879 [astro-ph.HE]} \BibitemShut {NoStop}%
\bibitem [{\citenamefont {Agazie}\ \emph {et~al.}(2023{\natexlab{a}})\citenamefont {Agazie} \emph {et~al.}}]{NANOGrav:2023gor}%
  \BibitemOpen
  \bibfield  {author} {\bibinfo {author} {\bibfnamefont {G.}~\bibnamefont {Agazie}} \emph {et~al.} (\bibinfo {collaboration} {NANOGrav}),\ }\href {\doibase 10.3847/2041-8213/acdac6} {\bibfield  {journal} {\bibinfo  {journal} {Astrophys. J. Lett.}\ }\textbf {\bibinfo {volume} {951}},\ \bibinfo {pages} {L8} (\bibinfo {year} {2023}{\natexlab{a}})},\ \Eprint {http://arxiv.org/abs/2306.16213} {arXiv:2306.16213 [astro-ph.HE]} \BibitemShut {NoStop}%
\bibitem [{\citenamefont {Afzal}\ \emph {et~al.}(2023)\citenamefont {Afzal} \emph {et~al.}}]{NANOGrav:2023hvm}%
  \BibitemOpen
  \bibfield  {author} {\bibinfo {author} {\bibfnamefont {A.}~\bibnamefont {Afzal}} \emph {et~al.} (\bibinfo {collaboration} {NANOGrav}),\ }\href {\doibase 10.3847/2041-8213/acdc91} {\bibfield  {journal} {\bibinfo  {journal} {Astrophys. J. Lett.}\ }\textbf {\bibinfo {volume} {951}},\ \bibinfo {pages} {L11} (\bibinfo {year} {2023})},\ \bibinfo {note} {[Erratum: Astrophys.J.Lett. 971, L27 (2024), Erratum: Astrophys.J. 971, L27 (2024)]},\ \Eprint {http://arxiv.org/abs/2306.16219} {arXiv:2306.16219 [astro-ph.HE]} \BibitemShut {NoStop}%
\bibitem [{\citenamefont {Agazie}\ \emph {et~al.}(2023{\natexlab{b}})\citenamefont {Agazie} \emph {et~al.}}]{NANOGrav:2023pdq}%
  \BibitemOpen
  \bibfield  {author} {\bibinfo {author} {\bibfnamefont {G.}~\bibnamefont {Agazie}} \emph {et~al.} (\bibinfo {collaboration} {NANOGrav}),\ }\href {\doibase 10.3847/2041-8213/ace18a} {\bibfield  {journal} {\bibinfo  {journal} {Astrophys. J. Lett.}\ }\textbf {\bibinfo {volume} {951}},\ \bibinfo {pages} {L50} (\bibinfo {year} {2023}{\natexlab{b}})},\ \Eprint {http://arxiv.org/abs/2306.16222} {arXiv:2306.16222 [astro-ph.HE]} \BibitemShut {NoStop}%
\bibitem [{\citenamefont {Seoane}\ \emph {et~al.}(2023)\citenamefont {Seoane} \emph {et~al.}}]{LISA:2022yao}%
  \BibitemOpen
  \bibfield  {author} {\bibinfo {author} {\bibfnamefont {P.~A.}\ \bibnamefont {Seoane}} \emph {et~al.} (\bibinfo {collaboration} {LISA}),\ }\href {\doibase 10.1007/s41114-022-00041-y} {\bibfield  {journal} {\bibinfo  {journal} {Living Rev. Rel.}\ }\textbf {\bibinfo {volume} {26}},\ \bibinfo {pages} {2} (\bibinfo {year} {2023})},\ \Eprint {http://arxiv.org/abs/2203.06016} {arXiv:2203.06016 [gr-qc]} \BibitemShut {NoStop}%
\bibitem [{\citenamefont {Amaro-Seoane}\ \emph {et~al.}(2022)\citenamefont {Amaro-Seoane} \emph {et~al.}}]{LISA:2022kgy}%
  \BibitemOpen
  \bibfield  {author} {\bibinfo {author} {\bibfnamefont {P.}~\bibnamefont {Amaro-Seoane}} \emph {et~al.} (\bibinfo {collaboration} {LISA}),\ }\href {\doibase 10.1007/s41114-022-00042-6} {\bibfield  {journal} {\bibinfo  {journal} {Living Rev. Rel.}\ }\textbf {\bibinfo {volume} {25}},\ \bibinfo {pages} {4} (\bibinfo {year} {2022})},\ \Eprint {http://arxiv.org/abs/1702.00786} {arXiv:1702.00786 [astro-ph.IM]} \BibitemShut {NoStop}%
\bibitem [{\citenamefont {Abbott}\ \emph {et~al.}(2016{\natexlab{a}})\citenamefont {Abbott} \emph {et~al.}}]{TheLIGOScientific:2016src}%
  \BibitemOpen
  \bibfield  {author} {\bibinfo {author} {\bibfnamefont {B.~P.}\ \bibnamefont {Abbott}} \emph {et~al.} (\bibinfo {collaboration} {LIGO Scientific, Virgo}),\ }\href {\doibase 10.1103/PhysRevLett.116.061102} {\bibfield  {journal} {\bibinfo  {journal} {Phys. Rev. Lett.}\ }\textbf {\bibinfo {volume} {116}},\ \bibinfo {pages} {061102} (\bibinfo {year} {2016}{\natexlab{a}})},\ \Eprint {http://arxiv.org/abs/1602.03837} {arXiv:1602.03837 [gr-qc]} \BibitemShut {NoStop}%
\bibitem [{\citenamefont {Abbott}\ \emph {et~al.}(2016{\natexlab{b}})\citenamefont {Abbott} \emph {et~al.}}]{TheLIGOScientific:2016pea}%
  \BibitemOpen
  \bibfield  {author} {\bibinfo {author} {\bibfnamefont {B.~P.}\ \bibnamefont {Abbott}} \emph {et~al.} (\bibinfo {collaboration} {LIGO Scientific, Virgo}),\ }\href {\doibase 10.1103/PhysRevLett.116.241103} {\bibfield  {journal} {\bibinfo  {journal} {Phys. Rev. Lett.}\ }\textbf {\bibinfo {volume} {116}},\ \bibinfo {pages} {241103} (\bibinfo {year} {2016}{\natexlab{b}})},\ \Eprint {http://arxiv.org/abs/1606.04855} {arXiv:1606.04855 [gr-qc]} \BibitemShut {NoStop}%
\bibitem [{\citenamefont {Abbott}\ \emph {et~al.}(2017{\natexlab{b}})\citenamefont {Abbott} \emph {et~al.}}]{TheLIGOScientific:2016qqj}%
  \BibitemOpen
  \bibfield  {author} {\bibinfo {author} {\bibfnamefont {B.~P.}\ \bibnamefont {Abbott}} \emph {et~al.} (\bibinfo {collaboration} {LIGO Scientific, Virgo}),\ }\href {\doibase 10.1103/PhysRevLett.118.221101} {\bibfield  {journal} {\bibinfo  {journal} {Phys. Rev. Lett.}\ }\textbf {\bibinfo {volume} {118}},\ \bibinfo {pages} {221101} (\bibinfo {year} {2017}{\natexlab{b}})},\ \Eprint {http://arxiv.org/abs/1706.01812} {arXiv:1706.01812 [gr-qc]} \BibitemShut {NoStop}%
\bibitem [{\citenamefont {Kokkotas}\ and\ \citenamefont {Schmidt}(1999)}]{Kokkotas:1999bd}%
  \BibitemOpen
  \bibfield  {author} {\bibinfo {author} {\bibfnamefont {K.~D.}\ \bibnamefont {Kokkotas}}\ and\ \bibinfo {author} {\bibfnamefont {B.~G.}\ \bibnamefont {Schmidt}},\ }\href {\doibase 10.12942/lrr-1999-2} {\bibfield  {journal} {\bibinfo  {journal} {Living Rev. Rel.}\ }\textbf {\bibinfo {volume} {2}},\ \bibinfo {pages} {2} (\bibinfo {year} {1999})},\ \Eprint {http://arxiv.org/abs/gr-qc/9909058} {arXiv:gr-qc/9909058} \BibitemShut {NoStop}%
\bibitem [{\citenamefont {Nollert}(1999)}]{Nollert:1999ji}%
  \BibitemOpen
  \bibfield  {author} {\bibinfo {author} {\bibfnamefont {H.-P.}\ \bibnamefont {Nollert}},\ }\href {\doibase 10.1088/0264-9381/16/12/201} {\bibfield  {journal} {\bibinfo  {journal} {Class. Quant. Grav.}\ }\textbf {\bibinfo {volume} {16}},\ \bibinfo {pages} {R159} (\bibinfo {year} {1999})}\BibitemShut {NoStop}%
\bibitem [{\citenamefont {Konoplya}\ and\ \citenamefont {Zhidenko}(2011)}]{Konoplya:2011qq}%
  \BibitemOpen
  \bibfield  {author} {\bibinfo {author} {\bibfnamefont {R.~A.}\ \bibnamefont {Konoplya}}\ and\ \bibinfo {author} {\bibfnamefont {A.}~\bibnamefont {Zhidenko}},\ }\href {\doibase 10.1103/RevModPhys.83.793} {\bibfield  {journal} {\bibinfo  {journal} {Rev. Mod. Phys.}\ }\textbf {\bibinfo {volume} {83}},\ \bibinfo {pages} {793} (\bibinfo {year} {2011})},\ \Eprint {http://arxiv.org/abs/1102.4014} {arXiv:1102.4014 [gr-qc]} \BibitemShut {NoStop}%
\bibitem [{\citenamefont {Giesler}\ \emph {et~al.}(2019)\citenamefont {Giesler}, \citenamefont {Isi}, \citenamefont {Scheel},\ and\ \citenamefont {Teukolsky}}]{Giesler:2019uxc}%
  \BibitemOpen
  \bibfield  {author} {\bibinfo {author} {\bibfnamefont {M.}~\bibnamefont {Giesler}}, \bibinfo {author} {\bibfnamefont {M.}~\bibnamefont {Isi}}, \bibinfo {author} {\bibfnamefont {M.~A.}\ \bibnamefont {Scheel}}, \ and\ \bibinfo {author} {\bibfnamefont {S.}~\bibnamefont {Teukolsky}},\ }\href {\doibase 10.1103/PhysRevX.9.041060} {\bibfield  {journal} {\bibinfo  {journal} {Phys. Rev. X}\ }\textbf {\bibinfo {volume} {9}},\ \bibinfo {pages} {041060} (\bibinfo {year} {2019})},\ \Eprint {http://arxiv.org/abs/1903.08284} {arXiv:1903.08284 [gr-qc]} \BibitemShut {NoStop}%
\bibitem [{\citenamefont {Giesler}\ \emph {et~al.}(2024)\citenamefont {Giesler} \emph {et~al.}}]{Giesler:2024hcr}%
  \BibitemOpen
  \bibfield  {author} {\bibinfo {author} {\bibfnamefont {M.}~\bibnamefont {Giesler}} \emph {et~al.},\ }\href@noop {} {\  (\bibinfo {year} {2024})},\ \Eprint {http://arxiv.org/abs/2411.11269} {arXiv:2411.11269 [gr-qc]} \BibitemShut {NoStop}%
\bibitem [{\citenamefont {Konoplya}\ and\ \citenamefont {Zhidenko}(2023{\natexlab{a}})}]{Konoplya:2023kem}%
  \BibitemOpen
  \bibfield  {author} {\bibinfo {author} {\bibfnamefont {R.~A.}\ \bibnamefont {Konoplya}}\ and\ \bibinfo {author} {\bibfnamefont {A.}~\bibnamefont {Zhidenko}},\ }\href {\doibase 10.1103/PhysRevD.108.104014} {\bibfield  {journal} {\bibinfo  {journal} {Phys. Rev. D}\ }\textbf {\bibinfo {volume} {108}},\ \bibinfo {pages} {104014} (\bibinfo {year} {2023}{\natexlab{a}})},\ \Eprint {http://arxiv.org/abs/2309.09330} {arXiv:2309.09330 [gr-qc]} \BibitemShut {NoStop}%
\bibitem [{\citenamefont {Konoplya}(2024{\natexlab{a}})}]{Konoplya:2023ahd}%
  \BibitemOpen
  \bibfield  {author} {\bibinfo {author} {\bibfnamefont {R.~A.}\ \bibnamefont {Konoplya}},\ }\href {\doibase 10.1016/j.physletb.2023.138266} {\bibfield  {journal} {\bibinfo  {journal} {Phys. Lett. B}\ }\textbf {\bibinfo {volume} {848}},\ \bibinfo {pages} {138266} (\bibinfo {year} {2024}{\natexlab{a}})},\ \Eprint {http://arxiv.org/abs/2311.09279} {arXiv:2311.09279 [gr-qc]} \BibitemShut {NoStop}%
\bibitem [{\citenamefont {Amaro-Seoane}\ \emph {et~al.}(2017)\citenamefont {Amaro-Seoane} \emph {et~al.}}]{LISA:2017pwj}%
  \BibitemOpen
  \bibfield  {author} {\bibinfo {author} {\bibfnamefont {P.}~\bibnamefont {Amaro-Seoane}} \emph {et~al.} (\bibinfo {collaboration} {LISA}),\ }\href@noop {} {\bibfield  {journal} {\bibinfo  {journal} {arXiv ePrints}\ } (\bibinfo {year} {2017})},\ \Eprint {http://arxiv.org/abs/1702.00786} {arXiv:1702.00786 [astro-ph.IM]} \BibitemShut {NoStop}%
\bibitem [{\citenamefont {Regge}\ and\ \citenamefont {Wheeler}(1957)}]{Regge:1957td}%
  \BibitemOpen
  \bibfield  {author} {\bibinfo {author} {\bibfnamefont {T.}~\bibnamefont {Regge}}\ and\ \bibinfo {author} {\bibfnamefont {J.~A.}\ \bibnamefont {Wheeler}},\ }\href {\doibase 10.1103/PhysRev.108.1063} {\bibfield  {journal} {\bibinfo  {journal} {Phys. Rev.}\ }\textbf {\bibinfo {volume} {108}},\ \bibinfo {pages} {1063} (\bibinfo {year} {1957})}\BibitemShut {NoStop}%
\bibitem [{\citenamefont {Leaver}(1985)}]{Leaver:1985ax}%
  \BibitemOpen
  \bibfield  {author} {\bibinfo {author} {\bibfnamefont {E.~W.}\ \bibnamefont {Leaver}},\ }\href {\doibase 10.1098/rspa.1985.0119} {\bibfield  {journal} {\bibinfo  {journal} {Proc. Roy. Soc. Lond. A}\ }\textbf {\bibinfo {volume} {402}},\ \bibinfo {pages} {285} (\bibinfo {year} {1985})}\BibitemShut {NoStop}%
\bibitem [{\citenamefont {Schutz}\ and\ \citenamefont {Will}(1985)}]{Schutz:1985km}%
  \BibitemOpen
  \bibfield  {author} {\bibinfo {author} {\bibfnamefont {B.~F.}\ \bibnamefont {Schutz}}\ and\ \bibinfo {author} {\bibfnamefont {C.~M.}\ \bibnamefont {Will}},\ }\href {\doibase 10.1086/184453} {\bibfield  {journal} {\bibinfo  {journal} {Astrophys. J. Lett.}\ }\textbf {\bibinfo {volume} {291}},\ \bibinfo {pages} {L33} (\bibinfo {year} {1985})}\BibitemShut {NoStop}%
\bibitem [{\citenamefont {Iyer}\ and\ \citenamefont {Will}(1987)}]{Iyer:1986np}%
  \BibitemOpen
  \bibfield  {author} {\bibinfo {author} {\bibfnamefont {S.}~\bibnamefont {Iyer}}\ and\ \bibinfo {author} {\bibfnamefont {C.~M.}\ \bibnamefont {Will}},\ }\href {\doibase 10.1103/PhysRevD.35.3621} {\bibfield  {journal} {\bibinfo  {journal} {Phys. Rev. D}\ }\textbf {\bibinfo {volume} {35}},\ \bibinfo {pages} {3621} (\bibinfo {year} {1987})}\BibitemShut {NoStop}%
\bibitem [{\citenamefont {Konoplya}\ \emph {et~al.}(2019{\natexlab{a}})\citenamefont {Konoplya}, \citenamefont {Zhidenko},\ and\ \citenamefont {Zinhailo}}]{Konoplya:2019hlu}%
  \BibitemOpen
  \bibfield  {author} {\bibinfo {author} {\bibfnamefont {R.~A.}\ \bibnamefont {Konoplya}}, \bibinfo {author} {\bibfnamefont {A.}~\bibnamefont {Zhidenko}}, \ and\ \bibinfo {author} {\bibfnamefont {A.~F.}\ \bibnamefont {Zinhailo}},\ }\href {\doibase 10.1088/1361-6382/ab2e25} {\bibfield  {journal} {\bibinfo  {journal} {Class. Quant. Grav.}\ }\textbf {\bibinfo {volume} {36}},\ \bibinfo {pages} {155002} (\bibinfo {year} {2019}{\natexlab{a}})},\ \Eprint {http://arxiv.org/abs/1904.10333} {arXiv:1904.10333 [gr-qc]} \BibitemShut {NoStop}%
\bibitem [{\citenamefont {Konoplya}(2003{\natexlab{a}})}]{Konoplya:2003ii}%
  \BibitemOpen
  \bibfield  {author} {\bibinfo {author} {\bibfnamefont {R.~A.}\ \bibnamefont {Konoplya}},\ }\href {\doibase 10.1103/PhysRevD.68.024018} {\bibfield  {journal} {\bibinfo  {journal} {Phys. Rev. D}\ }\textbf {\bibinfo {volume} {68}},\ \bibinfo {pages} {024018} (\bibinfo {year} {2003}{\natexlab{a}})},\ \Eprint {http://arxiv.org/abs/gr-qc/0303052} {arXiv:gr-qc/0303052} \BibitemShut {NoStop}%
\bibitem [{\citenamefont {Konoplya}(2004{\natexlab{a}})}]{Konoplya:2004ip}%
  \BibitemOpen
  \bibfield  {author} {\bibinfo {author} {\bibfnamefont {R.~A.}\ \bibnamefont {Konoplya}},\ }\href@noop {} {\bibfield  {journal} {\bibinfo  {journal} {J. Phys. Stud.}\ }\textbf {\bibinfo {volume} {8}},\ \bibinfo {pages} {93} (\bibinfo {year} {2004}{\natexlab{a}})}\BibitemShut {NoStop}%
\bibitem [{\citenamefont {Matyjasek}\ and\ \citenamefont {Opala}(2017)}]{Matyjasek:2017psv}%
  \BibitemOpen
  \bibfield  {author} {\bibinfo {author} {\bibfnamefont {J.}~\bibnamefont {Matyjasek}}\ and\ \bibinfo {author} {\bibfnamefont {M.}~\bibnamefont {Opala}},\ }\href {\doibase 10.1103/PhysRevD.96.024011} {\bibfield  {journal} {\bibinfo  {journal} {Phys. Rev. D}\ }\textbf {\bibinfo {volume} {96}},\ \bibinfo {pages} {024011} (\bibinfo {year} {2017})},\ \Eprint {http://arxiv.org/abs/1704.00361} {arXiv:1704.00361 [gr-qc]} \BibitemShut {NoStop}%
\bibitem [{\citenamefont {Gundlach}\ \emph {et~al.}(1994)\citenamefont {Gundlach}, \citenamefont {Price},\ and\ \citenamefont {Pullin}}]{Gundlach:1993tp}%
  \BibitemOpen
  \bibfield  {author} {\bibinfo {author} {\bibfnamefont {C.}~\bibnamefont {Gundlach}}, \bibinfo {author} {\bibfnamefont {R.~H.}\ \bibnamefont {Price}}, \ and\ \bibinfo {author} {\bibfnamefont {J.}~\bibnamefont {Pullin}},\ }\href {\doibase 10.1103/PhysRevD.49.883} {\bibfield  {journal} {\bibinfo  {journal} {Phys. Rev. D}\ }\textbf {\bibinfo {volume} {49}},\ \bibinfo {pages} {883} (\bibinfo {year} {1994})},\ \Eprint {http://arxiv.org/abs/gr-qc/9307009} {arXiv:gr-qc/9307009} \BibitemShut {NoStop}%
\bibitem [{\citenamefont {Konoplya}\ and\ \citenamefont {Zhidenko}(2023{\natexlab{b}})}]{Konoplya:2022zav}%
  \BibitemOpen
  \bibfield  {author} {\bibinfo {author} {\bibfnamefont {R.~A.}\ \bibnamefont {Konoplya}}\ and\ \bibinfo {author} {\bibfnamefont {A.}~\bibnamefont {Zhidenko}},\ }\href {\doibase 10.1103/PhysRevD.107.044009} {\bibfield  {journal} {\bibinfo  {journal} {Phys. Rev. D}\ }\textbf {\bibinfo {volume} {107}},\ \bibinfo {pages} {044009} (\bibinfo {year} {2023}{\natexlab{b}})},\ \Eprint {http://arxiv.org/abs/2211.02997} {arXiv:2211.02997 [gr-qc]} \BibitemShut {NoStop}%
\bibitem [{\citenamefont {Fortuna}\ and\ \citenamefont {Vega}(2023)}]{Fortuna:2020obg}%
  \BibitemOpen
  \bibfield  {author} {\bibinfo {author} {\bibfnamefont {S.}~\bibnamefont {Fortuna}}\ and\ \bibinfo {author} {\bibfnamefont {I.}~\bibnamefont {Vega}},\ }\href {\doibase 10.1140/epjc/s10052-023-12350-9} {\bibfield  {journal} {\bibinfo  {journal} {Eur. Phys. J. C}\ }\textbf {\bibinfo {volume} {83}},\ \bibinfo {pages} {1170} (\bibinfo {year} {2023})},\ \Eprint {http://arxiv.org/abs/2003.06232} {arXiv:2003.06232 [gr-qc]} \BibitemShut {NoStop}%
\bibitem [{\citenamefont {Konoplya}\ and\ \citenamefont {Stashko}(2025)}]{Konoplya:2025hgp}%
  \BibitemOpen
  \bibfield  {author} {\bibinfo {author} {\bibfnamefont {R.~A.}\ \bibnamefont {Konoplya}}\ and\ \bibinfo {author} {\bibfnamefont {O.~S.}\ \bibnamefont {Stashko}},\ }\href@noop {} {\  (\bibinfo {year} {2025})},\ \Eprint {http://arxiv.org/abs/2502.05689} {arXiv:2502.05689 [gr-qc]} \BibitemShut {NoStop}%
\bibitem [{\citenamefont {Leaver}(1990)}]{Leaver:1990zz}%
  \BibitemOpen
  \bibfield  {author} {\bibinfo {author} {\bibfnamefont {E.~W.}\ \bibnamefont {Leaver}},\ }\href {\doibase 10.1103/PhysRevD.41.2986} {\bibfield  {journal} {\bibinfo  {journal} {Phys. Rev. D}\ }\textbf {\bibinfo {volume} {41}},\ \bibinfo {pages} {2986} (\bibinfo {year} {1990})}\BibitemShut {NoStop}%
\bibitem [{\citenamefont {Konoplya}\ and\ \citenamefont {Zhidenko}(2006)}]{Konoplya:2006br}%
  \BibitemOpen
  \bibfield  {author} {\bibinfo {author} {\bibfnamefont {R.~A.}\ \bibnamefont {Konoplya}}\ and\ \bibinfo {author} {\bibfnamefont {A.}~\bibnamefont {Zhidenko}},\ }\href {\doibase 10.1103/PhysRevD.73.124040} {\bibfield  {journal} {\bibinfo  {journal} {Phys. Rev. D}\ }\textbf {\bibinfo {volume} {73}},\ \bibinfo {pages} {124040} (\bibinfo {year} {2006})},\ \Eprint {http://arxiv.org/abs/gr-qc/0605013} {arXiv:gr-qc/0605013} \BibitemShut {NoStop}%
\bibitem [{\citenamefont {Kanti}\ \emph {et~al.}(2006)\citenamefont {Kanti}, \citenamefont {Konoplya},\ and\ \citenamefont {Zhidenko}}]{Kanti:2006ua}%
  \BibitemOpen
  \bibfield  {author} {\bibinfo {author} {\bibfnamefont {P.}~\bibnamefont {Kanti}}, \bibinfo {author} {\bibfnamefont {R.~A.}\ \bibnamefont {Konoplya}}, \ and\ \bibinfo {author} {\bibfnamefont {A.}~\bibnamefont {Zhidenko}},\ }\href {\doibase 10.1103/PhysRevD.74.064008} {\bibfield  {journal} {\bibinfo  {journal} {Phys. Rev. D}\ }\textbf {\bibinfo {volume} {74}},\ \bibinfo {pages} {064008} (\bibinfo {year} {2006})},\ \Eprint {http://arxiv.org/abs/gr-qc/0607048} {arXiv:gr-qc/0607048} \BibitemShut {NoStop}%
\bibitem [{\citenamefont {Konoplya}\ and\ \citenamefont {Zhidenko}(2007{\natexlab{a}})}]{Konoplya:2007zx}%
  \BibitemOpen
  \bibfield  {author} {\bibinfo {author} {\bibfnamefont {R.~A.}\ \bibnamefont {Konoplya}}\ and\ \bibinfo {author} {\bibfnamefont {A.}~\bibnamefont {Zhidenko}},\ }\href {\doibase 10.1103/PhysRevD.76.084018} {\bibfield  {journal} {\bibinfo  {journal} {Phys. Rev. D}\ }\textbf {\bibinfo {volume} {76}},\ \bibinfo {pages} {084018} (\bibinfo {year} {2007}{\natexlab{a}})},\ \bibinfo {note} {[Erratum: Phys.Rev.D 90, 029901 (2014)]},\ \Eprint {http://arxiv.org/abs/0707.1890} {arXiv:0707.1890 [hep-th]} \BibitemShut {NoStop}%
\bibitem [{\citenamefont {Kokkotas}\ \emph {et~al.}(2011)\citenamefont {Kokkotas}, \citenamefont {Konoplya},\ and\ \citenamefont {Zhidenko}}]{Kokkotas:2010zd}%
  \BibitemOpen
  \bibfield  {author} {\bibinfo {author} {\bibfnamefont {K.~D.}\ \bibnamefont {Kokkotas}}, \bibinfo {author} {\bibfnamefont {R.~A.}\ \bibnamefont {Konoplya}}, \ and\ \bibinfo {author} {\bibfnamefont {A.}~\bibnamefont {Zhidenko}},\ }\href {\doibase 10.1103/PhysRevD.83.024031} {\bibfield  {journal} {\bibinfo  {journal} {Phys. Rev. D}\ }\textbf {\bibinfo {volume} {83}},\ \bibinfo {pages} {024031} (\bibinfo {year} {2011})},\ \Eprint {http://arxiv.org/abs/1011.1843} {arXiv:1011.1843 [gr-qc]} \BibitemShut {NoStop}%
\bibitem [{\citenamefont {Onozawa}(1997)}]{Onozawa:1996ux}%
  \BibitemOpen
  \bibfield  {author} {\bibinfo {author} {\bibfnamefont {H.}~\bibnamefont {Onozawa}},\ }\href {\doibase 10.1103/PhysRevD.55.3593} {\bibfield  {journal} {\bibinfo  {journal} {Phys. Rev. D}\ }\textbf {\bibinfo {volume} {55}},\ \bibinfo {pages} {3593} (\bibinfo {year} {1997})},\ \Eprint {http://arxiv.org/abs/gr-qc/9610048} {arXiv:gr-qc/9610048} \BibitemShut {NoStop}%
\bibitem [{\citenamefont {Andersson}\ and\ \citenamefont {Onozawa}(1996)}]{Andersson:1996xw}%
  \BibitemOpen
  \bibfield  {author} {\bibinfo {author} {\bibfnamefont {N.}~\bibnamefont {Andersson}}\ and\ \bibinfo {author} {\bibfnamefont {H.}~\bibnamefont {Onozawa}},\ }\href {\doibase 10.1103/PhysRevD.54.7470} {\bibfield  {journal} {\bibinfo  {journal} {Phys. Rev. D}\ }\textbf {\bibinfo {volume} {54}},\ \bibinfo {pages} {7470} (\bibinfo {year} {1996})},\ \Eprint {http://arxiv.org/abs/gr-qc/9607054} {arXiv:gr-qc/9607054} \BibitemShut {NoStop}%
\bibitem [{\citenamefont {Cardoso}\ \emph {et~al.}(2009)\citenamefont {Cardoso}, \citenamefont {Miranda}, \citenamefont {Berti}, \citenamefont {Witek},\ and\ \citenamefont {Zanchin}}]{Cardoso:2008bp}%
  \BibitemOpen
  \bibfield  {author} {\bibinfo {author} {\bibfnamefont {V.}~\bibnamefont {Cardoso}}, \bibinfo {author} {\bibfnamefont {A.~S.}\ \bibnamefont {Miranda}}, \bibinfo {author} {\bibfnamefont {E.}~\bibnamefont {Berti}}, \bibinfo {author} {\bibfnamefont {H.}~\bibnamefont {Witek}}, \ and\ \bibinfo {author} {\bibfnamefont {V.~T.}\ \bibnamefont {Zanchin}},\ }\href {\doibase 10.1103/PhysRevD.79.064016} {\bibfield  {journal} {\bibinfo  {journal} {Phys. Rev. D}\ }\textbf {\bibinfo {volume} {79}},\ \bibinfo {pages} {064016} (\bibinfo {year} {2009})},\ \Eprint {http://arxiv.org/abs/0812.1806} {arXiv:0812.1806 [hep-th]} \BibitemShut {NoStop}%
\bibitem [{\citenamefont {Konoplya}\ and\ \citenamefont {Stuchlík}(2017)}]{Konoplya:2017wot}%
  \BibitemOpen
  \bibfield  {author} {\bibinfo {author} {\bibfnamefont {R.~A.}\ \bibnamefont {Konoplya}}\ and\ \bibinfo {author} {\bibfnamefont {Z.}~\bibnamefont {Stuchlík}},\ }\href {\doibase 10.1016/j.physletb.2017.06.015} {\bibfield  {journal} {\bibinfo  {journal} {Phys. Lett. B}\ }\textbf {\bibinfo {volume} {771}},\ \bibinfo {pages} {597} (\bibinfo {year} {2017})},\ \Eprint {http://arxiv.org/abs/1705.05928} {arXiv:1705.05928 [gr-qc]} \BibitemShut {NoStop}%
\bibitem [{\citenamefont {Konoplya}(2023{\natexlab{a}})}]{Konoplya:2022gjp}%
  \BibitemOpen
  \bibfield  {author} {\bibinfo {author} {\bibfnamefont {R.~A.}\ \bibnamefont {Konoplya}},\ }\href {\doibase 10.1016/j.physletb.2023.137674} {\bibfield  {journal} {\bibinfo  {journal} {Phys. Lett. B}\ }\textbf {\bibinfo {volume} {838}},\ \bibinfo {pages} {137674} (\bibinfo {year} {2023}{\natexlab{a}})},\ \Eprint {http://arxiv.org/abs/2210.08373} {arXiv:2210.08373 [gr-qc]} \BibitemShut {NoStop}%
\bibitem [{\citenamefont {Bolokhov}(2024{\natexlab{a}})}]{Bolokhov:2023dxq}%
  \BibitemOpen
  \bibfield  {author} {\bibinfo {author} {\bibfnamefont {S.~V.}\ \bibnamefont {Bolokhov}},\ }\href {\doibase 10.1016/j.physletb.2024.138879} {\bibfield  {journal} {\bibinfo  {journal} {Phys. Lett. B}\ }\textbf {\bibinfo {volume} {856}},\ \bibinfo {pages} {138879} (\bibinfo {year} {2024}{\natexlab{a}})},\ \Eprint {http://arxiv.org/abs/2310.12326} {arXiv:2310.12326 [gr-qc]} \BibitemShut {NoStop}%
\bibitem [{\citenamefont {Ianniccari}\ \emph {et~al.}(2024)\citenamefont {Ianniccari}, \citenamefont {Iovino}, \citenamefont {Kehagias}, \citenamefont {Perrone},\ and\ \citenamefont {Riotto}}]{Ianniccari:2024eza}%
  \BibitemOpen
  \bibfield  {author} {\bibinfo {author} {\bibfnamefont {A.}~\bibnamefont {Ianniccari}}, \bibinfo {author} {\bibfnamefont {A.~J.}\ \bibnamefont {Iovino}}, \bibinfo {author} {\bibfnamefont {A.}~\bibnamefont {Kehagias}}, \bibinfo {author} {\bibfnamefont {D.}~\bibnamefont {Perrone}}, \ and\ \bibinfo {author} {\bibfnamefont {A.}~\bibnamefont {Riotto}},\ }\href {\doibase 10.1103/PhysRevLett.133.081401} {\bibfield  {journal} {\bibinfo  {journal} {Phys. Rev. Lett.}\ }\textbf {\bibinfo {volume} {133}},\ \bibinfo {pages} {081401} (\bibinfo {year} {2024})}\BibitemShut {NoStop}%
\bibitem [{\citenamefont {Hod}(2024)}]{Hod:2024ihh}%
  \BibitemOpen
  \bibfield  {author} {\bibinfo {author} {\bibfnamefont {S.}~\bibnamefont {Hod}},\ }\href {\doibase 10.1103/PhysRevD.110.064036} {\bibfield  {journal} {\bibinfo  {journal} {Phys. Rev. D}\ }\textbf {\bibinfo {volume} {110}},\ \bibinfo {pages} {064036} (\bibinfo {year} {2024})},\ \Eprint {http://arxiv.org/abs/2409.07517} {arXiv:2409.07517 [gr-qc]} \BibitemShut {NoStop}%
\bibitem [{\citenamefont {Glampedakis}\ and\ \citenamefont {Silva}(2019)}]{Glampedakis:2019dqh}%
  \BibitemOpen
  \bibfield  {author} {\bibinfo {author} {\bibfnamefont {K.}~\bibnamefont {Glampedakis}}\ and\ \bibinfo {author} {\bibfnamefont {H.~O.}\ \bibnamefont {Silva}},\ }\href {\doibase 10.1103/PhysRevD.100.044040} {\bibfield  {journal} {\bibinfo  {journal} {Phys. Rev. D}\ }\textbf {\bibinfo {volume} {100}},\ \bibinfo {pages} {044040} (\bibinfo {year} {2019})},\ \Eprint {http://arxiv.org/abs/1906.05455} {arXiv:1906.05455 [gr-qc]} \BibitemShut {NoStop}%
\bibitem [{\citenamefont {Silva}\ and\ \citenamefont {Glampedakis}(2020)}]{Silva:2019scu}%
  \BibitemOpen
  \bibfield  {author} {\bibinfo {author} {\bibfnamefont {H.~O.}\ \bibnamefont {Silva}}\ and\ \bibinfo {author} {\bibfnamefont {K.}~\bibnamefont {Glampedakis}},\ }\href {\doibase 10.1103/PhysRevD.101.044051} {\bibfield  {journal} {\bibinfo  {journal} {Phys. Rev. D}\ }\textbf {\bibinfo {volume} {101}},\ \bibinfo {pages} {044051} (\bibinfo {year} {2020})},\ \Eprint {http://arxiv.org/abs/1912.09286} {arXiv:1912.09286 [gr-qc]} \BibitemShut {NoStop}%
\bibitem [{\citenamefont {Bryant}\ \emph {et~al.}(2021)\citenamefont {Bryant}, \citenamefont {Silva}, \citenamefont {Yagi},\ and\ \citenamefont {Glampedakis}}]{Bryant:2021xdh}%
  \BibitemOpen
  \bibfield  {author} {\bibinfo {author} {\bibfnamefont {A.}~\bibnamefont {Bryant}}, \bibinfo {author} {\bibfnamefont {H.~O.}\ \bibnamefont {Silva}}, \bibinfo {author} {\bibfnamefont {K.}~\bibnamefont {Yagi}}, \ and\ \bibinfo {author} {\bibfnamefont {K.}~\bibnamefont {Glampedakis}},\ }\href {\doibase 10.1103/PhysRevD.104.044051} {\bibfield  {journal} {\bibinfo  {journal} {Phys. Rev. D}\ }\textbf {\bibinfo {volume} {104}},\ \bibinfo {pages} {044051} (\bibinfo {year} {2021})},\ \Eprint {http://arxiv.org/abs/2106.09657} {arXiv:2106.09657 [gr-qc]} \BibitemShut {NoStop}%
\bibitem [{\citenamefont {Zhidenko}(2008)}]{Zhidenko:2008fp}%
  \BibitemOpen
  \bibfield  {author} {\bibinfo {author} {\bibfnamefont {A.}~\bibnamefont {Zhidenko}},\ }\href {\doibase 10.1103/PhysRevD.78.024007} {\bibfield  {journal} {\bibinfo  {journal} {Phys. Rev. D}\ }\textbf {\bibinfo {volume} {78}},\ \bibinfo {pages} {024007} (\bibinfo {year} {2008})},\ \Eprint {http://arxiv.org/abs/0802.2262} {arXiv:0802.2262 [gr-qc]} \BibitemShut {NoStop}%
\bibitem [{\citenamefont {Konoplya}\ and\ \citenamefont {Abdalla}(2005)}]{Konoplya:2005sy}%
  \BibitemOpen
  \bibfield  {author} {\bibinfo {author} {\bibfnamefont {R.~A.}\ \bibnamefont {Konoplya}}\ and\ \bibinfo {author} {\bibfnamefont {E.}~\bibnamefont {Abdalla}},\ }\href {\doibase 10.1103/PhysRevD.71.084015} {\bibfield  {journal} {\bibinfo  {journal} {Phys. Rev. D}\ }\textbf {\bibinfo {volume} {71}},\ \bibinfo {pages} {084015} (\bibinfo {year} {2005})},\ \Eprint {http://arxiv.org/abs/hep-th/0503029} {arXiv:hep-th/0503029} \BibitemShut {NoStop}%
\bibitem [{\citenamefont {Konoplya}(2002{\natexlab{a}})}]{Konoplya:2001ji}%
  \BibitemOpen
  \bibfield  {author} {\bibinfo {author} {\bibfnamefont {R.~A.}\ \bibnamefont {Konoplya}},\ }\href {\doibase 10.1023/A:1015347628961} {\bibfield  {journal} {\bibinfo  {journal} {Gen. Rel. Grav.}\ }\textbf {\bibinfo {volume} {34}},\ \bibinfo {pages} {329} (\bibinfo {year} {2002}{\natexlab{a}})},\ \Eprint {http://arxiv.org/abs/gr-qc/0109096} {arXiv:gr-qc/0109096} \BibitemShut {NoStop}%
\bibitem [{\citenamefont {Chen}\ \emph {et~al.}(2023)\citenamefont {Chen}, \citenamefont {Chen}, \citenamefont {Ho},\ and\ \citenamefont {Tseng}}]{Chen:2022nlw}%
  \BibitemOpen
  \bibfield  {author} {\bibinfo {author} {\bibfnamefont {C.-Y.}\ \bibnamefont {Chen}}, \bibinfo {author} {\bibfnamefont {Y.-J.}\ \bibnamefont {Chen}}, \bibinfo {author} {\bibfnamefont {M.-Y.}\ \bibnamefont {Ho}}, \ and\ \bibinfo {author} {\bibfnamefont {Y.-H.}\ \bibnamefont {Tseng}},\ }\href {\doibase 10.1016/j.physletb.2023.138153} {\bibfield  {journal} {\bibinfo  {journal} {Phys. Lett. B}\ }\textbf {\bibinfo {volume} {845}},\ \bibinfo {pages} {138153} (\bibinfo {year} {2023})},\ \Eprint {http://arxiv.org/abs/2212.10028} {arXiv:2212.10028 [gr-qc]} \BibitemShut {NoStop}%
\bibitem [{\citenamefont {Allahyari}\ \emph {et~al.}(2019)\citenamefont {Allahyari}, \citenamefont {Firouzjahi},\ and\ \citenamefont {Mashhoon}}]{Allahyari:2018cmg}%
  \BibitemOpen
  \bibfield  {author} {\bibinfo {author} {\bibfnamefont {A.}~\bibnamefont {Allahyari}}, \bibinfo {author} {\bibfnamefont {H.}~\bibnamefont {Firouzjahi}}, \ and\ \bibinfo {author} {\bibfnamefont {B.}~\bibnamefont {Mashhoon}},\ }\href {\doibase 10.1103/PhysRevD.99.044005} {\bibfield  {journal} {\bibinfo  {journal} {Phys. Rev. D}\ }\textbf {\bibinfo {volume} {99}},\ \bibinfo {pages} {044005} (\bibinfo {year} {2019})},\ \Eprint {http://arxiv.org/abs/1812.03376} {arXiv:1812.03376 [gr-qc]} \BibitemShut {NoStop}%
\bibitem [{\citenamefont {Bolokhov}(2024{\natexlab{b}})}]{Bolokhov:2023bwm}%
  \BibitemOpen
  \bibfield  {author} {\bibinfo {author} {\bibfnamefont {S.~V.}\ \bibnamefont {Bolokhov}},\ }\href {\doibase 10.1103/PhysRevD.110.024010} {\bibfield  {journal} {\bibinfo  {journal} {Phys. Rev. D}\ }\textbf {\bibinfo {volume} {110}},\ \bibinfo {pages} {024010} (\bibinfo {year} {2024}{\natexlab{b}})},\ \Eprint {http://arxiv.org/abs/2311.05503} {arXiv:2311.05503 [gr-qc]} \BibitemShut {NoStop}%
\bibitem [{\citenamefont {Dubinsky}(2024{\natexlab{a}})}]{Dubinsky:2024aeu}%
  \BibitemOpen
  \bibfield  {author} {\bibinfo {author} {\bibfnamefont {A.}~\bibnamefont {Dubinsky}},\ }\href {\doibase 10.1016/j.dark.2024.101657} {\bibfield  {journal} {\bibinfo  {journal} {Physics of the Dark Universe}\ }\textbf {\bibinfo {volume} {46}},\ \bibinfo {pages} {101657} (\bibinfo {year} {2024}{\natexlab{a}})},\ \Eprint {http://arxiv.org/abs/2405.08262} {arXiv:2405.08262 [gr-qc]} \BibitemShut {NoStop}%
\bibitem [{\citenamefont {Dubinsky}(2024{\natexlab{b}})}]{Dubinsky:2024gwo}%
  \BibitemOpen
  \bibfield  {author} {\bibinfo {author} {\bibfnamefont {A.}~\bibnamefont {Dubinsky}},\ }\href {\doibase 10.1142/S0217732324501086} {\bibfield  {journal} {\bibinfo  {journal} {Mod. Phys. Lett. A}\ }\textbf {\bibinfo {volume} {39}},\ \bibinfo {pages} {2450108} (\bibinfo {year} {2024}{\natexlab{b}})},\ \Eprint {http://arxiv.org/abs/2404.18004} {arXiv:2404.18004 [gr-qc]} \BibitemShut {NoStop}%
\bibitem [{\citenamefont {Konoplya}\ and\ \citenamefont {Zhidenko}(2023{\natexlab{c}})}]{Konoplya:2023moy}%
  \BibitemOpen
  \bibfield  {author} {\bibinfo {author} {\bibfnamefont {R.~A.}\ \bibnamefont {Konoplya}}\ and\ \bibinfo {author} {\bibfnamefont {A.}~\bibnamefont {Zhidenko}},\ }\href {\doibase 10.1088/1361-6382/ad0a52} {\bibfield  {journal} {\bibinfo  {journal} {Class. Quant. Grav.}\ }\textbf {\bibinfo {volume} {40}},\ \bibinfo {pages} {245005} (\bibinfo {year} {2023}{\natexlab{c}})},\ \Eprint {http://arxiv.org/abs/2309.02560} {arXiv:2309.02560 [gr-qc]} \BibitemShut {NoStop}%
\bibitem [{\citenamefont {Dubinsky}(2025)}]{Dubinsky:2024rvf}%
  \BibitemOpen
  \bibfield  {author} {\bibinfo {author} {\bibfnamefont {A.}~\bibnamefont {Dubinsky}},\ }\href {\doibase 10.1016/j.physletb.2025.139251} {\bibfield  {journal} {\bibinfo  {journal} {Phys. Lett. B}\ }\textbf {\bibinfo {volume} {861}},\ \bibinfo {pages} {139251} (\bibinfo {year} {2025})},\ \Eprint {http://arxiv.org/abs/2409.16569} {arXiv:2409.16569 [gr-qc]} \BibitemShut {NoStop}%
\bibitem [{\citenamefont {Malik}(2024{\natexlab{a}})}]{Malik:2024sxv}%
  \BibitemOpen
  \bibfield  {author} {\bibinfo {author} {\bibfnamefont {Z.}~\bibnamefont {Malik}},\ }\href {\doibase 10.1007/s10773-024-05660-5} {\bibfield  {journal} {\bibinfo  {journal} {Int. J. Theor. Phys.}\ }\textbf {\bibinfo {volume} {63}},\ \bibinfo {pages} {128} (\bibinfo {year} {2024}{\natexlab{a}})}\BibitemShut {NoStop}%
\bibitem [{\citenamefont {Malik}(2024{\natexlab{b}})}]{Malik:2024tuf}%
  \BibitemOpen
  \bibfield  {author} {\bibinfo {author} {\bibfnamefont {Z.}~\bibnamefont {Malik}},\ }\href {\doibase 10.1209/0295-5075/ad7885} {\bibfield  {journal} {\bibinfo  {journal} {EPL}\ }\textbf {\bibinfo {volume} {147}},\ \bibinfo {pages} {69001} (\bibinfo {year} {2024}{\natexlab{b}})},\ \Eprint {http://arxiv.org/abs/2410.04306} {arXiv:2410.04306 [gr-qc]} \BibitemShut {NoStop}%
\bibitem [{\citenamefont {Malik}(2024{\natexlab{c}})}]{Malik:2024voy}%
  \BibitemOpen
  \bibfield  {author} {\bibinfo {author} {\bibfnamefont {Z.}~\bibnamefont {Malik}},\ }\href {\doibase 10.1142/S0217751X24500246} {\bibfield  {journal} {\bibinfo  {journal} {Int. J. Mod. Phys. A}\ }\textbf {\bibinfo {volume} {39}},\ \bibinfo {pages} {2450024} (\bibinfo {year} {2024}{\natexlab{c}})}\BibitemShut {NoStop}%
\bibitem [{\citenamefont {Malik}(2024{\natexlab{d}})}]{2753764}%
  \BibitemOpen
  \bibfield  {author} {\bibinfo {author} {\bibfnamefont {Z.}~\bibnamefont {Malik}},\ }\href {\doibase 10.13140/RG.2.2.27879.83363} {\  (\bibinfo {year} {2024}{\natexlab{d}}),\ 10.13140/RG.2.2.27879.83363}\BibitemShut {NoStop}%
\bibitem [{\citenamefont {Malik}(2024{\natexlab{e}})}]{Malik:2023bxc}%
  \BibitemOpen
  \bibfield  {author} {\bibinfo {author} {\bibfnamefont {Z.}~\bibnamefont {Malik}},\ }\href {\doibase 10.1007/s10773-024-05737-1} {\bibfield  {journal} {\bibinfo  {journal} {Int. J. Theor. Phys.}\ }\textbf {\bibinfo {volume} {63}},\ \bibinfo {pages} {199} (\bibinfo {year} {2024}{\natexlab{e}})},\ \Eprint {http://arxiv.org/abs/2308.10412} {arXiv:2308.10412 [gr-qc]} \BibitemShut {NoStop}%
\bibitem [{\citenamefont {Banados}\ \emph {et~al.}(1992)\citenamefont {Banados}, \citenamefont {Teitelboim},\ and\ \citenamefont {Zanelli}}]{Banados:1992wn}%
  \BibitemOpen
  \bibfield  {author} {\bibinfo {author} {\bibfnamefont {M.}~\bibnamefont {Banados}}, \bibinfo {author} {\bibfnamefont {C.}~\bibnamefont {Teitelboim}}, \ and\ \bibinfo {author} {\bibfnamefont {J.}~\bibnamefont {Zanelli}},\ }\href {\doibase 10.1103/PhysRevLett.69.1849} {\bibfield  {journal} {\bibinfo  {journal} {Phys. Rev. Lett.}\ }\textbf {\bibinfo {volume} {69}},\ \bibinfo {pages} {1849} (\bibinfo {year} {1992})},\ \Eprint {http://arxiv.org/abs/hep-th/9204099} {arXiv:hep-th/9204099} \BibitemShut {NoStop}%
\bibitem [{\citenamefont {Kovtun}\ \emph {et~al.}(2005)\citenamefont {Kovtun}, \citenamefont {Son},\ and\ \citenamefont {Starinets}}]{Kovtun:2004de}%
  \BibitemOpen
  \bibfield  {author} {\bibinfo {author} {\bibfnamefont {P.}~\bibnamefont {Kovtun}}, \bibinfo {author} {\bibfnamefont {D.~T.}\ \bibnamefont {Son}}, \ and\ \bibinfo {author} {\bibfnamefont {A.~O.}\ \bibnamefont {Starinets}},\ }\href {\doibase 10.1103/PhysRevLett.94.111601} {\bibfield  {journal} {\bibinfo  {journal} {Phys. Rev. Lett.}\ }\textbf {\bibinfo {volume} {94}},\ \bibinfo {pages} {111601} (\bibinfo {year} {2005})},\ \Eprint {http://arxiv.org/abs/hep-th/0405231} {arXiv:hep-th/0405231} \BibitemShut {NoStop}%
\bibitem [{\citenamefont {Policastro}\ \emph {et~al.}(2001)\citenamefont {Policastro}, \citenamefont {Son},\ and\ \citenamefont {Starinets}}]{Policastro:2001yc}%
  \BibitemOpen
  \bibfield  {author} {\bibinfo {author} {\bibfnamefont {G.}~\bibnamefont {Policastro}}, \bibinfo {author} {\bibfnamefont {D.~T.}\ \bibnamefont {Son}}, \ and\ \bibinfo {author} {\bibfnamefont {A.~O.}\ \bibnamefont {Starinets}},\ }\href {\doibase 10.1103/PhysRevLett.87.081601} {\bibfield  {journal} {\bibinfo  {journal} {Phys. Rev. Lett.}\ }\textbf {\bibinfo {volume} {87}},\ \bibinfo {pages} {081601} (\bibinfo {year} {2001})},\ \Eprint {http://arxiv.org/abs/hep-th/0104066} {arXiv:hep-th/0104066} \BibitemShut {NoStop}%
\bibitem [{\citenamefont {Horowitz}\ and\ \citenamefont {Hubeny}(2000)}]{Horowitz:1999jd}%
  \BibitemOpen
  \bibfield  {author} {\bibinfo {author} {\bibfnamefont {G.~T.}\ \bibnamefont {Horowitz}}\ and\ \bibinfo {author} {\bibfnamefont {V.~E.}\ \bibnamefont {Hubeny}},\ }\href {\doibase 10.1103/PhysRevD.62.024027} {\bibfield  {journal} {\bibinfo  {journal} {Phys. Rev. D}\ }\textbf {\bibinfo {volume} {62}},\ \bibinfo {pages} {024027} (\bibinfo {year} {2000})},\ \Eprint {http://arxiv.org/abs/hep-th/9909056} {arXiv:hep-th/9909056} \BibitemShut {NoStop}%
\bibitem [{\citenamefont {Birmingham}\ \emph {et~al.}(2002)\citenamefont {Birmingham}, \citenamefont {Sachs},\ and\ \citenamefont {Solodukhin}}]{Birmingham:2001pj}%
  \BibitemOpen
  \bibfield  {author} {\bibinfo {author} {\bibfnamefont {D.}~\bibnamefont {Birmingham}}, \bibinfo {author} {\bibfnamefont {I.}~\bibnamefont {Sachs}}, \ and\ \bibinfo {author} {\bibfnamefont {S.~N.}\ \bibnamefont {Solodukhin}},\ }\href {\doibase 10.1103/PhysRevLett.88.151301} {\bibfield  {journal} {\bibinfo  {journal} {Phys. Rev. Lett.}\ }\textbf {\bibinfo {volume} {88}},\ \bibinfo {pages} {151301} (\bibinfo {year} {2002})},\ \Eprint {http://arxiv.org/abs/hep-th/0112055} {arXiv:hep-th/0112055} \BibitemShut {NoStop}%
\bibitem [{\citenamefont {Konoplya}(2004{\natexlab{b}})}]{Konoplya:2004ik}%
  \BibitemOpen
  \bibfield  {author} {\bibinfo {author} {\bibfnamefont {R.~A.}\ \bibnamefont {Konoplya}},\ }\href {\doibase 10.1103/PhysRevD.70.047503} {\bibfield  {journal} {\bibinfo  {journal} {Phys. Rev. D}\ }\textbf {\bibinfo {volume} {70}},\ \bibinfo {pages} {047503} (\bibinfo {year} {2004}{\natexlab{b}})},\ \Eprint {http://arxiv.org/abs/hep-th/0406100} {arXiv:hep-th/0406100} \BibitemShut {NoStop}%
\bibitem [{\citenamefont {Becar}\ \emph {et~al.}(2014{\natexlab{a}})\citenamefont {Becar}, \citenamefont {Gonzalez},\ and\ \citenamefont {Vasquez}}]{Becar:2013qba}%
  \BibitemOpen
  \bibfield  {author} {\bibinfo {author} {\bibfnamefont {R.}~\bibnamefont {Becar}}, \bibinfo {author} {\bibfnamefont {P.~A.}\ \bibnamefont {Gonzalez}}, \ and\ \bibinfo {author} {\bibfnamefont {Y.}~\bibnamefont {Vasquez}},\ }\href {\doibase 10.1103/PhysRevD.89.023001} {\bibfield  {journal} {\bibinfo  {journal} {Phys. Rev. D}\ }\textbf {\bibinfo {volume} {89}},\ \bibinfo {pages} {023001} (\bibinfo {year} {2014}{\natexlab{a}})},\ \Eprint {http://arxiv.org/abs/1306.5974} {arXiv:1306.5974 [gr-qc]} \BibitemShut {NoStop}%
\bibitem [{\citenamefont {Gonz\'alez}\ and\ \citenamefont {V\'asquez}(2014)}]{Gonzalez:2014voa}%
  \BibitemOpen
  \bibfield  {author} {\bibinfo {author} {\bibfnamefont {P.~A.}\ \bibnamefont {Gonz\'alez}}\ and\ \bibinfo {author} {\bibfnamefont {Y.}~\bibnamefont {V\'asquez}},\ }\href {\doibase 10.1140/epjc/s10052-014-2969-1} {\bibfield  {journal} {\bibinfo  {journal} {Eur. Phys. J. C}\ }\textbf {\bibinfo {volume} {74}},\ \bibinfo {pages} {2969} (\bibinfo {year} {2014})},\ \Eprint {http://arxiv.org/abs/1404.5371} {arXiv:1404.5371 [gr-qc]} \BibitemShut {NoStop}%
\bibitem [{\citenamefont {Konoplya}\ and\ \citenamefont {Zhidenko}(2020{\natexlab{a}})}]{Konoplya:2020ibi}%
  \BibitemOpen
  \bibfield  {author} {\bibinfo {author} {\bibfnamefont {R.~A.}\ \bibnamefont {Konoplya}}\ and\ \bibinfo {author} {\bibfnamefont {A.}~\bibnamefont {Zhidenko}},\ }\href {\doibase 10.1103/PhysRevD.102.064004} {\bibfield  {journal} {\bibinfo  {journal} {Phys. Rev. D}\ }\textbf {\bibinfo {volume} {102}},\ \bibinfo {pages} {064004} (\bibinfo {year} {2020}{\natexlab{a}})},\ \Eprint {http://arxiv.org/abs/2003.12171} {arXiv:2003.12171 [gr-qc]} \BibitemShut {NoStop}%
\bibitem [{\citenamefont {Skvortsova}(2024{\natexlab{a}})}]{Skvortsova:2023zca}%
  \BibitemOpen
  \bibfield  {author} {\bibinfo {author} {\bibfnamefont {M.}~\bibnamefont {Skvortsova}},\ }\href {\doibase 10.1134/S0202289324010110} {\bibfield  {journal} {\bibinfo  {journal} {Grav. Cosmol.}\ }\textbf {\bibinfo {volume} {30}},\ \bibinfo {pages} {68} (\bibinfo {year} {2024}{\natexlab{a}})},\ \Eprint {http://arxiv.org/abs/2311.02729} {arXiv:2311.02729 [gr-qc]} \BibitemShut {NoStop}%
\bibitem [{\citenamefont {Becar}\ \emph {et~al.}(2014{\natexlab{b}})\citenamefont {Becar}, \citenamefont {Gonzalez},\ and\ \citenamefont {Vasquez}}]{Becar:2014jia}%
  \BibitemOpen
  \bibfield  {author} {\bibinfo {author} {\bibfnamefont {R.}~\bibnamefont {Becar}}, \bibinfo {author} {\bibfnamefont {P.~A.}\ \bibnamefont {Gonzalez}}, \ and\ \bibinfo {author} {\bibfnamefont {Y.}~\bibnamefont {Vasquez}},\ }\href {\doibase 10.1140/epjc/s10052-014-2940-1} {\bibfield  {journal} {\bibinfo  {journal} {Eur. Phys. J. C}\ }\textbf {\bibinfo {volume} {74}},\ \bibinfo {pages} {2940} (\bibinfo {year} {2014}{\natexlab{b}})},\ \Eprint {http://arxiv.org/abs/1405.1509} {arXiv:1405.1509 [gr-qc]} \BibitemShut {NoStop}%
\bibitem [{\citenamefont {B\'ecar}\ \emph {et~al.}(2014)\citenamefont {B\'ecar}, \citenamefont {Gonz\'alez},\ and\ \citenamefont {V\'asquez}}]{Becar:2014aka}%
  \BibitemOpen
  \bibfield  {author} {\bibinfo {author} {\bibfnamefont {R.}~\bibnamefont {B\'ecar}}, \bibinfo {author} {\bibfnamefont {P.~A.}\ \bibnamefont {Gonz\'alez}}, \ and\ \bibinfo {author} {\bibfnamefont {Y.}~\bibnamefont {V\'asquez}},\ }\href {\doibase 10.1140/epjc/s10052-014-3028-7} {\bibfield  {journal} {\bibinfo  {journal} {Eur. Phys. J. C}\ }\textbf {\bibinfo {volume} {74}},\ \bibinfo {pages} {3028} (\bibinfo {year} {2014})},\ \Eprint {http://arxiv.org/abs/1404.6023} {arXiv:1404.6023 [gr-qc]} \BibitemShut {NoStop}%
\bibitem [{\citenamefont {Cardoso}\ and\ \citenamefont {Cavaglia}(2006)}]{Cardoso:2006yj}%
  \BibitemOpen
  \bibfield  {author} {\bibinfo {author} {\bibfnamefont {V.}~\bibnamefont {Cardoso}}\ and\ \bibinfo {author} {\bibfnamefont {M.}~\bibnamefont {Cavaglia}},\ }\href {\doibase 10.1103/PhysRevD.74.024027} {\bibfield  {journal} {\bibinfo  {journal} {Phys. Rev. D}\ }\textbf {\bibinfo {volume} {74}},\ \bibinfo {pages} {024027} (\bibinfo {year} {2006})},\ \Eprint {http://arxiv.org/abs/gr-qc/0604101} {arXiv:gr-qc/0604101} \BibitemShut {NoStop}%
\bibitem [{\citenamefont {Chandrasekhar}(1983)}]{Chandrasekhar:1983book}%
  \BibitemOpen
  \bibfield  {author} {\bibinfo {author} {\bibfnamefont {S.}~\bibnamefont {Chandrasekhar}},\ }\href@noop {} {\emph {\bibinfo {title} {{The Mathematical Theory of Black Holes}}}}\ (\bibinfo  {publisher} {Oxford University Press},\ \bibinfo {year} {1983})\BibitemShut {NoStop}%
\bibitem [{\citenamefont {Maassen van~den Brink}(2000{\natexlab{a}})}]{MaassenvandenBrink:2000zh}%
  \BibitemOpen
  \bibfield  {author} {\bibinfo {author} {\bibfnamefont {A.}~\bibnamefont {Maassen van~den Brink}},\ }\href@noop {} {\bibfield  {journal} {\bibinfo  {journal} {Phys. Rev. D}\ }\textbf {\bibinfo {volume} {62}},\ \bibinfo {pages} {064009} (\bibinfo {year} {2000}{\natexlab{a}})},\ \Eprint {http://arxiv.org/abs/gr-qc/0001032} {arXiv:gr-qc/0001032} \BibitemShut {NoStop}%
\bibitem [{\citenamefont {Chandrasekhar}(1984{\natexlab{a}})}]{Chandrasekhar:1984zz}%
  \BibitemOpen
  \bibfield  {author} {\bibinfo {author} {\bibfnamefont {S.}~\bibnamefont {Chandrasekhar}},\ }\href@noop {} {\bibfield  {journal} {\bibinfo  {journal} {Proc. Roy. Soc. Lond. A}\ }\textbf {\bibinfo {volume} {392}},\ \bibinfo {pages} {1} (\bibinfo {year} {1984}{\natexlab{a}})}\BibitemShut {NoStop}%
\bibitem [{\citenamefont {Glampedakis}\ and\ \citenamefont {Silva}(2017)}]{Glampedakis:2017rar}%
  \BibitemOpen
  \bibfield  {author} {\bibinfo {author} {\bibfnamefont {K.}~\bibnamefont {Glampedakis}}\ and\ \bibinfo {author} {\bibfnamefont {H.~O.}\ \bibnamefont {Silva}},\ }\href {\doibase 10.1103/PhysRevD.96.064054} {\bibfield  {journal} {\bibinfo  {journal} {Phys. Rev. D}\ }\textbf {\bibinfo {volume} {96}},\ \bibinfo {pages} {064054} (\bibinfo {year} {2017})},\ \Eprint {http://arxiv.org/abs/1702.06459} {arXiv:1702.06459 [gr-qc]} \BibitemShut {NoStop}%
\bibitem [{\citenamefont {Sasaki}\ and\ \citenamefont {Nakamura}(1982)}]{Sasaki:1981sx}%
  \BibitemOpen
  \bibfield  {author} {\bibinfo {author} {\bibfnamefont {M.}~\bibnamefont {Sasaki}}\ and\ \bibinfo {author} {\bibfnamefont {T.}~\bibnamefont {Nakamura}},\ }\href@noop {} {\bibfield  {journal} {\bibinfo  {journal} {Phys. Lett. A}\ }\textbf {\bibinfo {volume} {89}},\ \bibinfo {pages} {68} (\bibinfo {year} {1982})}\BibitemShut {NoStop}%
\bibitem [{\citenamefont {Konoplya}\ and\ \citenamefont {Zhidenko}(2008)}]{Konoplya:2008ix}%
  \BibitemOpen
  \bibfield  {author} {\bibinfo {author} {\bibfnamefont {R.~A.}\ \bibnamefont {Konoplya}}\ and\ \bibinfo {author} {\bibfnamefont {A.}~\bibnamefont {Zhidenko}},\ }\href {\doibase 10.1103/PhysRevD.77.104004} {\bibfield  {journal} {\bibinfo  {journal} {Phys. Rev. D}\ }\textbf {\bibinfo {volume} {77}},\ \bibinfo {pages} {104004} (\bibinfo {year} {2008})},\ \Eprint {http://arxiv.org/abs/0802.0267} {arXiv:0802.0267 [hep-th]} \BibitemShut {NoStop}%
\bibitem [{\citenamefont {Cuyubamba}\ \emph {et~al.}(2016)\citenamefont {Cuyubamba}, \citenamefont {Konoplya},\ and\ \citenamefont {Zhidenko}}]{Cuyubamba:2016cug}%
  \BibitemOpen
  \bibfield  {author} {\bibinfo {author} {\bibfnamefont {M.~A.}\ \bibnamefont {Cuyubamba}}, \bibinfo {author} {\bibfnamefont {R.~A.}\ \bibnamefont {Konoplya}}, \ and\ \bibinfo {author} {\bibfnamefont {A.}~\bibnamefont {Zhidenko}},\ }\href {\doibase 10.1103/PhysRevD.93.104053} {\bibfield  {journal} {\bibinfo  {journal} {Phys. Rev. D}\ }\textbf {\bibinfo {volume} {93}},\ \bibinfo {pages} {104053} (\bibinfo {year} {2016})},\ \Eprint {http://arxiv.org/abs/1604.03604} {arXiv:1604.03604 [gr-qc]} \BibitemShut {NoStop}%
\bibitem [{\citenamefont {Konoplya}\ and\ \citenamefont {Zhidenko}(2017{\natexlab{a}})}]{Konoplya:2017ymp}%
  \BibitemOpen
  \bibfield  {author} {\bibinfo {author} {\bibfnamefont {R.~A.}\ \bibnamefont {Konoplya}}\ and\ \bibinfo {author} {\bibfnamefont {A.}~\bibnamefont {Zhidenko}},\ }\href {\doibase 10.1103/PhysRevD.95.104005} {\bibfield  {journal} {\bibinfo  {journal} {Phys. Rev. D}\ }\textbf {\bibinfo {volume} {95}},\ \bibinfo {pages} {104005} (\bibinfo {year} {2017}{\natexlab{a}})},\ \Eprint {http://arxiv.org/abs/1701.01652} {arXiv:1701.01652 [hep-th]} \BibitemShut {NoStop}%
\bibitem [{\citenamefont {González}\ \emph {et~al.}(2017)\citenamefont {González}, \citenamefont {Konoplya},\ and\ \citenamefont {Vásquez}}]{Gonzalez:2017gwa}%
  \BibitemOpen
  \bibfield  {author} {\bibinfo {author} {\bibfnamefont {P.~A.}\ \bibnamefont {González}}, \bibinfo {author} {\bibfnamefont {R.~A.}\ \bibnamefont {Konoplya}}, \ and\ \bibinfo {author} {\bibfnamefont {Y.}~\bibnamefont {Vásquez}},\ }\href {\doibase 10.1103/PhysRevD.95.124012} {\bibfield  {journal} {\bibinfo  {journal} {Phys. Rev. D}\ }\textbf {\bibinfo {volume} {95}},\ \bibinfo {pages} {124012} (\bibinfo {year} {2017})},\ \Eprint {http://arxiv.org/abs/1703.06215} {arXiv:1703.06215 [gr-qc]} \BibitemShut {NoStop}%
\bibitem [{\citenamefont {Konoplya}\ and\ \citenamefont {Zhidenko}(2017{\natexlab{b}})}]{Konoplya:2017zwo}%
  \BibitemOpen
  \bibfield  {author} {\bibinfo {author} {\bibfnamefont {R.~A.}\ \bibnamefont {Konoplya}}\ and\ \bibinfo {author} {\bibfnamefont {A.}~\bibnamefont {Zhidenko}},\ }\href {\doibase 10.1007/JHEP09(2017)139} {\bibfield  {journal} {\bibinfo  {journal} {JHEP}\ }\textbf {\bibinfo {volume} {09}},\ \bibinfo {pages} {139} (\bibinfo {year} {2017}{\natexlab{b}})},\ \Eprint {http://arxiv.org/abs/1705.07732} {arXiv:1705.07732 [hep-th]} \BibitemShut {NoStop}%
\bibitem [{\citenamefont {Cuyubamba}\ \emph {et~al.}(2018)\citenamefont {Cuyubamba}, \citenamefont {Konoplya},\ and\ \citenamefont {Zhidenko}}]{Cuyubamba:2018jdl}%
  \BibitemOpen
  \bibfield  {author} {\bibinfo {author} {\bibfnamefont {M.~A.}\ \bibnamefont {Cuyubamba}}, \bibinfo {author} {\bibfnamefont {R.~A.}\ \bibnamefont {Konoplya}}, \ and\ \bibinfo {author} {\bibfnamefont {A.}~\bibnamefont {Zhidenko}},\ }\href {\doibase 10.1103/PhysRevD.98.044040} {\bibfield  {journal} {\bibinfo  {journal} {Phys. Rev. D}\ }\textbf {\bibinfo {volume} {98}},\ \bibinfo {pages} {044040} (\bibinfo {year} {2018})},\ \Eprint {http://arxiv.org/abs/1804.11170} {arXiv:1804.11170 [gr-qc]} \BibitemShut {NoStop}%
\bibitem [{\citenamefont {Konoplya}\ and\ \citenamefont {Zinhailo}(2020)}]{Konoplya:2020bxa}%
  \BibitemOpen
  \bibfield  {author} {\bibinfo {author} {\bibfnamefont {R.~A.}\ \bibnamefont {Konoplya}}\ and\ \bibinfo {author} {\bibfnamefont {A.~F.}\ \bibnamefont {Zinhailo}},\ }\href {\doibase 10.1140/epjc/s10052-020-08639-8} {\bibfield  {journal} {\bibinfo  {journal} {Eur. Phys. J. C}\ }\textbf {\bibinfo {volume} {80}},\ \bibinfo {pages} {1049} (\bibinfo {year} {2020})},\ \Eprint {http://arxiv.org/abs/2003.01188} {arXiv:2003.01188 [gr-qc]} \BibitemShut {NoStop}%
\bibitem [{\citenamefont {Starinets}(2002)}]{Starinets:2002br}%
  \BibitemOpen
  \bibfield  {author} {\bibinfo {author} {\bibfnamefont {A.~O.}\ \bibnamefont {Starinets}},\ }\href {\doibase 10.1103/PhysRevD.66.124013} {\bibfield  {journal} {\bibinfo  {journal} {Phys. Rev. D}\ }\textbf {\bibinfo {volume} {66}},\ \bibinfo {pages} {124013} (\bibinfo {year} {2002})},\ \Eprint {http://arxiv.org/abs/hep-th/0207133} {arXiv:hep-th/0207133} \BibitemShut {NoStop}%
\bibitem [{\citenamefont {Konoplya}\ and\ \citenamefont {Zhidenko}(2020{\natexlab{b}})}]{Konoplya:2020juj}%
  \BibitemOpen
  \bibfield  {author} {\bibinfo {author} {\bibfnamefont {R.~A.}\ \bibnamefont {Konoplya}}\ and\ \bibinfo {author} {\bibfnamefont {A.}~\bibnamefont {Zhidenko}},\ }\href {\doibase 10.1016/j.dark.2020.100697} {\bibfield  {journal} {\bibinfo  {journal} {Phys. Dark Univ.}\ }\textbf {\bibinfo {volume} {30}},\ \bibinfo {pages} {100697} (\bibinfo {year} {2020}{\natexlab{b}})},\ \Eprint {http://arxiv.org/abs/2003.12492} {arXiv:2003.12492 [gr-qc]} \BibitemShut {NoStop}%
\bibitem [{\citenamefont {Cardoso}\ \emph {et~al.}(2003)\citenamefont {Cardoso}, \citenamefont {Konoplya},\ and\ \citenamefont {Lemos}}]{Cardoso:2003cj}%
  \BibitemOpen
  \bibfield  {author} {\bibinfo {author} {\bibfnamefont {V.}~\bibnamefont {Cardoso}}, \bibinfo {author} {\bibfnamefont {R.}~\bibnamefont {Konoplya}}, \ and\ \bibinfo {author} {\bibfnamefont {J.~P.~S.}\ \bibnamefont {Lemos}},\ }\href {\doibase 10.1103/PhysRevD.68.044024} {\bibfield  {journal} {\bibinfo  {journal} {Phys. Rev. D}\ }\textbf {\bibinfo {volume} {68}},\ \bibinfo {pages} {044024} (\bibinfo {year} {2003})},\ \Eprint {http://arxiv.org/abs/gr-qc/0305037} {arXiv:gr-qc/0305037} \BibitemShut {NoStop}%
\bibitem [{\citenamefont {Konoplya}\ and\ \citenamefont {Zhidenko}(2014)}]{Konoplya:2013sba}%
  \BibitemOpen
  \bibfield  {author} {\bibinfo {author} {\bibfnamefont {R.~A.}\ \bibnamefont {Konoplya}}\ and\ \bibinfo {author} {\bibfnamefont {A.}~\bibnamefont {Zhidenko}},\ }\href {\doibase 10.1103/PhysRevD.89.024011} {\bibfield  {journal} {\bibinfo  {journal} {Phys. Rev. D}\ }\textbf {\bibinfo {volume} {89}},\ \bibinfo {pages} {024011} (\bibinfo {year} {2014})},\ \Eprint {http://arxiv.org/abs/1309.7667} {arXiv:1309.7667 [hep-th]} \BibitemShut {NoStop}%
\bibitem [{\citenamefont {Nunez}\ and\ \citenamefont {Starinets}(2003)}]{Nunez:2003eq}%
  \BibitemOpen
  \bibfield  {author} {\bibinfo {author} {\bibfnamefont {A.}~\bibnamefont {Nunez}}\ and\ \bibinfo {author} {\bibfnamefont {A.~O.}\ \bibnamefont {Starinets}},\ }\href {\doibase 10.1103/PhysRevD.67.124013} {\bibfield  {journal} {\bibinfo  {journal} {Phys. Rev. D}\ }\textbf {\bibinfo {volume} {67}},\ \bibinfo {pages} {124013} (\bibinfo {year} {2003})},\ \Eprint {http://arxiv.org/abs/hep-th/0302026} {arXiv:hep-th/0302026} \BibitemShut {NoStop}%
\bibitem [{\citenamefont {Konoplya}(2024{\natexlab{b}})}]{Konoplya:2024ptj}%
  \BibitemOpen
  \bibfield  {author} {\bibinfo {author} {\bibfnamefont {R.~A.}\ \bibnamefont {Konoplya}},\ }\href {\doibase 10.1103/PhysRevD.109.104018} {\bibfield  {journal} {\bibinfo  {journal} {Phys. Rev. D}\ }\textbf {\bibinfo {volume} {109}},\ \bibinfo {pages} {104018} (\bibinfo {year} {2024}{\natexlab{b}})},\ \Eprint {http://arxiv.org/abs/2401.17106} {arXiv:2401.17106 [gr-qc]} \BibitemShut {NoStop}%
\bibitem [{\citenamefont {Konoplya}\ and\ \citenamefont {Zhidenko}(2009)}]{Konoplya:2008au}%
  \BibitemOpen
  \bibfield  {author} {\bibinfo {author} {\bibfnamefont {R.~A.}\ \bibnamefont {Konoplya}}\ and\ \bibinfo {author} {\bibfnamefont {A.}~\bibnamefont {Zhidenko}},\ }\href {\doibase 10.1103/PhysRevLett.103.161101} {\bibfield  {journal} {\bibinfo  {journal} {Phys. Rev. Lett.}\ }\textbf {\bibinfo {volume} {103}},\ \bibinfo {pages} {161101} (\bibinfo {year} {2009})},\ \Eprint {http://arxiv.org/abs/0809.2822} {arXiv:0809.2822 [hep-th]} \BibitemShut {NoStop}%
\bibitem [{\citenamefont {Konoplya}\ \emph {et~al.}(2025)\citenamefont {Konoplya}, \citenamefont {Khrabustovskyi}, \citenamefont {K\v{r}\'\i{}\v{z}},\ and\ \citenamefont {Zhidenko}}]{Konoplya:2025mvj}%
  \BibitemOpen
  \bibfield  {author} {\bibinfo {author} {\bibfnamefont {R.~A.}\ \bibnamefont {Konoplya}}, \bibinfo {author} {\bibfnamefont {A.}~\bibnamefont {Khrabustovskyi}}, \bibinfo {author} {\bibfnamefont {J.}~\bibnamefont {K\v{r}\'\i{}\v{z}}}, \ and\ \bibinfo {author} {\bibfnamefont {A.}~\bibnamefont {Zhidenko}},\ }\href@noop {} {\  (\bibinfo {year} {2025})},\ \Eprint {http://arxiv.org/abs/2501.16134} {arXiv:2501.16134 [gr-qc]} \BibitemShut {NoStop}%
\bibitem [{\citenamefont {Konoplya}\ \emph {et~al.}(2018{\natexlab{a}})\citenamefont {Konoplya}, \citenamefont {Stuchl\'\i{}k},\ and\ \citenamefont {Zhidenko}}]{Konoplya:2018qov}%
  \BibitemOpen
  \bibfield  {author} {\bibinfo {author} {\bibfnamefont {R.~A.}\ \bibnamefont {Konoplya}}, \bibinfo {author} {\bibfnamefont {Z.}~\bibnamefont {Stuchl\'\i{}k}}, \ and\ \bibinfo {author} {\bibfnamefont {A.}~\bibnamefont {Zhidenko}},\ }\href {\doibase 10.1103/PhysRevD.98.104033} {\bibfield  {journal} {\bibinfo  {journal} {Phys. Rev. D}\ }\textbf {\bibinfo {volume} {98}},\ \bibinfo {pages} {104033} (\bibinfo {year} {2018}{\natexlab{a}})},\ \Eprint {http://arxiv.org/abs/1808.03346} {arXiv:1808.03346 [gr-qc]} \BibitemShut {NoStop}%
\bibitem [{\citenamefont {Konoplya}\ \emph {et~al.}(2008)\citenamefont {Konoplya}, \citenamefont {Murata}, \citenamefont {Soda},\ and\ \citenamefont {Zhidenko}}]{Konoplya:2008yy}%
  \BibitemOpen
  \bibfield  {author} {\bibinfo {author} {\bibfnamefont {R.~A.}\ \bibnamefont {Konoplya}}, \bibinfo {author} {\bibfnamefont {K.}~\bibnamefont {Murata}}, \bibinfo {author} {\bibfnamefont {J.}~\bibnamefont {Soda}}, \ and\ \bibinfo {author} {\bibfnamefont {A.}~\bibnamefont {Zhidenko}},\ }\href {\doibase 10.1103/PhysRevD.78.084012} {\bibfield  {journal} {\bibinfo  {journal} {Phys. Rev. D}\ }\textbf {\bibinfo {volume} {78}},\ \bibinfo {pages} {084012} (\bibinfo {year} {2008})},\ \Eprint {http://arxiv.org/abs/0807.1897} {arXiv:0807.1897 [hep-th]} \BibitemShut {NoStop}%
\bibitem [{\citenamefont {Konoplya}\ \emph {et~al.}(2019{\natexlab{b}})\citenamefont {Konoplya}, \citenamefont {Posada}, \citenamefont {Stuchl\'\i{}k},\ and\ \citenamefont {Zhidenko}}]{Konoplya:2019nzp}%
  \BibitemOpen
  \bibfield  {author} {\bibinfo {author} {\bibfnamefont {R.~A.}\ \bibnamefont {Konoplya}}, \bibinfo {author} {\bibfnamefont {C.}~\bibnamefont {Posada}}, \bibinfo {author} {\bibfnamefont {Z.}~\bibnamefont {Stuchl\'\i{}k}}, \ and\ \bibinfo {author} {\bibfnamefont {A.}~\bibnamefont {Zhidenko}},\ }\href {\doibase 10.1103/PhysRevD.100.044027} {\bibfield  {journal} {\bibinfo  {journal} {Phys. Rev. D}\ }\textbf {\bibinfo {volume} {100}},\ \bibinfo {pages} {044027} (\bibinfo {year} {2019}{\natexlab{b}})},\ \Eprint {http://arxiv.org/abs/1905.08097} {arXiv:1905.08097 [gr-qc]} \BibitemShut {NoStop}%
\bibitem [{\citenamefont {Konoplya}\ and\ \citenamefont {Zhidenko}(2022{\natexlab{a}})}]{Konoplya:2022xid}%
  \BibitemOpen
  \bibfield  {author} {\bibinfo {author} {\bibfnamefont {R.~A.}\ \bibnamefont {Konoplya}}\ and\ \bibinfo {author} {\bibfnamefont {A.}~\bibnamefont {Zhidenko}},\ }\href {\doibase 10.1103/PhysRevD.106.124004} {\bibfield  {journal} {\bibinfo  {journal} {Phys. Rev. D}\ }\textbf {\bibinfo {volume} {106}},\ \bibinfo {pages} {124004} (\bibinfo {year} {2022}{\natexlab{a}})},\ \Eprint {http://arxiv.org/abs/2209.12058} {arXiv:2209.12058 [gr-qc]} \BibitemShut {NoStop}%
\bibitem [{\citenamefont {Konoplya}(2003{\natexlab{b}})}]{Konoplya:2003dd}%
  \BibitemOpen
  \bibfield  {author} {\bibinfo {author} {\bibfnamefont {R.~A.}\ \bibnamefont {Konoplya}},\ }\href {\doibase 10.1103/PhysRevD.68.124017} {\bibfield  {journal} {\bibinfo  {journal} {Phys. Rev. D}\ }\textbf {\bibinfo {volume} {68}},\ \bibinfo {pages} {124017} (\bibinfo {year} {2003}{\natexlab{b}})},\ \Eprint {http://arxiv.org/abs/hep-th/0309030} {arXiv:hep-th/0309030} \BibitemShut {NoStop}%
\bibitem [{\citenamefont {Chandrasekhar}(1984{\natexlab{b}})}]{Chandrasekhar:1984}%
  \BibitemOpen
  \bibfield  {author} {\bibinfo {author} {\bibfnamefont {S.}~\bibnamefont {Chandrasekhar}},\ }\href@noop {} {\bibfield  {journal} {\bibinfo  {journal} {Proc. Roy. Soc. Lond. A}\ }\textbf {\bibinfo {volume} {392}},\ \bibinfo {pages} {1} (\bibinfo {year} {1984}{\natexlab{b}})}\BibitemShut {NoStop}%
\bibitem [{\citenamefont {Maassen van~den Brink}(2000{\natexlab{b}})}]{MaassenvandenBrink:2000iwh}%
  \BibitemOpen
  \bibfield  {author} {\bibinfo {author} {\bibfnamefont {A.}~\bibnamefont {Maassen van~den Brink}},\ }\href@noop {} {\bibfield  {journal} {\bibinfo  {journal} {Phys. Rev. D}\ }\textbf {\bibinfo {volume} {62}},\ \bibinfo {pages} {064009} (\bibinfo {year} {2000}{\natexlab{b}})},\ \Eprint {http://arxiv.org/abs/gr-qc/0001032} {gr-qc/0001032} \BibitemShut {NoStop}%
\bibitem [{\citenamefont {Takahashi}\ and\ \citenamefont {Soda}(2012)}]{Takahashi:2011du}%
  \BibitemOpen
  \bibfield  {author} {\bibinfo {author} {\bibfnamefont {T.}~\bibnamefont {Takahashi}}\ and\ \bibinfo {author} {\bibfnamefont {J.}~\bibnamefont {Soda}},\ }\href {\doibase 10.1088/0264-9381/29/3/035008} {\bibfield  {journal} {\bibinfo  {journal} {Class. Quant. Grav.}\ }\textbf {\bibinfo {volume} {29}},\ \bibinfo {pages} {035008} (\bibinfo {year} {2012})},\ \Eprint {http://arxiv.org/abs/1108.5041} {arXiv:1108.5041 [hep-th]} \BibitemShut {NoStop}%
\bibitem [{\citenamefont {Takahashi}(2011)}]{Takahashi:2011qda}%
  \BibitemOpen
  \bibfield  {author} {\bibinfo {author} {\bibfnamefont {T.}~\bibnamefont {Takahashi}},\ }\href {\doibase 10.1143/PTP.125.1289} {\bibfield  {journal} {\bibinfo  {journal} {Prog. Theor. Phys.}\ }\textbf {\bibinfo {volume} {125}},\ \bibinfo {pages} {1289} (\bibinfo {year} {2011})},\ \Eprint {http://arxiv.org/abs/1102.1785} {arXiv:1102.1785 [gr-qc]} \BibitemShut {NoStop}%
\bibitem [{\citenamefont {Takahashi}\ and\ \citenamefont {Soda}(2010)}]{Takahashi:2010gz}%
  \BibitemOpen
  \bibfield  {author} {\bibinfo {author} {\bibfnamefont {T.}~\bibnamefont {Takahashi}}\ and\ \bibinfo {author} {\bibfnamefont {J.}~\bibnamefont {Soda}},\ }\href {\doibase 10.1143/PTP.124.711} {\bibfield  {journal} {\bibinfo  {journal} {Prog. Theor. Phys.}\ }\textbf {\bibinfo {volume} {124}},\ \bibinfo {pages} {711} (\bibinfo {year} {2010})},\ \Eprint {http://arxiv.org/abs/1008.1618} {arXiv:1008.1618 [gr-qc]} \BibitemShut {NoStop}%
\bibitem [{\citenamefont {Dotti}\ and\ \citenamefont {Gleiser}(2005{\natexlab{a}})}]{Dotti:2004sh}%
  \BibitemOpen
  \bibfield  {author} {\bibinfo {author} {\bibfnamefont {G.}~\bibnamefont {Dotti}}\ and\ \bibinfo {author} {\bibfnamefont {R.~J.}\ \bibnamefont {Gleiser}},\ }\href {\doibase 10.1088/0264-9381/22/1/L01} {\bibfield  {journal} {\bibinfo  {journal} {Class. Quant. Grav.}\ }\textbf {\bibinfo {volume} {22}},\ \bibinfo {pages} {L1} (\bibinfo {year} {2005}{\natexlab{a}})},\ \Eprint {http://arxiv.org/abs/gr-qc/0409005} {arXiv:gr-qc/0409005} \BibitemShut {NoStop}%
\bibitem [{\citenamefont {Dotti}\ and\ \citenamefont {Gleiser}(2005{\natexlab{b}})}]{Dotti:2005sq}%
  \BibitemOpen
  \bibfield  {author} {\bibinfo {author} {\bibfnamefont {G.}~\bibnamefont {Dotti}}\ and\ \bibinfo {author} {\bibfnamefont {R.~J.}\ \bibnamefont {Gleiser}},\ }\href {\doibase 10.1103/PhysRevD.72.044018} {\bibfield  {journal} {\bibinfo  {journal} {Phys. Rev. D}\ }\textbf {\bibinfo {volume} {72}},\ \bibinfo {pages} {044018} (\bibinfo {year} {2005}{\natexlab{b}})},\ \Eprint {http://arxiv.org/abs/gr-qc/0503117} {arXiv:gr-qc/0503117} \BibitemShut {NoStop}%
\bibitem [{\citenamefont {Gleiser}\ and\ \citenamefont {Dotti}(2005)}]{Gleiser:2005ra}%
  \BibitemOpen
  \bibfield  {author} {\bibinfo {author} {\bibfnamefont {R.~J.}\ \bibnamefont {Gleiser}}\ and\ \bibinfo {author} {\bibfnamefont {G.}~\bibnamefont {Dotti}},\ }\href {\doibase 10.1103/PhysRevD.72.124002} {\bibfield  {journal} {\bibinfo  {journal} {Phys. Rev. D}\ }\textbf {\bibinfo {volume} {72}},\ \bibinfo {pages} {124002} (\bibinfo {year} {2005})},\ \Eprint {http://arxiv.org/abs/gr-qc/0510069} {arXiv:gr-qc/0510069} \BibitemShut {NoStop}%
\bibitem [{\citenamefont {Gonzalez}\ \emph {et~al.}(2012)\citenamefont {Gonzalez}, \citenamefont {Saavedra},\ and\ \citenamefont {Vasquez}}]{Gonzalez:2012de}%
  \BibitemOpen
  \bibfield  {author} {\bibinfo {author} {\bibfnamefont {P.~A.}\ \bibnamefont {Gonzalez}}, \bibinfo {author} {\bibfnamefont {J.}~\bibnamefont {Saavedra}}, \ and\ \bibinfo {author} {\bibfnamefont {Y.}~\bibnamefont {Vasquez}},\ }\href {\doibase 10.1142/S021827181250054X} {\bibfield  {journal} {\bibinfo  {journal} {Int. J. Mod. Phys. D}\ }\textbf {\bibinfo {volume} {21}},\ \bibinfo {pages} {1250054} (\bibinfo {year} {2012})},\ \Eprint {http://arxiv.org/abs/1201.4521} {arXiv:1201.4521 [gr-qc]} \BibitemShut {NoStop}%
\bibitem [{\citenamefont {Becar}\ \emph {et~al.}(2013)\citenamefont {Becar}, \citenamefont {Gonzalez},\ and\ \citenamefont {Vasquez}}]{Becar:2012bj}%
  \BibitemOpen
  \bibfield  {author} {\bibinfo {author} {\bibfnamefont {R.}~\bibnamefont {Becar}}, \bibinfo {author} {\bibfnamefont {P.~A.}\ \bibnamefont {Gonzalez}}, \ and\ \bibinfo {author} {\bibfnamefont {Y.}~\bibnamefont {Vasquez}},\ }\href {\doibase 10.1142/S0218271813500077} {\bibfield  {journal} {\bibinfo  {journal} {Int. J. Mod. Phys. D}\ }\textbf {\bibinfo {volume} {22}},\ \bibinfo {pages} {1350007} (\bibinfo {year} {2013})},\ \Eprint {http://arxiv.org/abs/1210.7561} {arXiv:1210.7561 [gr-qc]} \BibitemShut {NoStop}%
\bibitem [{\citenamefont {Catalan}\ \emph {et~al.}(2014)\citenamefont {Catalan}, \citenamefont {Cisternas}, \citenamefont {Gonzalez},\ and\ \citenamefont {Vasquez}}]{Catalan:2013eza}%
  \BibitemOpen
  \bibfield  {author} {\bibinfo {author} {\bibfnamefont {M.}~\bibnamefont {Catalan}}, \bibinfo {author} {\bibfnamefont {E.}~\bibnamefont {Cisternas}}, \bibinfo {author} {\bibfnamefont {P.~A.}\ \bibnamefont {Gonzalez}}, \ and\ \bibinfo {author} {\bibfnamefont {Y.}~\bibnamefont {Vasquez}},\ }\href {\doibase 10.1140/epjc/s10052-014-2813-7} {\bibfield  {journal} {\bibinfo  {journal} {Eur. Phys. J. C}\ }\textbf {\bibinfo {volume} {74}},\ \bibinfo {pages} {2813} (\bibinfo {year} {2014})},\ \Eprint {http://arxiv.org/abs/1312.6451} {arXiv:1312.6451 [gr-qc]} \BibitemShut {NoStop}%
\bibitem [{\citenamefont {Zinhailo}(2019)}]{Zinhailo:2019rwd}%
  \BibitemOpen
  \bibfield  {author} {\bibinfo {author} {\bibfnamefont {A.~F.}\ \bibnamefont {Zinhailo}},\ }\href {\doibase 10.1140/epjc/s10052-019-7425-9} {\bibfield  {journal} {\bibinfo  {journal} {Eur. Phys. J. C}\ }\textbf {\bibinfo {volume} {79}},\ \bibinfo {pages} {912} (\bibinfo {year} {2019})},\ \Eprint {http://arxiv.org/abs/1909.12664} {arXiv:1909.12664 [gr-qc]} \BibitemShut {NoStop}%
\bibitem [{\citenamefont {B\'ecar}\ \emph {et~al.}(2024)\citenamefont {B\'ecar}, \citenamefont {Gonz\'alez}, \citenamefont {Papantonopoulos},\ and\ \citenamefont {V\'asquez}}]{Becar:2023zbl}%
  \BibitemOpen
  \bibfield  {author} {\bibinfo {author} {\bibfnamefont {R.}~\bibnamefont {B\'ecar}}, \bibinfo {author} {\bibfnamefont {P.~A.}\ \bibnamefont {Gonz\'alez}}, \bibinfo {author} {\bibfnamefont {E.}~\bibnamefont {Papantonopoulos}}, \ and\ \bibinfo {author} {\bibfnamefont {Y.}~\bibnamefont {V\'asquez}},\ }\href {\doibase 10.1140/epjc/s10052-024-12553-8} {\bibfield  {journal} {\bibinfo  {journal} {Eur. Phys. J. C}\ }\textbf {\bibinfo {volume} {84}},\ \bibinfo {pages} {329} (\bibinfo {year} {2024})},\ \Eprint {http://arxiv.org/abs/2310.00857} {arXiv:2310.00857 [gr-qc]} \BibitemShut {NoStop}%
\bibitem [{\citenamefont {Konoplya}\ and\ \citenamefont {Zhidenko}(2007{\natexlab{b}})}]{Konoplya:2006ar}%
  \BibitemOpen
  \bibfield  {author} {\bibinfo {author} {\bibfnamefont {R.~A.}\ \bibnamefont {Konoplya}}\ and\ \bibinfo {author} {\bibfnamefont {A.}~\bibnamefont {Zhidenko}},\ }\href {\doibase 10.1016/j.physletb.2007.03.018} {\bibfield  {journal} {\bibinfo  {journal} {Phys. Lett. B}\ }\textbf {\bibinfo {volume} {648}},\ \bibinfo {pages} {236} (\bibinfo {year} {2007}{\natexlab{b}})},\ \Eprint {http://arxiv.org/abs/hep-th/0611226} {arXiv:hep-th/0611226} \BibitemShut {NoStop}%
\bibitem [{\citenamefont {Chen}\ and\ \citenamefont {Kotla\v{r}\'\i{}k}(2023)}]{Chen:2023akf}%
  \BibitemOpen
  \bibfield  {author} {\bibinfo {author} {\bibfnamefont {C.-Y.}\ \bibnamefont {Chen}}\ and\ \bibinfo {author} {\bibfnamefont {P.}~\bibnamefont {Kotla\v{r}\'\i{}k}},\ }\href {\doibase 10.1103/PhysRevD.108.064052} {\bibfield  {journal} {\bibinfo  {journal} {Phys. Rev. D}\ }\textbf {\bibinfo {volume} {108}},\ \bibinfo {pages} {064052} (\bibinfo {year} {2023})},\ \Eprint {http://arxiv.org/abs/2307.07360} {arXiv:2307.07360 [gr-qc]} \BibitemShut {NoStop}%
\bibitem [{\citenamefont {Konoplya}\ and\ \citenamefont {Zhidenko}(2007{\natexlab{c}})}]{Konoplya:2006rv}%
  \BibitemOpen
  \bibfield  {author} {\bibinfo {author} {\bibfnamefont {R.~A.}\ \bibnamefont {Konoplya}}\ and\ \bibinfo {author} {\bibfnamefont {A.}~\bibnamefont {Zhidenko}},\ }\href {\doibase 10.1016/j.physletb.2006.11.036} {\bibfield  {journal} {\bibinfo  {journal} {Phys. Lett. B}\ }\textbf {\bibinfo {volume} {644}},\ \bibinfo {pages} {186} (\bibinfo {year} {2007}{\natexlab{c}})},\ \Eprint {http://arxiv.org/abs/gr-qc/0605082} {arXiv:gr-qc/0605082} \BibitemShut {NoStop}%
\bibitem [{\citenamefont {Dubinsky}\ and\ \citenamefont {Zinhailo}(2024{\natexlab{a}})}]{Dubinsky:2024hmn}%
  \BibitemOpen
  \bibfield  {author} {\bibinfo {author} {\bibfnamefont {A.}~\bibnamefont {Dubinsky}}\ and\ \bibinfo {author} {\bibfnamefont {A.}~\bibnamefont {Zinhailo}},\ }\href {\doibase 10.1140/epjc/s10052-024-13206-6} {\bibfield  {journal} {\bibinfo  {journal} {Eur. Phys. J. C}\ }\textbf {\bibinfo {volume} {84}},\ \bibinfo {pages} {847} (\bibinfo {year} {2024}{\natexlab{a}})},\ \Eprint {http://arxiv.org/abs/2404.01834} {arXiv:2404.01834 [gr-qc]} \BibitemShut {NoStop}%
\bibitem [{\citenamefont {Dubinsky}(2024{\natexlab{c}})}]{Dubinsky:2024fvi}%
  \BibitemOpen
  \bibfield  {author} {\bibinfo {author} {\bibfnamefont {A.}~\bibnamefont {Dubinsky}},\ }\href {\doibase 10.13140/RG.2.2.35132.24961} {\  (\bibinfo {year} {2024}{\natexlab{c}}),\ 10.13140/RG.2.2.35132.24961},\ \Eprint {http://arxiv.org/abs/2405.13552} {arXiv:2405.13552 [gr-qc]} \BibitemShut {NoStop}%
\bibitem [{\citenamefont {Al-Badawi}(2023)}]{Al-Badawi:2023lvx}%
  \BibitemOpen
  \bibfield  {author} {\bibinfo {author} {\bibfnamefont {A.}~\bibnamefont {Al-Badawi}},\ }\href {\doibase 10.1140/epjc/s10052-023-11804-4} {\bibfield  {journal} {\bibinfo  {journal} {Eur. Phys. J. C}\ }\textbf {\bibinfo {volume} {83}},\ \bibinfo {pages} {620} (\bibinfo {year} {2023})},\ \Eprint {http://arxiv.org/abs/2307.07974} {arXiv:2307.07974 [gr-qc]} \BibitemShut {NoStop}%
\bibitem [{\citenamefont {Konoplya}\ \emph {et~al.}(2020{\natexlab{a}})\citenamefont {Konoplya}, \citenamefont {Zinhailo},\ and\ \citenamefont {Stuchlik}}]{Konoplya:2020jgt}%
  \BibitemOpen
  \bibfield  {author} {\bibinfo {author} {\bibfnamefont {R.~A.}\ \bibnamefont {Konoplya}}, \bibinfo {author} {\bibfnamefont {A.~F.}\ \bibnamefont {Zinhailo}}, \ and\ \bibinfo {author} {\bibfnamefont {Z.}~\bibnamefont {Stuchlik}},\ }\href {\doibase 10.1103/PhysRevD.102.044023} {\bibfield  {journal} {\bibinfo  {journal} {Phys. Rev. D}\ }\textbf {\bibinfo {volume} {102}},\ \bibinfo {pages} {044023} (\bibinfo {year} {2020}{\natexlab{a}})},\ \Eprint {http://arxiv.org/abs/2006.10462} {arXiv:2006.10462 [gr-qc]} \BibitemShut {NoStop}%
\bibitem [{\citenamefont {Guo}\ \emph {et~al.}(2024)\citenamefont {Guo}, \citenamefont {Tan},\ and\ \citenamefont {Liu}}]{Guo:2023nkd}%
  \BibitemOpen
  \bibfield  {author} {\bibinfo {author} {\bibfnamefont {W.-D.}\ \bibnamefont {Guo}}, \bibinfo {author} {\bibfnamefont {Q.}~\bibnamefont {Tan}}, \ and\ \bibinfo {author} {\bibfnamefont {Y.-X.}\ \bibnamefont {Liu}},\ }\href {\doibase 10.1088/1475-7516/2024/07/008} {\bibfield  {journal} {\bibinfo  {journal} {JCAP}\ }\textbf {\bibinfo {volume} {07}},\ \bibinfo {pages} {008} (\bibinfo {year} {2024})},\ \Eprint {http://arxiv.org/abs/2312.16605} {arXiv:2312.16605 [gr-qc]} \BibitemShut {NoStop}%
\bibitem [{\citenamefont {Skvortsova}(2024{\natexlab{b}})}]{Skvortsova:2024atk}%
  \BibitemOpen
  \bibfield  {author} {\bibinfo {author} {\bibfnamefont {M.}~\bibnamefont {Skvortsova}},\ }\href {\doibase 10.1002/prop.202400132} {\bibfield  {journal} {\bibinfo  {journal} {Fortsch. Phys.}\ }\textbf {\bibinfo {volume} {72}},\ \bibinfo {pages} {2400132} (\bibinfo {year} {2024}{\natexlab{b}})},\ \Eprint {http://arxiv.org/abs/2405.06390} {arXiv:2405.06390 [gr-qc]} \BibitemShut {NoStop}%
\bibitem [{\citenamefont {Kodama}\ \emph {et~al.}(2010)\citenamefont {Kodama}, \citenamefont {Konoplya},\ and\ \citenamefont {Zhidenko}}]{Kodama:2009bf}%
  \BibitemOpen
  \bibfield  {author} {\bibinfo {author} {\bibfnamefont {H.}~\bibnamefont {Kodama}}, \bibinfo {author} {\bibfnamefont {R.~A.}\ \bibnamefont {Konoplya}}, \ and\ \bibinfo {author} {\bibfnamefont {A.}~\bibnamefont {Zhidenko}},\ }\href {\doibase 10.1103/PhysRevD.81.044007} {\bibfield  {journal} {\bibinfo  {journal} {Phys. Rev. D}\ }\textbf {\bibinfo {volume} {81}},\ \bibinfo {pages} {044007} (\bibinfo {year} {2010})},\ \Eprint {http://arxiv.org/abs/0904.2154} {arXiv:0904.2154 [gr-qc]} \BibitemShut {NoStop}%
\bibitem [{\citenamefont {Skvortsova}(2024{\natexlab{c}})}]{Skvortsova:2023zmj}%
  \BibitemOpen
  \bibfield  {author} {\bibinfo {author} {\bibfnamefont {M.}~\bibnamefont {Skvortsova}},\ }\href {\doibase 10.1002/prop.202400036} {\bibfield  {journal} {\bibinfo  {journal} {Fortsch. Phys.}\ }\textbf {\bibinfo {volume} {72}},\ \bibinfo {pages} {2400036} (\bibinfo {year} {2024}{\natexlab{c}})},\ \Eprint {http://arxiv.org/abs/2311.11650} {arXiv:2311.11650 [gr-qc]} \BibitemShut {NoStop}%
\bibitem [{\citenamefont {Skvortsova}(2025)}]{Skvortsova:2024eqi}%
  \BibitemOpen
  \bibfield  {author} {\bibinfo {author} {\bibfnamefont {M.}~\bibnamefont {Skvortsova}},\ }\href {\doibase 10.1209/0295-5075/adaee2} {\bibfield  {journal} {\bibinfo  {journal} {EPL}\ }\textbf {\bibinfo {volume} {149}},\ \bibinfo {pages} {59001} (\bibinfo {year} {2025})},\ \Eprint {http://arxiv.org/abs/2503.03650} {arXiv:2503.03650 [gr-qc]} \BibitemShut {NoStop}%
\bibitem [{\citenamefont {Churilova}\ \emph {et~al.}(2021)\citenamefont {Churilova}, \citenamefont {Konoplya}, \citenamefont {Stuchlik},\ and\ \citenamefont {Zhidenko}}]{Churilova:2021tgn}%
  \BibitemOpen
  \bibfield  {author} {\bibinfo {author} {\bibfnamefont {M.~S.}\ \bibnamefont {Churilova}}, \bibinfo {author} {\bibfnamefont {R.~A.}\ \bibnamefont {Konoplya}}, \bibinfo {author} {\bibfnamefont {Z.}~\bibnamefont {Stuchlik}}, \ and\ \bibinfo {author} {\bibfnamefont {A.}~\bibnamefont {Zhidenko}},\ }\href {\doibase 10.1088/1475-7516/2021/10/010} {\bibfield  {journal} {\bibinfo  {journal} {JCAP}\ }\textbf {\bibinfo {volume} {10}},\ \bibinfo {pages} {010} (\bibinfo {year} {2021})},\ \Eprint {http://arxiv.org/abs/2107.05977} {arXiv:2107.05977 [gr-qc]} \BibitemShut {NoStop}%
\bibitem [{\citenamefont {L\"utf\"uo\u{g}lu}(2025)}]{Lutfuoglu:2025hjy}%
  \BibitemOpen
  \bibfield  {author} {\bibinfo {author} {\bibfnamefont {B.~C.}\ \bibnamefont {L\"utf\"uo\u{g}lu}},\ }\href@noop {} {\  (\bibinfo {year} {2025})},\ \Eprint {http://arxiv.org/abs/2503.16087} {arXiv:2503.16087 [gr-qc]} \BibitemShut {NoStop}%
\bibitem [{\citenamefont {Hamil}\ and\ \citenamefont {L\"utf\"uo\u{g}lu}(2025)}]{Hamil:2025cms}%
  \BibitemOpen
  \bibfield  {author} {\bibinfo {author} {\bibfnamefont {B.}~\bibnamefont {Hamil}}\ and\ \bibinfo {author} {\bibfnamefont {B.~C.}\ \bibnamefont {L\"utf\"uo\u{g}lu}},\ }\href@noop {} {\  (\bibinfo {year} {2025})},\ \Eprint {http://arxiv.org/abs/2503.17474} {arXiv:2503.17474 [gr-qc]} \BibitemShut {NoStop}%
\bibitem [{\citenamefont {Gong}\ \emph {et~al.}(2024)\citenamefont {Gong}, \citenamefont {Li}, \citenamefont {Zhang}, \citenamefont {Fu},\ and\ \citenamefont {Wu}}]{Gong:2023ghh}%
  \BibitemOpen
  \bibfield  {author} {\bibinfo {author} {\bibfnamefont {H.}~\bibnamefont {Gong}}, \bibinfo {author} {\bibfnamefont {S.}~\bibnamefont {Li}}, \bibinfo {author} {\bibfnamefont {D.}~\bibnamefont {Zhang}}, \bibinfo {author} {\bibfnamefont {G.}~\bibnamefont {Fu}}, \ and\ \bibinfo {author} {\bibfnamefont {J.-P.}\ \bibnamefont {Wu}},\ }\href {\doibase 10.1103/PhysRevD.110.044040} {\bibfield  {journal} {\bibinfo  {journal} {Phys. Rev. D}\ }\textbf {\bibinfo {volume} {110}},\ \bibinfo {pages} {044040} (\bibinfo {year} {2024})},\ \Eprint {http://arxiv.org/abs/2312.17639} {arXiv:2312.17639 [gr-qc]} \BibitemShut {NoStop}%
\bibitem [{\citenamefont {Mashhoon}(1982)}]{Mashhoon:1982im}%
  \BibitemOpen
  \bibfield  {author} {\bibinfo {author} {\bibfnamefont {B.}~\bibnamefont {Mashhoon}},\ }in\ \href@noop {} {\emph {\bibinfo {booktitle} {{3rd Marcel Grossmann Meeting on the Recent Developments of General Relativity}}}}\ (\bibinfo {year} {1982})\BibitemShut {NoStop}%
\bibitem [{\citenamefont {Ferrari}\ and\ \citenamefont {Mashhoon}(1984)}]{Ferrari:1984zz}%
  \BibitemOpen
  \bibfield  {author} {\bibinfo {author} {\bibfnamefont {V.}~\bibnamefont {Ferrari}}\ and\ \bibinfo {author} {\bibfnamefont {B.}~\bibnamefont {Mashhoon}},\ }\href {\doibase 10.1103/PhysRevD.30.295} {\bibfield  {journal} {\bibinfo  {journal} {Phys. Rev. D}\ }\textbf {\bibinfo {volume} {30}},\ \bibinfo {pages} {295} (\bibinfo {year} {1984})}\BibitemShut {NoStop}%
\bibitem [{\citenamefont {Konoplya}(2005)}]{Konoplya:2004xx}%
  \BibitemOpen
  \bibfield  {author} {\bibinfo {author} {\bibfnamefont {R.}~\bibnamefont {Konoplya}},\ }\href {\doibase 10.1103/PhysRevD.71.024038} {\bibfield  {journal} {\bibinfo  {journal} {Phys. Rev. D}\ }\textbf {\bibinfo {volume} {71}},\ \bibinfo {pages} {024038} (\bibinfo {year} {2005})},\ \Eprint {http://arxiv.org/abs/hep-th/0410057} {arXiv:hep-th/0410057} \BibitemShut {NoStop}%
\bibitem [{\citenamefont {Abdalla}\ \emph {et~al.}(2005)\citenamefont {Abdalla}, \citenamefont {Konoplya},\ and\ \citenamefont {Molina}}]{Abdalla:2005hu}%
  \BibitemOpen
  \bibfield  {author} {\bibinfo {author} {\bibfnamefont {E.}~\bibnamefont {Abdalla}}, \bibinfo {author} {\bibfnamefont {R.~A.}\ \bibnamefont {Konoplya}}, \ and\ \bibinfo {author} {\bibfnamefont {C.}~\bibnamefont {Molina}},\ }\href {\doibase 10.1103/PhysRevD.72.084006} {\bibfield  {journal} {\bibinfo  {journal} {Phys. Rev. D}\ }\textbf {\bibinfo {volume} {72}},\ \bibinfo {pages} {084006} (\bibinfo {year} {2005})},\ \Eprint {http://arxiv.org/abs/hep-th/0507100} {arXiv:hep-th/0507100} \BibitemShut {NoStop}%
\bibitem [{\citenamefont {Konoplya}\ and\ \citenamefont {Zhidenko}(2020{\natexlab{c}})}]{Konoplya:2020der}%
  \BibitemOpen
  \bibfield  {author} {\bibinfo {author} {\bibfnamefont {R.~A.}\ \bibnamefont {Konoplya}}\ and\ \bibinfo {author} {\bibfnamefont {A.}~\bibnamefont {Zhidenko}},\ }\href {\doibase 10.1016/j.physletb.2020.135607} {\bibfield  {journal} {\bibinfo  {journal} {Phys. Lett. B}\ }\textbf {\bibinfo {volume} {807}},\ \bibinfo {pages} {135607} (\bibinfo {year} {2020}{\natexlab{c}})},\ \Eprint {http://arxiv.org/abs/2005.02225} {arXiv:2005.02225 [gr-qc]} \BibitemShut {NoStop}%
\bibitem [{\citenamefont {Konoplya}\ \emph {et~al.}(2019{\natexlab{c}})\citenamefont {Konoplya}, \citenamefont {Zinhailo},\ and\ \citenamefont {Stuchl\'\i{}k}}]{Konoplya:2019hml}%
  \BibitemOpen
  \bibfield  {author} {\bibinfo {author} {\bibfnamefont {R.~A.}\ \bibnamefont {Konoplya}}, \bibinfo {author} {\bibfnamefont {A.~F.}\ \bibnamefont {Zinhailo}}, \ and\ \bibinfo {author} {\bibfnamefont {Z.}~\bibnamefont {Stuchl\'\i{}k}},\ }\href {\doibase 10.1103/PhysRevD.99.124042} {\bibfield  {journal} {\bibinfo  {journal} {Phys. Rev. D}\ }\textbf {\bibinfo {volume} {99}},\ \bibinfo {pages} {124042} (\bibinfo {year} {2019}{\natexlab{c}})},\ \Eprint {http://arxiv.org/abs/1903.03483} {arXiv:1903.03483 [gr-qc]} \BibitemShut {NoStop}%
\bibitem [{\citenamefont {Paul}(2024)}]{Paul:2023eep}%
  \BibitemOpen
  \bibfield  {author} {\bibinfo {author} {\bibfnamefont {P.}~\bibnamefont {Paul}},\ }\href {\doibase 10.1140/epjc/s10052-024-12563-6} {\bibfield  {journal} {\bibinfo  {journal} {Eur. Phys. J. C}\ }\textbf {\bibinfo {volume} {84}},\ \bibinfo {pages} {218} (\bibinfo {year} {2024})},\ \Eprint {http://arxiv.org/abs/2312.16479} {arXiv:2312.16479 [gr-qc]} \BibitemShut {NoStop}%
\bibitem [{\citenamefont {Kokkotas}\ \emph {et~al.}(2017{\natexlab{a}})\citenamefont {Kokkotas}, \citenamefont {Konoplya},\ and\ \citenamefont {Zhidenko}}]{Kokkotas:2017zwt}%
  \BibitemOpen
  \bibfield  {author} {\bibinfo {author} {\bibfnamefont {K.}~\bibnamefont {Kokkotas}}, \bibinfo {author} {\bibfnamefont {R.~A.}\ \bibnamefont {Konoplya}}, \ and\ \bibinfo {author} {\bibfnamefont {A.}~\bibnamefont {Zhidenko}},\ }\href {\doibase 10.1103/PhysRevD.96.064007} {\bibfield  {journal} {\bibinfo  {journal} {Phys. Rev. D}\ }\textbf {\bibinfo {volume} {96}},\ \bibinfo {pages} {064007} (\bibinfo {year} {2017}{\natexlab{a}})},\ \Eprint {http://arxiv.org/abs/1705.09875} {arXiv:1705.09875 [gr-qc]} \BibitemShut {NoStop}%
\bibitem [{\citenamefont {Zinhailo}(2018)}]{Zinhailo:2018ska}%
  \BibitemOpen
  \bibfield  {author} {\bibinfo {author} {\bibfnamefont {A.~F.}\ \bibnamefont {Zinhailo}},\ }\href {\doibase 10.1140/epjc/s10052-018-6467-8} {\bibfield  {journal} {\bibinfo  {journal} {Eur. Phys. J. C}\ }\textbf {\bibinfo {volume} {78}},\ \bibinfo {pages} {992} (\bibinfo {year} {2018})},\ \Eprint {http://arxiv.org/abs/1809.03913} {arXiv:1809.03913 [gr-qc]} \BibitemShut {NoStop}%
\bibitem [{\citenamefont {Kokkotas}\ \emph {et~al.}(2015)\citenamefont {Kokkotas}, \citenamefont {Konoplya},\ and\ \citenamefont {Zhidenko}}]{Kokkotas:2015uma}%
  \BibitemOpen
  \bibfield  {author} {\bibinfo {author} {\bibfnamefont {K.~D.}\ \bibnamefont {Kokkotas}}, \bibinfo {author} {\bibfnamefont {R.~A.}\ \bibnamefont {Konoplya}}, \ and\ \bibinfo {author} {\bibfnamefont {A.}~\bibnamefont {Zhidenko}},\ }\href {\doibase 10.1103/PhysRevD.92.064022} {\bibfield  {journal} {\bibinfo  {journal} {Phys. Rev. D}\ }\textbf {\bibinfo {volume} {92}},\ \bibinfo {pages} {064022} (\bibinfo {year} {2015})},\ \Eprint {http://arxiv.org/abs/1507.05649} {arXiv:1507.05649 [gr-qc]} \BibitemShut {NoStop}%
\bibitem [{\citenamefont {Malybayev}\ \emph {et~al.}(2021)\citenamefont {Malybayev}, \citenamefont {Boshkayev},\ and\ \citenamefont {Ivashchuk}}]{Malybayev:2021lfq}%
  \BibitemOpen
  \bibfield  {author} {\bibinfo {author} {\bibfnamefont {A.~N.}\ \bibnamefont {Malybayev}}, \bibinfo {author} {\bibfnamefont {K.~A.}\ \bibnamefont {Boshkayev}}, \ and\ \bibinfo {author} {\bibfnamefont {V.~D.}\ \bibnamefont {Ivashchuk}},\ }\href {\doibase 10.1140/epjc/s10052-021-09252-z} {\bibfield  {journal} {\bibinfo  {journal} {Eur. Phys. J. C}\ }\textbf {\bibinfo {volume} {81}},\ \bibinfo {pages} {475} (\bibinfo {year} {2021})},\ \Eprint {http://arxiv.org/abs/2103.10920} {arXiv:2103.10920 [gr-qc]} \BibitemShut {NoStop}%
\bibitem [{\citenamefont {Konoplya}\ and\ \citenamefont {Stashko}(2024)}]{Konoplya:2024lch}%
  \BibitemOpen
  \bibfield  {author} {\bibinfo {author} {\bibfnamefont {R.~A.}\ \bibnamefont {Konoplya}}\ and\ \bibinfo {author} {\bibfnamefont {O.~S.}\ \bibnamefont {Stashko}},\ }\href@noop {} {\  (\bibinfo {year} {2024})},\ \Eprint {http://arxiv.org/abs/2408.02578} {arXiv:2408.02578 [gr-qc]} \BibitemShut {NoStop}%
\bibitem [{\citenamefont {Malik}(2025{\natexlab{a}})}]{Malik:2024elk}%
  \BibitemOpen
  \bibfield  {author} {\bibinfo {author} {\bibfnamefont {Z.}~\bibnamefont {Malik}},\ }\href {\doibase 10.1007/s10773-024-05847-w} {\bibfield  {journal} {\bibinfo  {journal} {Int. J. Theor. Phys.}\ }\textbf {\bibinfo {volume} {64}},\ \bibinfo {pages} {30} (\bibinfo {year} {2025}{\natexlab{a}})}\BibitemShut {NoStop}%
\bibitem [{\citenamefont {Kanti}\ and\ \citenamefont {Konoplya}(2006)}]{Kanti:2005xa}%
  \BibitemOpen
  \bibfield  {author} {\bibinfo {author} {\bibfnamefont {P.}~\bibnamefont {Kanti}}\ and\ \bibinfo {author} {\bibfnamefont {R.~A.}\ \bibnamefont {Konoplya}},\ }\href {\doibase 10.1103/PhysRevD.73.044002} {\bibfield  {journal} {\bibinfo  {journal} {Phys. Rev. D}\ }\textbf {\bibinfo {volume} {73}},\ \bibinfo {pages} {044002} (\bibinfo {year} {2006})},\ \Eprint {http://arxiv.org/abs/hep-th/0512257} {arXiv:hep-th/0512257} \BibitemShut {NoStop}%
\bibitem [{\citenamefont {Malik}(2024{\natexlab{f}})}]{Malik:2024itg}%
  \BibitemOpen
  \bibfield  {author} {\bibinfo {author} {\bibfnamefont {Z.}~\bibnamefont {Malik}},\ }\href {\doibase 10.13140/RG.2.2.13059.72487} {\  (\bibinfo {year} {2024}{\natexlab{f}}),\ 10.13140/RG.2.2.13059.72487}\BibitemShut {NoStop}%
\bibitem [{\citenamefont {Konoplya}\ and\ \citenamefont {Liu}(2012)}]{Konoplya:2012vh}%
  \BibitemOpen
  \bibfield  {author} {\bibinfo {author} {\bibfnamefont {R.~A.}\ \bibnamefont {Konoplya}}\ and\ \bibinfo {author} {\bibfnamefont {Y.-C.}\ \bibnamefont {Liu}},\ }\href {\doibase 10.1103/PhysRevD.86.084007} {\bibfield  {journal} {\bibinfo  {journal} {Phys. Rev. D}\ }\textbf {\bibinfo {volume} {86}},\ \bibinfo {pages} {084007} (\bibinfo {year} {2012})},\ \Eprint {http://arxiv.org/abs/1205.5311} {arXiv:1205.5311 [gr-qc]} \BibitemShut {NoStop}%
\bibitem [{\citenamefont {Zhidenko}(2004)}]{Zhidenko:2003wq}%
  \BibitemOpen
  \bibfield  {author} {\bibinfo {author} {\bibfnamefont {A.}~\bibnamefont {Zhidenko}},\ }\href {\doibase 10.1088/0264-9381/21/1/019} {\bibfield  {journal} {\bibinfo  {journal} {Class. Quant. Grav.}\ }\textbf {\bibinfo {volume} {21}},\ \bibinfo {pages} {273} (\bibinfo {year} {2004})},\ \Eprint {http://arxiv.org/abs/gr-qc/0307012} {arXiv:gr-qc/0307012} \BibitemShut {NoStop}%
\bibitem [{\citenamefont {Konoplya}(2020)}]{Konoplya:2019xmn}%
  \BibitemOpen
  \bibfield  {author} {\bibinfo {author} {\bibfnamefont {R.~A.}\ \bibnamefont {Konoplya}},\ }\href {\doibase 10.1016/j.physletb.2020.135363} {\bibfield  {journal} {\bibinfo  {journal} {Phys. Lett. B}\ }\textbf {\bibinfo {volume} {804}},\ \bibinfo {pages} {135363} (\bibinfo {year} {2020})},\ \Eprint {http://arxiv.org/abs/1912.10582} {arXiv:1912.10582 [gr-qc]} \BibitemShut {NoStop}%
\bibitem [{\citenamefont {Konoplya}\ and\ \citenamefont {Zhidenko}(2020{\natexlab{d}})}]{Konoplya:2020hyk}%
  \BibitemOpen
  \bibfield  {author} {\bibinfo {author} {\bibfnamefont {R.~A.}\ \bibnamefont {Konoplya}}\ and\ \bibinfo {author} {\bibfnamefont {A.}~\bibnamefont {Zhidenko}},\ }\href {\doibase 10.1103/PhysRevD.101.124004} {\bibfield  {journal} {\bibinfo  {journal} {Phys. Rev. D}\ }\textbf {\bibinfo {volume} {101}},\ \bibinfo {pages} {124004} (\bibinfo {year} {2020}{\natexlab{d}})},\ \Eprint {http://arxiv.org/abs/2001.06100} {arXiv:2001.06100 [gr-qc]} \BibitemShut {NoStop}%
\bibitem [{\citenamefont {Churilova}(2019)}]{Churilova:2019jqx}%
  \BibitemOpen
  \bibfield  {author} {\bibinfo {author} {\bibfnamefont {M.~S.}\ \bibnamefont {Churilova}},\ }\href {\doibase 10.1140/epjc/s10052-019-7146-0} {\bibfield  {journal} {\bibinfo  {journal} {Eur. Phys. J. C}\ }\textbf {\bibinfo {volume} {79}},\ \bibinfo {pages} {629} (\bibinfo {year} {2019})},\ \Eprint {http://arxiv.org/abs/1905.04536} {arXiv:1905.04536 [gr-qc]} \BibitemShut {NoStop}%
\bibitem [{\citenamefont {Konoplya}\ and\ \citenamefont {Zhidenko}(2023{\natexlab{d}})}]{Konoplya:2023owh}%
  \BibitemOpen
  \bibfield  {author} {\bibinfo {author} {\bibfnamefont {R.~A.}\ \bibnamefont {Konoplya}}\ and\ \bibinfo {author} {\bibfnamefont {A.}~\bibnamefont {Zhidenko}},\ }\href {\doibase 10.1088/1475-7516/2023/08/008} {\bibfield  {journal} {\bibinfo  {journal} {JCAP}\ }\textbf {\bibinfo {volume} {08}},\ \bibinfo {pages} {008} (\bibinfo {year} {2023}{\natexlab{d}})},\ \Eprint {http://arxiv.org/abs/2303.03130} {arXiv:2303.03130 [gr-qc]} \BibitemShut {NoStop}%
\bibitem [{\citenamefont {Konoplya}(2021)}]{Konoplya:2021ube}%
  \BibitemOpen
  \bibfield  {author} {\bibinfo {author} {\bibfnamefont {R.~A.}\ \bibnamefont {Konoplya}},\ }\href {\doibase 10.1016/j.physletb.2021.136734} {\bibfield  {journal} {\bibinfo  {journal} {Phys. Lett. B}\ }\textbf {\bibinfo {volume} {823}},\ \bibinfo {pages} {136734} (\bibinfo {year} {2021})},\ \Eprint {http://arxiv.org/abs/2109.01640} {arXiv:2109.01640 [gr-qc]} \BibitemShut {NoStop}%
\bibitem [{\citenamefont {Konoplya}\ and\ \citenamefont {Zhidenko}(2022{\natexlab{b}})}]{Konoplya:2022hbl}%
  \BibitemOpen
  \bibfield  {author} {\bibinfo {author} {\bibfnamefont {R.~A.}\ \bibnamefont {Konoplya}}\ and\ \bibinfo {author} {\bibfnamefont {A.}~\bibnamefont {Zhidenko}},\ }\href {\doibase 10.3847/1538-4357/ac76bc} {\bibfield  {journal} {\bibinfo  {journal} {Astrophys. J.}\ }\textbf {\bibinfo {volume} {933}},\ \bibinfo {pages} {166} (\bibinfo {year} {2022}{\natexlab{b}})},\ \Eprint {http://arxiv.org/abs/2202.02205} {arXiv:2202.02205 [gr-qc]} \BibitemShut {NoStop}%
\bibitem [{\citenamefont {Konoplya}(2023{\natexlab{b}})}]{Konoplya:2023ppx}%
  \BibitemOpen
  \bibfield  {author} {\bibinfo {author} {\bibfnamefont {R.~A.}\ \bibnamefont {Konoplya}},\ }\href {\doibase 10.1088/1475-7516/2023/07/001} {\bibfield  {journal} {\bibinfo  {journal} {JCAP}\ }\textbf {\bibinfo {volume} {07}},\ \bibinfo {pages} {001} (\bibinfo {year} {2023}{\natexlab{b}})},\ \Eprint {http://arxiv.org/abs/2305.09187} {arXiv:2305.09187 [gr-qc]} \BibitemShut {NoStop}%
\bibitem [{\citenamefont {Konoplya}\ and\ \citenamefont {Zhidenko}(2024{\natexlab{a}})}]{Konoplya:2024hfg}%
  \BibitemOpen
  \bibfield  {author} {\bibinfo {author} {\bibfnamefont {R.~A.}\ \bibnamefont {Konoplya}}\ and\ \bibinfo {author} {\bibfnamefont {A.}~\bibnamefont {Zhidenko}},\ }\href {\doibase 10.1103/PhysRevD.109.104005} {\bibfield  {journal} {\bibinfo  {journal} {Phys. Rev. D}\ }\textbf {\bibinfo {volume} {109}},\ \bibinfo {pages} {104005} (\bibinfo {year} {2024}{\natexlab{a}})},\ \Eprint {http://arxiv.org/abs/2403.07848} {arXiv:2403.07848 [gr-qc]} \BibitemShut {NoStop}%
\bibitem [{\citenamefont {Bolokhov}\ and\ \citenamefont {Ivashchuk}(2022)}]{Bolokhov:2022rqv}%
  \BibitemOpen
  \bibfield  {author} {\bibinfo {author} {\bibfnamefont {S.~V.}\ \bibnamefont {Bolokhov}}\ and\ \bibinfo {author} {\bibfnamefont {V.~D.}\ \bibnamefont {Ivashchuk}},\ }\href {\doibase 10.1140/epjc/s10052-022-10578-5} {\bibfield  {journal} {\bibinfo  {journal} {Eur. Phys. J. C}\ }\textbf {\bibinfo {volume} {82}},\ \bibinfo {pages} {624} (\bibinfo {year} {2022})},\ \Eprint {http://arxiv.org/abs/2201.03104} {arXiv:2201.03104 [gr-qc]} \BibitemShut {NoStop}%
\bibitem [{\citenamefont {Konoplya}\ \emph {et~al.}(2022)\citenamefont {Konoplya}, \citenamefont {Zinhailo}, \citenamefont {Kunz}, \citenamefont {Stuchlik},\ and\ \citenamefont {Zhidenko}}]{Konoplya:2022hll}%
  \BibitemOpen
  \bibfield  {author} {\bibinfo {author} {\bibfnamefont {R.~A.}\ \bibnamefont {Konoplya}}, \bibinfo {author} {\bibfnamefont {A.~F.}\ \bibnamefont {Zinhailo}}, \bibinfo {author} {\bibfnamefont {J.}~\bibnamefont {Kunz}}, \bibinfo {author} {\bibfnamefont {Z.}~\bibnamefont {Stuchlik}}, \ and\ \bibinfo {author} {\bibfnamefont {A.}~\bibnamefont {Zhidenko}},\ }\href {\doibase 10.1088/1475-7516/2022/10/091} {\bibfield  {journal} {\bibinfo  {journal} {JCAP}\ }\textbf {\bibinfo {volume} {10}},\ \bibinfo {pages} {091} (\bibinfo {year} {2022})},\ \Eprint {http://arxiv.org/abs/2206.14714} {arXiv:2206.14714 [gr-qc]} \BibitemShut {NoStop}%
\bibitem [{\citenamefont {Skvortsova}(2024{\natexlab{d}})}]{Skvortsova:2024wly}%
  \BibitemOpen
  \bibfield  {author} {\bibinfo {author} {\bibfnamefont {M.}~\bibnamefont {Skvortsova}},\ }\href {\doibase 10.1134/S020228932470018X} {\bibfield  {journal} {\bibinfo  {journal} {Grav. Cosmol.}\ }\textbf {\bibinfo {volume} {30}},\ \bibinfo {pages} {279} (\bibinfo {year} {2024}{\natexlab{d}})},\ \Eprint {http://arxiv.org/abs/2405.15807} {arXiv:2405.15807 [gr-qc]} \BibitemShut {NoStop}%
\bibitem [{\citenamefont {Toshmatov}\ \emph {et~al.}(2015)\citenamefont {Toshmatov}, \citenamefont {Abdujabbarov}, \citenamefont {Stuchlík},\ and\ \citenamefont {Ahmedov}}]{Toshmatov:2015wga}%
  \BibitemOpen
  \bibfield  {author} {\bibinfo {author} {\bibfnamefont {B.}~\bibnamefont {Toshmatov}}, \bibinfo {author} {\bibfnamefont {A.}~\bibnamefont {Abdujabbarov}}, \bibinfo {author} {\bibfnamefont {Z.}~\bibnamefont {Stuchlík}}, \ and\ \bibinfo {author} {\bibfnamefont {B.}~\bibnamefont {Ahmedov}},\ }\href {\doibase 10.1103/PhysRevD.91.083008} {\bibfield  {journal} {\bibinfo  {journal} {Phys. Rev. D}\ }\textbf {\bibinfo {volume} {91}},\ \bibinfo {pages} {083008} (\bibinfo {year} {2015})},\ \Eprint {http://arxiv.org/abs/1503.05737} {arXiv:1503.05737 [gr-qc]} \BibitemShut {NoStop}%
\bibitem [{\citenamefont {Konoplya}(2018)}]{Konoplya:2018ala}%
  \BibitemOpen
  \bibfield  {author} {\bibinfo {author} {\bibfnamefont {R.~A.}\ \bibnamefont {Konoplya}},\ }\href {\doibase 10.1016/j.physletb.2018.07.025} {\bibfield  {journal} {\bibinfo  {journal} {Phys. Lett. B}\ }\textbf {\bibinfo {volume} {784}},\ \bibinfo {pages} {43} (\bibinfo {year} {2018})},\ \Eprint {http://arxiv.org/abs/1805.04718} {arXiv:1805.04718 [gr-qc]} \BibitemShut {NoStop}%
\bibitem [{\citenamefont {Malik}(2024{\natexlab{g}})}]{Malik:2024wvs}%
  \BibitemOpen
  \bibfield  {author} {\bibinfo {author} {\bibfnamefont {Z.}~\bibnamefont {Malik}},\ }\href@noop {} {\  (\bibinfo {year} {2024}{\natexlab{g}})},\ \Eprint {http://arxiv.org/abs/2412.13385} {arXiv:2412.13385 [gr-qc]} \BibitemShut {NoStop}%
\bibitem [{\citenamefont {Jusufi}(2020{\natexlab{a}})}]{Jusufi:2019ltj}%
  \BibitemOpen
  \bibfield  {author} {\bibinfo {author} {\bibfnamefont {K.}~\bibnamefont {Jusufi}},\ }\href {\doibase 10.1103/PhysRevD.101.084055} {\bibfield  {journal} {\bibinfo  {journal} {Phys. Rev. D}\ }\textbf {\bibinfo {volume} {101}},\ \bibinfo {pages} {084055} (\bibinfo {year} {2020}{\natexlab{a}})},\ \Eprint {http://arxiv.org/abs/1912.13320} {arXiv:1912.13320 [gr-qc]} \BibitemShut {NoStop}%
\bibitem [{\citenamefont {Jusufi}(2020{\natexlab{b}})}]{Jusufi:2020dhz}%
  \BibitemOpen
  \bibfield  {author} {\bibinfo {author} {\bibfnamefont {K.}~\bibnamefont {Jusufi}},\ }\href {\doibase 10.1103/PhysRevD.101.124063} {\bibfield  {journal} {\bibinfo  {journal} {Phys. Rev. D}\ }\textbf {\bibinfo {volume} {101}},\ \bibinfo {pages} {124063} (\bibinfo {year} {2020}{\natexlab{b}})},\ \Eprint {http://arxiv.org/abs/2004.04664} {arXiv:2004.04664 [gr-qc]} \BibitemShut {NoStop}%
\bibitem [{\citenamefont {Malik}(2025{\natexlab{b}})}]{Malik:2024qsz}%
  \BibitemOpen
  \bibfield  {author} {\bibinfo {author} {\bibfnamefont {Z.}~\bibnamefont {Malik}},\ }\href {\doibase 10.1142/S0217751X2450132X} {\bibfield  {journal} {\bibinfo  {journal} {Int. J. Mod. Phys. A}\ }\textbf {\bibinfo {volume} {40}},\ \bibinfo {pages} {2450132} (\bibinfo {year} {2025}{\natexlab{b}})}\BibitemShut {NoStop}%
\bibitem [{\citenamefont {Konoplya}\ and\ \citenamefont {Zhidenko}(2017{\natexlab{c}})}]{Konoplya:2017lhs}%
  \BibitemOpen
  \bibfield  {author} {\bibinfo {author} {\bibfnamefont {R.~A.}\ \bibnamefont {Konoplya}}\ and\ \bibinfo {author} {\bibfnamefont {A.}~\bibnamefont {Zhidenko}},\ }\href {\doibase 10.1088/1475-7516/2017/05/050} {\bibfield  {journal} {\bibinfo  {journal} {JCAP}\ }\textbf {\bibinfo {volume} {05}},\ \bibinfo {pages} {050} (\bibinfo {year} {2017}{\natexlab{c}})},\ \Eprint {http://arxiv.org/abs/1705.01656} {arXiv:1705.01656 [hep-th]} \BibitemShut {NoStop}%
\bibitem [{\citenamefont {Malik}(2024{\natexlab{h}})}]{Malik:2024bmp}%
  \BibitemOpen
  \bibfield  {author} {\bibinfo {author} {\bibfnamefont {Z.}~\bibnamefont {Malik}},\ }\href {\doibase 10.1515/zna-2024-0153} {\bibfield  {journal} {\bibinfo  {journal} {Z. Naturforsch. A}\ }\textbf {\bibinfo {volume} {79}},\ \bibinfo {pages} {1063} (\bibinfo {year} {2024}{\natexlab{h}})}\BibitemShut {NoStop}%
\bibitem [{\citenamefont {Bolokhov}(2024{\natexlab{c}})}]{Bolokhov:2024ixe}%
  \BibitemOpen
  \bibfield  {author} {\bibinfo {author} {\bibfnamefont {S.~V.}\ \bibnamefont {Bolokhov}},\ }\href {\doibase 10.1140/epjc/s10052-024-12990-5} {\bibfield  {journal} {\bibinfo  {journal} {Eur. Phys. J. C}\ }\textbf {\bibinfo {volume} {84}},\ \bibinfo {pages} {634} (\bibinfo {year} {2024}{\natexlab{c}})},\ \Eprint {http://arxiv.org/abs/2404.09364} {arXiv:2404.09364 [gr-qc]} \BibitemShut {NoStop}%
\bibitem [{\citenamefont {Malik}(2024{\natexlab{i}})}]{Malik:2024zoo}%
  \BibitemOpen
  \bibfield  {author} {\bibinfo {author} {\bibfnamefont {Z.}~\bibnamefont {Malik}},\ }\href {\doibase 10.13140/RG.2.2.20827.58400} {\  (\bibinfo {year} {2024}{\natexlab{i}}),\ 10.13140/RG.2.2.20827.58400}\BibitemShut {NoStop}%
\bibitem [{\citenamefont {Dubinsky}\ and\ \citenamefont {Zinhailo}(2024{\natexlab{b}})}]{Dubinsky:2024nzo}%
  \BibitemOpen
  \bibfield  {author} {\bibinfo {author} {\bibfnamefont {A.}~\bibnamefont {Dubinsky}}\ and\ \bibinfo {author} {\bibfnamefont {A.~F.}\ \bibnamefont {Zinhailo}},\ }\href@noop {} {\  (\bibinfo {year} {2024}{\natexlab{b}})},\ \Eprint {http://arxiv.org/abs/2410.15232} {arXiv:2410.15232 [gr-qc]} \BibitemShut {NoStop}%
\bibitem [{\citenamefont {Konoplya}\ and\ \citenamefont {Molina}(2005)}]{Konoplya:2005et}%
  \BibitemOpen
  \bibfield  {author} {\bibinfo {author} {\bibfnamefont {R.~A.}\ \bibnamefont {Konoplya}}\ and\ \bibinfo {author} {\bibfnamefont {C.}~\bibnamefont {Molina}},\ }\href {\doibase 10.1103/PhysRevD.71.124009} {\bibfield  {journal} {\bibinfo  {journal} {Phys. Rev. D}\ }\textbf {\bibinfo {volume} {71}},\ \bibinfo {pages} {124009} (\bibinfo {year} {2005})},\ \Eprint {http://arxiv.org/abs/gr-qc/0504139} {arXiv:gr-qc/0504139} \BibitemShut {NoStop}%
\bibitem [{\citenamefont {Bronnikov}(1973)}]{Bronnikov:1973fh}%
  \BibitemOpen
  \bibfield  {author} {\bibinfo {author} {\bibfnamefont {K.~A.}\ \bibnamefont {Bronnikov}},\ }\href@noop {} {\bibfield  {journal} {\bibinfo  {journal} {Acta Phys. Polon. B}\ }\textbf {\bibinfo {volume} {4}},\ \bibinfo {pages} {251} (\bibinfo {year} {1973})}\BibitemShut {NoStop}%
\bibitem [{\citenamefont {Morris}\ and\ \citenamefont {Thorne}(1988)}]{Morris:1988cz}%
  \BibitemOpen
  \bibfield  {author} {\bibinfo {author} {\bibfnamefont {M.~S.}\ \bibnamefont {Morris}}\ and\ \bibinfo {author} {\bibfnamefont {K.~S.}\ \bibnamefont {Thorne}},\ }\href {\doibase 10.1119/1.15620} {\bibfield  {journal} {\bibinfo  {journal} {Am. J. Phys.}\ }\textbf {\bibinfo {volume} {56}},\ \bibinfo {pages} {395} (\bibinfo {year} {1988})}\BibitemShut {NoStop}%
\bibitem [{\citenamefont {Konoplya}\ and\ \citenamefont {Zhidenko}(2004)}]{Konoplya:2004uk}%
  \BibitemOpen
  \bibfield  {author} {\bibinfo {author} {\bibfnamefont {R.~A.}\ \bibnamefont {Konoplya}}\ and\ \bibinfo {author} {\bibfnamefont {A.}~\bibnamefont {Zhidenko}},\ }\href {\doibase 10.1088/1126-6708/2004/06/037} {\bibfield  {journal} {\bibinfo  {journal} {JHEP}\ }\textbf {\bibinfo {volume} {06}},\ \bibinfo {pages} {037} (\bibinfo {year} {2004})},\ \Eprint {http://arxiv.org/abs/hep-th/0402080} {arXiv:hep-th/0402080} \BibitemShut {NoStop}%
\bibitem [{\citenamefont {Dyatlov}(2012)}]{Dyatlov:2011jd}%
  \BibitemOpen
  \bibfield  {author} {\bibinfo {author} {\bibfnamefont {S.}~\bibnamefont {Dyatlov}},\ }\href {\doibase 10.1007/s00023-012-0159-y} {\bibfield  {journal} {\bibinfo  {journal} {Annales Henri Poincare}\ }\textbf {\bibinfo {volume} {13}},\ \bibinfo {pages} {1101} (\bibinfo {year} {2012})},\ \Eprint {http://arxiv.org/abs/1101.1260} {arXiv:1101.1260 [math.AP]} \BibitemShut {NoStop}%
\bibitem [{\citenamefont {Dyatlov}(2011)}]{Dyatlov:2011zz}%
  \BibitemOpen
  \bibfield  {author} {\bibinfo {author} {\bibfnamefont {S.}~\bibnamefont {Dyatlov}},\ }\href {\doibase 10.4310/MRL.2011.v18.n5.a19} {\bibfield  {journal} {\bibinfo  {journal} {Math. Res. Lett.}\ }\textbf {\bibinfo {volume} {18}},\ \bibinfo {pages} {1023} (\bibinfo {year} {2011})},\ \Eprint {http://arxiv.org/abs/1010.5201} {arXiv:1010.5201 [math.AP]} \BibitemShut {NoStop}%
\bibitem [{\citenamefont {Hintz}\ and\ \citenamefont {Vasy}(2018)}]{Hintz:2016gwb}%
  \BibitemOpen
  \bibfield  {author} {\bibinfo {author} {\bibfnamefont {P.}~\bibnamefont {Hintz}}\ and\ \bibinfo {author} {\bibfnamefont {A.}~\bibnamefont {Vasy}},\ }\href {\doibase 10.4310/ACTA.2018.v220.n1.a1} {\bibfield  {journal} {\bibinfo  {journal} {Acta Math.}\ }\textbf {\bibinfo {volume} {220}},\ \bibinfo {pages} {1} (\bibinfo {year} {2018})},\ \Eprint {http://arxiv.org/abs/1606.04014} {arXiv:1606.04014 [math.DG]} \BibitemShut {NoStop}%
\bibitem [{\citenamefont {Jansen}(2017)}]{Jansen:2017oag}%
  \BibitemOpen
  \bibfield  {author} {\bibinfo {author} {\bibfnamefont {A.}~\bibnamefont {Jansen}},\ }\href {\doibase 10.1140/epjp/i2017-11825-9} {\bibfield  {journal} {\bibinfo  {journal} {Eur. Phys. J. Plus}\ }\textbf {\bibinfo {volume} {132}},\ \bibinfo {pages} {546} (\bibinfo {year} {2017})},\ \Eprint {http://arxiv.org/abs/1709.09178} {arXiv:1709.09178 [gr-qc]} \BibitemShut {NoStop}%
\bibitem [{\citenamefont {Jing}(2004)}]{Jing:2003wq}%
  \BibitemOpen
  \bibfield  {author} {\bibinfo {author} {\bibfnamefont {J.-l.}\ \bibnamefont {Jing}},\ }\href {\doibase 10.1103/PhysRevD.69.084009} {\bibfield  {journal} {\bibinfo  {journal} {Phys. Rev. D}\ }\textbf {\bibinfo {volume} {69}},\ \bibinfo {pages} {084009} (\bibinfo {year} {2004})},\ \Eprint {http://arxiv.org/abs/gr-qc/0312079} {arXiv:gr-qc/0312079} \BibitemShut {NoStop}%
\bibitem [{\citenamefont {Arag\'on}\ \emph {et~al.}(2020)\citenamefont {Arag\'on}, \citenamefont {B\'ecar}, \citenamefont {Gonz\'alez},\ and\ \citenamefont {V\'asquez}}]{Aragon:2020qdc}%
  \BibitemOpen
  \bibfield  {author} {\bibinfo {author} {\bibfnamefont {A.}~\bibnamefont {Arag\'on}}, \bibinfo {author} {\bibfnamefont {R.}~\bibnamefont {B\'ecar}}, \bibinfo {author} {\bibfnamefont {P.~A.}\ \bibnamefont {Gonz\'alez}}, \ and\ \bibinfo {author} {\bibfnamefont {Y.}~\bibnamefont {V\'asquez}},\ }\href {\doibase 10.1140/epjc/s10052-020-8298-7} {\bibfield  {journal} {\bibinfo  {journal} {Eur. Phys. J. C}\ }\textbf {\bibinfo {volume} {80}},\ \bibinfo {pages} {773} (\bibinfo {year} {2020})},\ \Eprint {http://arxiv.org/abs/2004.05632} {arXiv:2004.05632 [gr-qc]} \BibitemShut {NoStop}%
\bibitem [{\citenamefont {Cardoso}\ \emph {et~al.}(2018)\citenamefont {Cardoso}, \citenamefont {Costa}, \citenamefont {Destounis}, \citenamefont {Hintz},\ and\ \citenamefont {Jansen}}]{Cardoso:2017soq}%
  \BibitemOpen
  \bibfield  {author} {\bibinfo {author} {\bibfnamefont {V.}~\bibnamefont {Cardoso}}, \bibinfo {author} {\bibfnamefont {J.~a.~L.}\ \bibnamefont {Costa}}, \bibinfo {author} {\bibfnamefont {K.}~\bibnamefont {Destounis}}, \bibinfo {author} {\bibfnamefont {P.}~\bibnamefont {Hintz}}, \ and\ \bibinfo {author} {\bibfnamefont {A.}~\bibnamefont {Jansen}},\ }\href {\doibase 10.1103/PhysRevLett.120.031103} {\bibfield  {journal} {\bibinfo  {journal} {Phys. Rev. Lett.}\ }\textbf {\bibinfo {volume} {120}},\ \bibinfo {pages} {031103} (\bibinfo {year} {2018})},\ \Eprint {http://arxiv.org/abs/1711.10502} {arXiv:1711.10502 [gr-qc]} \BibitemShut {NoStop}%
\bibitem [{\citenamefont {Dias}\ \emph {et~al.}(2018{\natexlab{a}})\citenamefont {Dias}, \citenamefont {Eperon}, \citenamefont {Reall},\ and\ \citenamefont {Santos}}]{Dias:2018ynt}%
  \BibitemOpen
  \bibfield  {author} {\bibinfo {author} {\bibfnamefont {O.~J.~C.}\ \bibnamefont {Dias}}, \bibinfo {author} {\bibfnamefont {F.~C.}\ \bibnamefont {Eperon}}, \bibinfo {author} {\bibfnamefont {H.~S.}\ \bibnamefont {Reall}}, \ and\ \bibinfo {author} {\bibfnamefont {J.~E.}\ \bibnamefont {Santos}},\ }\href {\doibase 10.1103/PhysRevD.97.104060} {\bibfield  {journal} {\bibinfo  {journal} {Phys. Rev. D}\ }\textbf {\bibinfo {volume} {97}},\ \bibinfo {pages} {104060} (\bibinfo {year} {2018}{\natexlab{a}})},\ \Eprint {http://arxiv.org/abs/1801.09694} {arXiv:1801.09694 [gr-qc]} \BibitemShut {NoStop}%
\bibitem [{\citenamefont {Dias}\ \emph {et~al.}(2018{\natexlab{b}})\citenamefont {Dias}, \citenamefont {Reall},\ and\ \citenamefont {Santos}}]{Dias:2018etb}%
  \BibitemOpen
  \bibfield  {author} {\bibinfo {author} {\bibfnamefont {O.~J.~C.}\ \bibnamefont {Dias}}, \bibinfo {author} {\bibfnamefont {H.~S.}\ \bibnamefont {Reall}}, \ and\ \bibinfo {author} {\bibfnamefont {J.~E.}\ \bibnamefont {Santos}},\ }\href {\doibase 10.1007/JHEP10(2018)001} {\bibfield  {journal} {\bibinfo  {journal} {JHEP}\ }\textbf {\bibinfo {volume} {10}},\ \bibinfo {pages} {001} (\bibinfo {year} {2018}{\natexlab{b}})},\ \Eprint {http://arxiv.org/abs/1808.02895} {arXiv:1808.02895 [gr-qc]} \BibitemShut {NoStop}%
\bibitem [{\citenamefont {Mo}\ \emph {et~al.}(2018)\citenamefont {Mo}, \citenamefont {Tian}, \citenamefont {Wang}, \citenamefont {Zhang},\ and\ \citenamefont {Zhong}}]{Mo:2018nnu}%
  \BibitemOpen
  \bibfield  {author} {\bibinfo {author} {\bibfnamefont {Y.}~\bibnamefont {Mo}}, \bibinfo {author} {\bibfnamefont {Y.}~\bibnamefont {Tian}}, \bibinfo {author} {\bibfnamefont {B.}~\bibnamefont {Wang}}, \bibinfo {author} {\bibfnamefont {H.}~\bibnamefont {Zhang}}, \ and\ \bibinfo {author} {\bibfnamefont {Z.}~\bibnamefont {Zhong}},\ }\href {\doibase 10.1103/PhysRevD.98.124025} {\bibfield  {journal} {\bibinfo  {journal} {Phys. Rev. D}\ }\textbf {\bibinfo {volume} {98}},\ \bibinfo {pages} {124025} (\bibinfo {year} {2018})},\ \Eprint {http://arxiv.org/abs/1808.03635} {arXiv:1808.03635 [gr-qc]} \BibitemShut {NoStop}%
\bibitem [{\citenamefont {Konoplya}\ and\ \citenamefont {Zhidenko}(2022{\natexlab{c}})}]{Konoplya:2022kld}%
  \BibitemOpen
  \bibfield  {author} {\bibinfo {author} {\bibfnamefont {R.~A.}\ \bibnamefont {Konoplya}}\ and\ \bibinfo {author} {\bibfnamefont {A.}~\bibnamefont {Zhidenko}},\ }\href {\doibase 10.1088/1475-7516/2022/11/028} {\bibfield  {journal} {\bibinfo  {journal} {JCAP}\ }\textbf {\bibinfo {volume} {11}},\ \bibinfo {pages} {028} (\bibinfo {year} {2022}{\natexlab{c}})},\ \Eprint {http://arxiv.org/abs/2210.04314} {arXiv:2210.04314 [gr-qc]} \BibitemShut {NoStop}%
\bibitem [{\citenamefont {Cardoso}\ and\ \citenamefont {Lemos}(2003)}]{Cardoso:2003sw}%
  \BibitemOpen
  \bibfield  {author} {\bibinfo {author} {\bibfnamefont {V.}~\bibnamefont {Cardoso}}\ and\ \bibinfo {author} {\bibfnamefont {J.~P.~S.}\ \bibnamefont {Lemos}},\ }\href {\doibase 10.1103/PhysRevD.67.084020} {\bibfield  {journal} {\bibinfo  {journal} {Phys. Rev. D}\ }\textbf {\bibinfo {volume} {67}},\ \bibinfo {pages} {084020} (\bibinfo {year} {2003})},\ \Eprint {http://arxiv.org/abs/gr-qc/0301078} {arXiv:gr-qc/0301078} \BibitemShut {NoStop}%
\bibitem [{\citenamefont {Moss}\ and\ \citenamefont {Norman}(2002)}]{Moss:2001ga}%
  \BibitemOpen
  \bibfield  {author} {\bibinfo {author} {\bibfnamefont {I.~G.}\ \bibnamefont {Moss}}\ and\ \bibinfo {author} {\bibfnamefont {J.~P.}\ \bibnamefont {Norman}},\ }\href {\doibase 10.1088/0264-9381/19/8/319} {\bibfield  {journal} {\bibinfo  {journal} {Class. Quant. Grav.}\ }\textbf {\bibinfo {volume} {19}},\ \bibinfo {pages} {2323} (\bibinfo {year} {2002})},\ \Eprint {http://arxiv.org/abs/gr-qc/0201016} {arXiv:gr-qc/0201016} \BibitemShut {NoStop}%
\bibitem [{\citenamefont {Molina}(2003)}]{Molina:2003ff}%
  \BibitemOpen
  \bibfield  {author} {\bibinfo {author} {\bibfnamefont {C.}~\bibnamefont {Molina}},\ }\href {\doibase 10.1103/PhysRevD.68.064007} {\bibfield  {journal} {\bibinfo  {journal} {Phys. Rev. D}\ }\textbf {\bibinfo {volume} {68}},\ \bibinfo {pages} {064007} (\bibinfo {year} {2003})},\ \Eprint {http://arxiv.org/abs/gr-qc/0304053} {arXiv:gr-qc/0304053} \BibitemShut {NoStop}%
\bibitem [{\citenamefont {Cardona}\ and\ \citenamefont {Molina}(2017)}]{Cardona:2017scd}%
  \BibitemOpen
  \bibfield  {author} {\bibinfo {author} {\bibfnamefont {A.~F.}\ \bibnamefont {Cardona}}\ and\ \bibinfo {author} {\bibfnamefont {C.}~\bibnamefont {Molina}},\ }\href {\doibase 10.1088/1361-6382/aa9428} {\bibfield  {journal} {\bibinfo  {journal} {Class. Quant. Grav.}\ }\textbf {\bibinfo {volume} {34}},\ \bibinfo {pages} {245002} (\bibinfo {year} {2017})},\ \Eprint {http://arxiv.org/abs/1711.00479} {arXiv:1711.00479 [gr-qc]} \BibitemShut {NoStop}%
\bibitem [{\citenamefont {Churilova}\ \emph {et~al.}(2022)\citenamefont {Churilova}, \citenamefont {Konoplya},\ and\ \citenamefont {Zhidenko}}]{Churilova:2021nnc}%
  \BibitemOpen
  \bibfield  {author} {\bibinfo {author} {\bibfnamefont {M.~S.}\ \bibnamefont {Churilova}}, \bibinfo {author} {\bibfnamefont {R.~A.}\ \bibnamefont {Konoplya}}, \ and\ \bibinfo {author} {\bibfnamefont {A.}~\bibnamefont {Zhidenko}},\ }\href {\doibase 10.1103/PhysRevD.105.084003} {\bibfield  {journal} {\bibinfo  {journal} {Phys. Rev. D}\ }\textbf {\bibinfo {volume} {105}},\ \bibinfo {pages} {084003} (\bibinfo {year} {2022})},\ \Eprint {http://arxiv.org/abs/2108.04858} {arXiv:2108.04858 [gr-qc]} \BibitemShut {NoStop}%
\bibitem [{\citenamefont {Ohashi}\ and\ \citenamefont {Sakagami}(2004)}]{Ohashi:2004wr}%
  \BibitemOpen
  \bibfield  {author} {\bibinfo {author} {\bibfnamefont {A.}~\bibnamefont {Ohashi}}\ and\ \bibinfo {author} {\bibfnamefont {M.-a.}\ \bibnamefont {Sakagami}},\ }\href {\doibase 10.1088/0264-9381/21/16/010} {\bibfield  {journal} {\bibinfo  {journal} {Class. Quant. Grav.}\ }\textbf {\bibinfo {volume} {21}},\ \bibinfo {pages} {3973} (\bibinfo {year} {2004})},\ \Eprint {http://arxiv.org/abs/gr-qc/0407009} {arXiv:gr-qc/0407009} \BibitemShut {NoStop}%
\bibitem [{\citenamefont {Konoplya}\ and\ \citenamefont {Zhidenko}(2005)}]{Konoplya:2004wg}%
  \BibitemOpen
  \bibfield  {author} {\bibinfo {author} {\bibfnamefont {R.~A.}\ \bibnamefont {Konoplya}}\ and\ \bibinfo {author} {\bibfnamefont {A.~V.}\ \bibnamefont {Zhidenko}},\ }\href {\doibase 10.1016/j.physletb.2005.01.078} {\bibfield  {journal} {\bibinfo  {journal} {Phys. Lett. B}\ }\textbf {\bibinfo {volume} {609}},\ \bibinfo {pages} {377} (\bibinfo {year} {2005})},\ \Eprint {http://arxiv.org/abs/gr-qc/0411059} {arXiv:gr-qc/0411059} \BibitemShut {NoStop}%
\bibitem [{\citenamefont {Zinhailo}(2024)}]{Zinhailo:2024jzt}%
  \BibitemOpen
  \bibfield  {author} {\bibinfo {author} {\bibfnamefont {A.~F.}\ \bibnamefont {Zinhailo}},\ }\href {\doibase 10.1016/j.physletb.2024.138682} {\bibfield  {journal} {\bibinfo  {journal} {Phys. Lett. B}\ }\textbf {\bibinfo {volume} {853}},\ \bibinfo {pages} {138682} (\bibinfo {year} {2024})},\ \Eprint {http://arxiv.org/abs/2403.06867} {arXiv:2403.06867 [gr-qc]} \BibitemShut {NoStop}%
\bibitem [{\citenamefont {Konoplya}\ and\ \citenamefont {Zhidenko}(2018)}]{Konoplya:2017tvu}%
  \BibitemOpen
  \bibfield  {author} {\bibinfo {author} {\bibfnamefont {R.~A.}\ \bibnamefont {Konoplya}}\ and\ \bibinfo {author} {\bibfnamefont {A.}~\bibnamefont {Zhidenko}},\ }\href {\doibase 10.1103/PhysRevD.97.084034} {\bibfield  {journal} {\bibinfo  {journal} {Phys. Rev. D}\ }\textbf {\bibinfo {volume} {97}},\ \bibinfo {pages} {084034} (\bibinfo {year} {2018})},\ \Eprint {http://arxiv.org/abs/1712.06667} {arXiv:1712.06667 [gr-qc]} \BibitemShut {NoStop}%
\bibitem [{\citenamefont {Konoplya}(2006)}]{Konoplya:2005hr}%
  \BibitemOpen
  \bibfield  {author} {\bibinfo {author} {\bibfnamefont {R.~A.}\ \bibnamefont {Konoplya}},\ }\href {\doibase 10.1103/PhysRevD.73.024009} {\bibfield  {journal} {\bibinfo  {journal} {Phys. Rev. D}\ }\textbf {\bibinfo {volume} {73}},\ \bibinfo {pages} {024009} (\bibinfo {year} {2006})},\ \Eprint {http://arxiv.org/abs/gr-qc/0509026} {arXiv:gr-qc/0509026} \BibitemShut {NoStop}%
\bibitem [{\citenamefont {Konoplya}\ and\ \citenamefont {Zhidenko}(2024{\natexlab{b}})}]{Konoplya:2023fmh}%
  \BibitemOpen
  \bibfield  {author} {\bibinfo {author} {\bibfnamefont {R.~A.}\ \bibnamefont {Konoplya}}\ and\ \bibinfo {author} {\bibfnamefont {A.}~\bibnamefont {Zhidenko}},\ }\href {\doibase 10.1016/j.physletb.2024.138685} {\bibfield  {journal} {\bibinfo  {journal} {Phys. Lett. B}\ }\textbf {\bibinfo {volume} {853}},\ \bibinfo {pages} {138685} (\bibinfo {year} {2024}{\natexlab{b}})},\ \Eprint {http://arxiv.org/abs/2307.01110} {arXiv:2307.01110 [gr-qc]} \BibitemShut {NoStop}%
\bibitem [{\citenamefont {Brito}\ \emph {et~al.}(2015)\citenamefont {Brito}, \citenamefont {Cardoso},\ and\ \citenamefont {Pani}}]{Brito:2015oca}%
  \BibitemOpen
  \bibfield  {author} {\bibinfo {author} {\bibfnamefont {R.}~\bibnamefont {Brito}}, \bibinfo {author} {\bibfnamefont {V.}~\bibnamefont {Cardoso}}, \ and\ \bibinfo {author} {\bibfnamefont {P.}~\bibnamefont {Pani}},\ }\href {\doibase 10.1007/978-3-319-19000-6} {\bibfield  {journal} {\bibinfo  {journal} {Lect. Notes Phys.}\ }\textbf {\bibinfo {volume} {906}},\ \bibinfo {pages} {pp.1} (\bibinfo {year} {2015})},\ \Eprint {http://arxiv.org/abs/1501.06570} {arXiv:1501.06570 [gr-qc]} \BibitemShut {NoStop}%
\bibitem [{\citenamefont {Annulli}\ \emph {et~al.}(2020)\citenamefont {Annulli}, \citenamefont {Cardoso},\ and\ \citenamefont {Vicente}}]{Annulli:2020lyc}%
  \BibitemOpen
  \bibfield  {author} {\bibinfo {author} {\bibfnamefont {L.}~\bibnamefont {Annulli}}, \bibinfo {author} {\bibfnamefont {V.}~\bibnamefont {Cardoso}}, \ and\ \bibinfo {author} {\bibfnamefont {R.}~\bibnamefont {Vicente}},\ }\href {\doibase 10.1103/PhysRevD.102.063022} {\bibfield  {journal} {\bibinfo  {journal} {Phys. Rev. D}\ }\textbf {\bibinfo {volume} {102}},\ \bibinfo {pages} {063022} (\bibinfo {year} {2020})},\ \Eprint {http://arxiv.org/abs/2009.00012} {arXiv:2009.00012 [gr-qc]} \BibitemShut {NoStop}%
\bibitem [{\citenamefont {Chung}\ \emph {et~al.}(2021)\citenamefont {Chung}, \citenamefont {Gais}, \citenamefont {Cheung},\ and\ \citenamefont {Li}}]{Chung:2021roh}%
  \BibitemOpen
  \bibfield  {author} {\bibinfo {author} {\bibfnamefont {A.~K.-W.}\ \bibnamefont {Chung}}, \bibinfo {author} {\bibfnamefont {J.}~\bibnamefont {Gais}}, \bibinfo {author} {\bibfnamefont {M.~H.-Y.}\ \bibnamefont {Cheung}}, \ and\ \bibinfo {author} {\bibfnamefont {T.~G.~F.}\ \bibnamefont {Li}},\ }\href {\doibase 10.1103/PhysRevD.104.084028} {\bibfield  {journal} {\bibinfo  {journal} {Phys. Rev. D}\ }\textbf {\bibinfo {volume} {104}},\ \bibinfo {pages} {084028} (\bibinfo {year} {2021})},\ \Eprint {http://arxiv.org/abs/2107.05492} {arXiv:2107.05492 [gr-qc]} \BibitemShut {NoStop}%
\bibitem [{\citenamefont {Konoplya}(2008)}]{Konoplya:2008hj}%
  \BibitemOpen
  \bibfield  {author} {\bibinfo {author} {\bibfnamefont {R.~A.}\ \bibnamefont {Konoplya}},\ }\href {\doibase 10.1016/j.physletb.2008.11.059} {\bibfield  {journal} {\bibinfo  {journal} {Phys. Lett. B}\ }\textbf {\bibinfo {volume} {666}},\ \bibinfo {pages} {283} (\bibinfo {year} {2008})},\ \Eprint {http://arxiv.org/abs/0801.0846} {arXiv:0801.0846 [hep-th]} \BibitemShut {NoStop}%
\bibitem [{\citenamefont {Konoplya}\ and\ \citenamefont {Fontana}(2008)}]{Konoplya:2007yy}%
  \BibitemOpen
  \bibfield  {author} {\bibinfo {author} {\bibfnamefont {R.~A.}\ \bibnamefont {Konoplya}}\ and\ \bibinfo {author} {\bibfnamefont {R.~D.~B.}\ \bibnamefont {Fontana}},\ }\href {\doibase 10.1016/j.physletb.2007.10.065} {\bibfield  {journal} {\bibinfo  {journal} {Phys. Lett. B}\ }\textbf {\bibinfo {volume} {659}},\ \bibinfo {pages} {375} (\bibinfo {year} {2008})},\ \Eprint {http://arxiv.org/abs/0707.1156} {arXiv:0707.1156 [hep-th]} \BibitemShut {NoStop}%
\bibitem [{\citenamefont {Davlataliev}\ \emph {et~al.}(2024)\citenamefont {Davlataliev}, \citenamefont {Narzilloev}, \citenamefont {Hussain}, \citenamefont {Abdujabbarov},\ and\ \citenamefont {Ahmedov}}]{Davlataliev:2024mjl}%
  \BibitemOpen
  \bibfield  {author} {\bibinfo {author} {\bibfnamefont {A.}~\bibnamefont {Davlataliev}}, \bibinfo {author} {\bibfnamefont {B.}~\bibnamefont {Narzilloev}}, \bibinfo {author} {\bibfnamefont {I.}~\bibnamefont {Hussain}}, \bibinfo {author} {\bibfnamefont {A.}~\bibnamefont {Abdujabbarov}}, \ and\ \bibinfo {author} {\bibfnamefont {B.}~\bibnamefont {Ahmedov}},\ }\href@noop {} {\  (\bibinfo {year} {2024})},\ \Eprint {http://arxiv.org/abs/2412.09464} {arXiv:2412.09464 [gr-qc]} \BibitemShut {NoStop}%
\bibitem [{\citenamefont {Seahra}\ \emph {et~al.}(2005)\citenamefont {Seahra}, \citenamefont {Clarkson},\ and\ \citenamefont {Maartens}}]{Seahra:2004fg}%
  \BibitemOpen
  \bibfield  {author} {\bibinfo {author} {\bibfnamefont {S.~S.}\ \bibnamefont {Seahra}}, \bibinfo {author} {\bibfnamefont {C.}~\bibnamefont {Clarkson}}, \ and\ \bibinfo {author} {\bibfnamefont {R.}~\bibnamefont {Maartens}},\ }\href {\doibase 10.1103/PhysRevLett.94.121302} {\bibfield  {journal} {\bibinfo  {journal} {Phys. Rev. Lett.}\ }\textbf {\bibinfo {volume} {94}},\ \bibinfo {pages} {121302} (\bibinfo {year} {2005})},\ \Eprint {http://arxiv.org/abs/gr-qc/0408032} {arXiv:gr-qc/0408032} \BibitemShut {NoStop}%
\bibitem [{\citenamefont {Koyama}\ and\ \citenamefont {Tomimatsu}(2001{\natexlab{a}})}]{Koyama:2001ee}%
  \BibitemOpen
  \bibfield  {author} {\bibinfo {author} {\bibfnamefont {H.}~\bibnamefont {Koyama}}\ and\ \bibinfo {author} {\bibfnamefont {A.}~\bibnamefont {Tomimatsu}},\ }\href {\doibase 10.1103/PhysRevD.64.044014} {\bibfield  {journal} {\bibinfo  {journal} {Phys. Rev. D}\ }\textbf {\bibinfo {volume} {64}},\ \bibinfo {pages} {044014} (\bibinfo {year} {2001}{\natexlab{a}})},\ \Eprint {http://arxiv.org/abs/gr-qc/0103086} {arXiv:gr-qc/0103086} \BibitemShut {NoStop}%
\bibitem [{\citenamefont {Koyama}\ and\ \citenamefont {Tomimatsu}(2001{\natexlab{b}})}]{Koyama:2000hj}%
  \BibitemOpen
  \bibfield  {author} {\bibinfo {author} {\bibfnamefont {H.}~\bibnamefont {Koyama}}\ and\ \bibinfo {author} {\bibfnamefont {A.}~\bibnamefont {Tomimatsu}},\ }\href {\doibase 10.1103/PhysRevD.63.064032} {\bibfield  {journal} {\bibinfo  {journal} {Phys. Rev. D}\ }\textbf {\bibinfo {volume} {63}},\ \bibinfo {pages} {064032} (\bibinfo {year} {2001}{\natexlab{b}})},\ \Eprint {http://arxiv.org/abs/gr-qc/0012022} {arXiv:gr-qc/0012022} \BibitemShut {NoStop}%
\bibitem [{\citenamefont {Konoplya}\ \emph {et~al.}(2007)\citenamefont {Konoplya}, \citenamefont {Zhidenko},\ and\ \citenamefont {Molina}}]{Konoplya:2006gq}%
  \BibitemOpen
  \bibfield  {author} {\bibinfo {author} {\bibfnamefont {R.~A.}\ \bibnamefont {Konoplya}}, \bibinfo {author} {\bibfnamefont {A.}~\bibnamefont {Zhidenko}}, \ and\ \bibinfo {author} {\bibfnamefont {C.}~\bibnamefont {Molina}},\ }\href {\doibase 10.1103/PhysRevD.75.084004} {\bibfield  {journal} {\bibinfo  {journal} {Phys. Rev. D}\ }\textbf {\bibinfo {volume} {75}},\ \bibinfo {pages} {084004} (\bibinfo {year} {2007})},\ \Eprint {http://arxiv.org/abs/gr-qc/0602047} {arXiv:gr-qc/0602047} \BibitemShut {NoStop}%
\bibitem [{\citenamefont {Rogatko}\ and\ \citenamefont {Szyplowska}(2007)}]{Rogatko:2007zz}%
  \BibitemOpen
  \bibfield  {author} {\bibinfo {author} {\bibfnamefont {M.}~\bibnamefont {Rogatko}}\ and\ \bibinfo {author} {\bibfnamefont {A.}~\bibnamefont {Szyplowska}},\ }\href {\doibase 10.1103/PhysRevD.76.044010} {\bibfield  {journal} {\bibinfo  {journal} {Phys. Rev. D}\ }\textbf {\bibinfo {volume} {76}},\ \bibinfo {pages} {044010} (\bibinfo {year} {2007})}\BibitemShut {NoStop}%
\bibitem [{\citenamefont {Gibbons}\ \emph {et~al.}(2008)\citenamefont {Gibbons}, \citenamefont {Rogatko},\ and\ \citenamefont {Szyplowska}}]{Gibbons:2008gg}%
  \BibitemOpen
  \bibfield  {author} {\bibinfo {author} {\bibfnamefont {G.~W.}\ \bibnamefont {Gibbons}}, \bibinfo {author} {\bibfnamefont {M.}~\bibnamefont {Rogatko}}, \ and\ \bibinfo {author} {\bibfnamefont {A.}~\bibnamefont {Szyplowska}},\ }\href {\doibase 10.1103/PhysRevD.77.064024} {\bibfield  {journal} {\bibinfo  {journal} {Phys. Rev. D}\ }\textbf {\bibinfo {volume} {77}},\ \bibinfo {pages} {064024} (\bibinfo {year} {2008})},\ \Eprint {http://arxiv.org/abs/0802.3259} {arXiv:0802.3259 [hep-th]} \BibitemShut {NoStop}%
\bibitem [{\citenamefont {Konoplya}\ \emph {et~al.}(2024)\citenamefont {Konoplya}, \citenamefont {Stuchl\'\i{}k},\ and\ \citenamefont {Zhidenko}}]{Konoplya:2024wds}%
  \BibitemOpen
  \bibfield  {author} {\bibinfo {author} {\bibfnamefont {R.~A.}\ \bibnamefont {Konoplya}}, \bibinfo {author} {\bibfnamefont {Z.}~\bibnamefont {Stuchl\'\i{}k}}, \ and\ \bibinfo {author} {\bibfnamefont {A.}~\bibnamefont {Zhidenko}},\ }\href@noop {} {\  (\bibinfo {year} {2024})},\ \Eprint {http://arxiv.org/abs/2411.09014} {arXiv:2411.09014 [gr-qc]} \BibitemShut {NoStop}%
\bibitem [{\citenamefont {Konoplya}\ and\ \citenamefont {Zhidenko}(2013)}]{Konoplya:2013rxa}%
  \BibitemOpen
  \bibfield  {author} {\bibinfo {author} {\bibfnamefont {R.~A.}\ \bibnamefont {Konoplya}}\ and\ \bibinfo {author} {\bibfnamefont {A.}~\bibnamefont {Zhidenko}},\ }\href {\doibase 10.1103/PhysRevD.88.024054} {\bibfield  {journal} {\bibinfo  {journal} {Phys. Rev. D}\ }\textbf {\bibinfo {volume} {88}},\ \bibinfo {pages} {024054} (\bibinfo {year} {2013})},\ \Eprint {http://arxiv.org/abs/1307.1812} {arXiv:1307.1812 [gr-qc]} \BibitemShut {NoStop}%
\bibitem [{\citenamefont {Moderski}\ and\ \citenamefont {Rogatko}(2001)}]{Moderski:2001tk}%
  \BibitemOpen
  \bibfield  {author} {\bibinfo {author} {\bibfnamefont {R.}~\bibnamefont {Moderski}}\ and\ \bibinfo {author} {\bibfnamefont {M.}~\bibnamefont {Rogatko}},\ }\href {\doibase 10.1103/PhysRevD.64.044024} {\bibfield  {journal} {\bibinfo  {journal} {Phys. Rev. D}\ }\textbf {\bibinfo {volume} {64}},\ \bibinfo {pages} {044024} (\bibinfo {year} {2001})},\ \Eprint {http://arxiv.org/abs/gr-qc/0105056} {arXiv:gr-qc/0105056} \BibitemShut {NoStop}%
\bibitem [{\citenamefont {Jing}(2005)}]{Jing:2004zb}%
  \BibitemOpen
  \bibfield  {author} {\bibinfo {author} {\bibfnamefont {J.}~\bibnamefont {Jing}},\ }\href {\doibase 10.1103/PhysRevD.72.027501} {\bibfield  {journal} {\bibinfo  {journal} {Phys. Rev. D}\ }\textbf {\bibinfo {volume} {72}},\ \bibinfo {pages} {027501} (\bibinfo {year} {2005})},\ \Eprint {http://arxiv.org/abs/gr-qc/0408090} {arXiv:gr-qc/0408090} \BibitemShut {NoStop}%
\bibitem [{\citenamefont {Churilova}\ \emph {et~al.}(2020)\citenamefont {Churilova}, \citenamefont {Konoplya},\ and\ \citenamefont {Zhidenko}}]{Churilova:2019qph}%
  \BibitemOpen
  \bibfield  {author} {\bibinfo {author} {\bibfnamefont {M.~S.}\ \bibnamefont {Churilova}}, \bibinfo {author} {\bibfnamefont {R.~A.}\ \bibnamefont {Konoplya}}, \ and\ \bibinfo {author} {\bibfnamefont {A.}~\bibnamefont {Zhidenko}},\ }\href {\doibase 10.1016/j.physletb.2020.135207} {\bibfield  {journal} {\bibinfo  {journal} {Phys. Lett. B}\ }\textbf {\bibinfo {volume} {802}},\ \bibinfo {pages} {135207} (\bibinfo {year} {2020})},\ \Eprint {http://arxiv.org/abs/1911.05246} {arXiv:1911.05246 [gr-qc]} \BibitemShut {NoStop}%
\bibitem [{\citenamefont {Rezzolla}\ and\ \citenamefont {Zhidenko}(2014)}]{Rezzolla:2014mua}%
  \BibitemOpen
  \bibfield  {author} {\bibinfo {author} {\bibfnamefont {L.}~\bibnamefont {Rezzolla}}\ and\ \bibinfo {author} {\bibfnamefont {A.}~\bibnamefont {Zhidenko}},\ }\href {\doibase 10.1103/PhysRevD.90.084009} {\bibfield  {journal} {\bibinfo  {journal} {Phys. Rev. D}\ }\textbf {\bibinfo {volume} {90}},\ \bibinfo {pages} {084009} (\bibinfo {year} {2014})},\ \Eprint {http://arxiv.org/abs/1407.3086} {arXiv:1407.3086 [gr-qc]} \BibitemShut {NoStop}%
\bibitem [{\citenamefont {Konoplya}\ \emph {et~al.}(2016)\citenamefont {Konoplya}, \citenamefont {Rezzolla},\ and\ \citenamefont {Zhidenko}}]{Konoplya:2016jvv}%
  \BibitemOpen
  \bibfield  {author} {\bibinfo {author} {\bibfnamefont {R.}~\bibnamefont {Konoplya}}, \bibinfo {author} {\bibfnamefont {L.}~\bibnamefont {Rezzolla}}, \ and\ \bibinfo {author} {\bibfnamefont {A.}~\bibnamefont {Zhidenko}},\ }\href {\doibase 10.1103/PhysRevD.93.064015} {\bibfield  {journal} {\bibinfo  {journal} {Phys. Rev. D}\ }\textbf {\bibinfo {volume} {93}},\ \bibinfo {pages} {064015} (\bibinfo {year} {2016})},\ \Eprint {http://arxiv.org/abs/1602.02378} {arXiv:1602.02378 [gr-qc]} \BibitemShut {NoStop}%
\bibitem [{\citenamefont {Younsi}\ \emph {et~al.}(2016)\citenamefont {Younsi}, \citenamefont {Zhidenko}, \citenamefont {Rezzolla}, \citenamefont {Konoplya},\ and\ \citenamefont {Mizuno}}]{Younsi:2016azx}%
  \BibitemOpen
  \bibfield  {author} {\bibinfo {author} {\bibfnamefont {Z.}~\bibnamefont {Younsi}}, \bibinfo {author} {\bibfnamefont {A.}~\bibnamefont {Zhidenko}}, \bibinfo {author} {\bibfnamefont {L.}~\bibnamefont {Rezzolla}}, \bibinfo {author} {\bibfnamefont {R.}~\bibnamefont {Konoplya}}, \ and\ \bibinfo {author} {\bibfnamefont {Y.}~\bibnamefont {Mizuno}},\ }\href {\doibase 10.1103/PhysRevD.94.084025} {\bibfield  {journal} {\bibinfo  {journal} {Phys. Rev. D}\ }\textbf {\bibinfo {volume} {94}},\ \bibinfo {pages} {084025} (\bibinfo {year} {2016})},\ \Eprint {http://arxiv.org/abs/1607.05767} {arXiv:1607.05767 [gr-qc]} \BibitemShut {NoStop}%
\bibitem [{\citenamefont {Konoplya}\ \emph {et~al.}(2020{\natexlab{b}})\citenamefont {Konoplya}, \citenamefont {Pappas},\ and\ \citenamefont {Stuchl\'\i{}k}}]{Konoplya:2020kqb}%
  \BibitemOpen
  \bibfield  {author} {\bibinfo {author} {\bibfnamefont {R.~A.}\ \bibnamefont {Konoplya}}, \bibinfo {author} {\bibfnamefont {T.~D.}\ \bibnamefont {Pappas}}, \ and\ \bibinfo {author} {\bibfnamefont {Z.}~\bibnamefont {Stuchl\'\i{}k}},\ }\href {\doibase 10.1103/PhysRevD.102.084043} {\bibfield  {journal} {\bibinfo  {journal} {Phys. Rev. D}\ }\textbf {\bibinfo {volume} {102}},\ \bibinfo {pages} {084043} (\bibinfo {year} {2020}{\natexlab{b}})},\ \Eprint {http://arxiv.org/abs/2007.14860} {arXiv:2007.14860 [gr-qc]} \BibitemShut {NoStop}%
\bibitem [{\citenamefont {Bronnikov}\ \emph {et~al.}(2021)\citenamefont {Bronnikov}, \citenamefont {Konoplya},\ and\ \citenamefont {Pappas}}]{Bronnikov:2021liv}%
  \BibitemOpen
  \bibfield  {author} {\bibinfo {author} {\bibfnamefont {K.~A.}\ \bibnamefont {Bronnikov}}, \bibinfo {author} {\bibfnamefont {R.~A.}\ \bibnamefont {Konoplya}}, \ and\ \bibinfo {author} {\bibfnamefont {T.~D.}\ \bibnamefont {Pappas}},\ }\href {\doibase 10.1103/PhysRevD.103.124062} {\bibfield  {journal} {\bibinfo  {journal} {Phys. Rev. D}\ }\textbf {\bibinfo {volume} {103}},\ \bibinfo {pages} {124062} (\bibinfo {year} {2021})},\ \Eprint {http://arxiv.org/abs/2102.10679} {arXiv:2102.10679 [gr-qc]} \BibitemShut {NoStop}%
\bibitem [{\citenamefont {Kocherlakota}\ and\ \citenamefont {Rezzolla}(2020)}]{Kocherlakota:2020kyu}%
  \BibitemOpen
  \bibfield  {author} {\bibinfo {author} {\bibfnamefont {P.}~\bibnamefont {Kocherlakota}}\ and\ \bibinfo {author} {\bibfnamefont {L.}~\bibnamefont {Rezzolla}},\ }\href {\doibase 10.1103/PhysRevD.102.064058} {\bibfield  {journal} {\bibinfo  {journal} {Phys. Rev. D}\ }\textbf {\bibinfo {volume} {102}},\ \bibinfo {pages} {064058} (\bibinfo {year} {2020})},\ \Eprint {http://arxiv.org/abs/2007.15593} {arXiv:2007.15593 [gr-qc]} \BibitemShut {NoStop}%
\bibitem [{\citenamefont {Zhang}(2024)}]{Zhang:2024rvk}%
  \BibitemOpen
  \bibfield  {author} {\bibinfo {author} {\bibfnamefont {S.-J.}\ \bibnamefont {Zhang}},\ }\href {\doibase 10.1103/PhysRevD.109.084066} {\bibfield  {journal} {\bibinfo  {journal} {Phys. Rev. D}\ }\textbf {\bibinfo {volume} {109}},\ \bibinfo {pages} {084066} (\bibinfo {year} {2024})},\ \Eprint {http://arxiv.org/abs/2402.15050} {arXiv:2402.15050 [gr-qc]} \BibitemShut {NoStop}%
\bibitem [{\citenamefont {Cassing}\ and\ \citenamefont {Rezzolla}(2023)}]{Cassing:2023bpt}%
  \BibitemOpen
  \bibfield  {author} {\bibinfo {author} {\bibfnamefont {M.}~\bibnamefont {Cassing}}\ and\ \bibinfo {author} {\bibfnamefont {L.}~\bibnamefont {Rezzolla}},\ }\href {\doibase 10.1093/mnras/stad1039} {\bibfield  {journal} {\bibinfo  {journal} {Mon. Not. Roy. Astron. Soc.}\ }\textbf {\bibinfo {volume} {522}},\ \bibinfo {pages} {2415} (\bibinfo {year} {2023})},\ \Eprint {http://arxiv.org/abs/2302.09135} {arXiv:2302.09135 [gr-qc]} \BibitemShut {NoStop}%
\bibitem [{\citenamefont {Li}\ \emph {et~al.}(2021)\citenamefont {Li}, \citenamefont {Abdujabbarov},\ and\ \citenamefont {Han}}]{Li:2021mnx}%
  \BibitemOpen
  \bibfield  {author} {\bibinfo {author} {\bibfnamefont {S.}~\bibnamefont {Li}}, \bibinfo {author} {\bibfnamefont {A.~A.}\ \bibnamefont {Abdujabbarov}}, \ and\ \bibinfo {author} {\bibfnamefont {W.-B.}\ \bibnamefont {Han}},\ }\href {\doibase 10.1140/epjc/s10052-021-09445-6} {\bibfield  {journal} {\bibinfo  {journal} {Eur. Phys. J. C}\ }\textbf {\bibinfo {volume} {81}},\ \bibinfo {pages} {649} (\bibinfo {year} {2021})},\ \Eprint {http://arxiv.org/abs/2103.08104} {arXiv:2103.08104 [gr-qc]} \BibitemShut {NoStop}%
\bibitem [{\citenamefont {Ma}\ and\ \citenamefont {Rezzolla}(2024)}]{Ma:2024kbu}%
  \BibitemOpen
  \bibfield  {author} {\bibinfo {author} {\bibfnamefont {Y.}~\bibnamefont {Ma}}\ and\ \bibinfo {author} {\bibfnamefont {L.}~\bibnamefont {Rezzolla}},\ }\href {\doibase 10.1103/PhysRevD.110.024032} {\bibfield  {journal} {\bibinfo  {journal} {Phys. Rev. D}\ }\textbf {\bibinfo {volume} {110}},\ \bibinfo {pages} {024032} (\bibinfo {year} {2024})},\ \Eprint {http://arxiv.org/abs/2404.06509} {arXiv:2404.06509 [gr-qc]} \BibitemShut {NoStop}%
\bibitem [{\citenamefont {Shashank}\ and\ \citenamefont {Bambi}(2022)}]{Shashank:2021giy}%
  \BibitemOpen
  \bibfield  {author} {\bibinfo {author} {\bibfnamefont {S.}~\bibnamefont {Shashank}}\ and\ \bibinfo {author} {\bibfnamefont {C.}~\bibnamefont {Bambi}},\ }\href {\doibase 10.1103/PhysRevD.105.104004} {\bibfield  {journal} {\bibinfo  {journal} {Phys. Rev. D}\ }\textbf {\bibinfo {volume} {105}},\ \bibinfo {pages} {104004} (\bibinfo {year} {2022})},\ \Eprint {http://arxiv.org/abs/2112.05388} {arXiv:2112.05388 [gr-qc]} \BibitemShut {NoStop}%
\bibitem [{\citenamefont {Konoplya}\ and\ \citenamefont {Zhidenko}(2021)}]{Konoplya:2021slg}%
  \BibitemOpen
  \bibfield  {author} {\bibinfo {author} {\bibfnamefont {R.~A.}\ \bibnamefont {Konoplya}}\ and\ \bibinfo {author} {\bibfnamefont {A.}~\bibnamefont {Zhidenko}},\ }\href {\doibase 10.1103/PhysRevD.103.104033} {\bibfield  {journal} {\bibinfo  {journal} {Phys. Rev. D}\ }\textbf {\bibinfo {volume} {103}},\ \bibinfo {pages} {104033} (\bibinfo {year} {2021})},\ \Eprint {http://arxiv.org/abs/2103.03855} {arXiv:2103.03855 [gr-qc]} \BibitemShut {NoStop}%
\bibitem [{\citenamefont {Kokkotas}\ \emph {et~al.}(2017{\natexlab{b}})\citenamefont {Kokkotas}, \citenamefont {Konoplya},\ and\ \citenamefont {Zhidenko}}]{Kokkotas:2017ymc}%
  \BibitemOpen
  \bibfield  {author} {\bibinfo {author} {\bibfnamefont {K.~D.}\ \bibnamefont {Kokkotas}}, \bibinfo {author} {\bibfnamefont {R.~A.}\ \bibnamefont {Konoplya}}, \ and\ \bibinfo {author} {\bibfnamefont {A.}~\bibnamefont {Zhidenko}},\ }\href {\doibase 10.1103/PhysRevD.96.064004} {\bibfield  {journal} {\bibinfo  {journal} {Phys. Rev. D}\ }\textbf {\bibinfo {volume} {96}},\ \bibinfo {pages} {064004} (\bibinfo {year} {2017}{\natexlab{b}})},\ \Eprint {http://arxiv.org/abs/1706.07460} {arXiv:1706.07460 [gr-qc]} \BibitemShut {NoStop}%
\bibitem [{\citenamefont {Yu}\ \emph {et~al.}(2021)\citenamefont {Yu}, \citenamefont {Jiang}, \citenamefont {Abdikamalov}, \citenamefont {Ayzenberg}, \citenamefont {Bambi}, \citenamefont {Liu}, \citenamefont {Nampalliwar},\ and\ \citenamefont {Tripathi}}]{Yu:2021xen}%
  \BibitemOpen
  \bibfield  {author} {\bibinfo {author} {\bibfnamefont {Z.}~\bibnamefont {Yu}}, \bibinfo {author} {\bibfnamefont {Q.}~\bibnamefont {Jiang}}, \bibinfo {author} {\bibfnamefont {A.~B.}\ \bibnamefont {Abdikamalov}}, \bibinfo {author} {\bibfnamefont {D.}~\bibnamefont {Ayzenberg}}, \bibinfo {author} {\bibfnamefont {C.}~\bibnamefont {Bambi}}, \bibinfo {author} {\bibfnamefont {H.}~\bibnamefont {Liu}}, \bibinfo {author} {\bibfnamefont {S.}~\bibnamefont {Nampalliwar}}, \ and\ \bibinfo {author} {\bibfnamefont {A.}~\bibnamefont {Tripathi}},\ }\href {\doibase 10.1103/PhysRevD.104.084035} {\bibfield  {journal} {\bibinfo  {journal} {Phys. Rev. D}\ }\textbf {\bibinfo {volume} {104}},\ \bibinfo {pages} {084035} (\bibinfo {year} {2021})},\ \Eprint {http://arxiv.org/abs/2106.11658} {arXiv:2106.11658 [astro-ph.HE]} \BibitemShut {NoStop}%
\bibitem [{\citenamefont {Konoplya}\ \emph {et~al.}(2021)\citenamefont {Konoplya}, \citenamefont {Kunz},\ and\ \citenamefont {Zhidenko}}]{Konoplya:2021qll}%
  \BibitemOpen
  \bibfield  {author} {\bibinfo {author} {\bibfnamefont {R.~A.}\ \bibnamefont {Konoplya}}, \bibinfo {author} {\bibfnamefont {J.}~\bibnamefont {Kunz}}, \ and\ \bibinfo {author} {\bibfnamefont {A.}~\bibnamefont {Zhidenko}},\ }\href {\doibase 10.1088/1475-7516/2021/12/002} {\bibfield  {journal} {\bibinfo  {journal} {JCAP}\ }\textbf {\bibinfo {volume} {12}},\ \bibinfo {pages} {002} (\bibinfo {year} {2021})},\ \Eprint {http://arxiv.org/abs/2102.10649} {arXiv:2102.10649 [gr-qc]} \BibitemShut {NoStop}%
\bibitem [{\citenamefont {Toshmatov}\ and\ \citenamefont {Ahmedov}(2023)}]{Toshmatov:2023anz}%
  \BibitemOpen
  \bibfield  {author} {\bibinfo {author} {\bibfnamefont {B.}~\bibnamefont {Toshmatov}}\ and\ \bibinfo {author} {\bibfnamefont {B.}~\bibnamefont {Ahmedov}},\ }\href {\doibase 10.1103/PhysRevD.108.084035} {\bibfield  {journal} {\bibinfo  {journal} {Phys. Rev. D}\ }\textbf {\bibinfo {volume} {108}},\ \bibinfo {pages} {084035} (\bibinfo {year} {2023})},\ \Eprint {http://arxiv.org/abs/2311.04602} {arXiv:2311.04602 [gr-qc]} \BibitemShut {NoStop}%
\bibitem [{\citenamefont {Konoplya}\ \emph {et~al.}(2020{\natexlab{c}})\citenamefont {Konoplya}, \citenamefont {Pappas},\ and\ \citenamefont {Zhidenko}}]{Konoplya:2019fpy}%
  \BibitemOpen
  \bibfield  {author} {\bibinfo {author} {\bibfnamefont {R.~A.}\ \bibnamefont {Konoplya}}, \bibinfo {author} {\bibfnamefont {T.}~\bibnamefont {Pappas}}, \ and\ \bibinfo {author} {\bibfnamefont {A.}~\bibnamefont {Zhidenko}},\ }\href {\doibase 10.1103/PhysRevD.101.044054} {\bibfield  {journal} {\bibinfo  {journal} {Phys. Rev. D}\ }\textbf {\bibinfo {volume} {101}},\ \bibinfo {pages} {044054} (\bibinfo {year} {2020}{\natexlab{c}})},\ \Eprint {http://arxiv.org/abs/1907.10112} {arXiv:1907.10112 [gr-qc]} \BibitemShut {NoStop}%
\bibitem [{\citenamefont {Nampalliwar}\ \emph {et~al.}(2020)\citenamefont {Nampalliwar}, \citenamefont {Xin}, \citenamefont {Srivastava}, \citenamefont {Abdikamalov}, \citenamefont {Ayzenberg}, \citenamefont {Bambi}, \citenamefont {Dauser}, \citenamefont {Garcia},\ and\ \citenamefont {Tripathi}}]{Nampalliwar:2019iti}%
  \BibitemOpen
  \bibfield  {author} {\bibinfo {author} {\bibfnamefont {S.}~\bibnamefont {Nampalliwar}}, \bibinfo {author} {\bibfnamefont {S.}~\bibnamefont {Xin}}, \bibinfo {author} {\bibfnamefont {S.}~\bibnamefont {Srivastava}}, \bibinfo {author} {\bibfnamefont {A.~B.}\ \bibnamefont {Abdikamalov}}, \bibinfo {author} {\bibfnamefont {D.}~\bibnamefont {Ayzenberg}}, \bibinfo {author} {\bibfnamefont {C.}~\bibnamefont {Bambi}}, \bibinfo {author} {\bibfnamefont {T.}~\bibnamefont {Dauser}}, \bibinfo {author} {\bibfnamefont {J.~A.}\ \bibnamefont {Garcia}}, \ and\ \bibinfo {author} {\bibfnamefont {A.}~\bibnamefont {Tripathi}},\ }\href {\doibase 10.1103/PhysRevD.102.124071} {\bibfield  {journal} {\bibinfo  {journal} {Phys. Rev. D}\ }\textbf {\bibinfo {volume} {102}},\ \bibinfo {pages} {124071} (\bibinfo {year} {2020})},\ \Eprint {http://arxiv.org/abs/1903.12119} {arXiv:1903.12119 [gr-qc]} \BibitemShut {NoStop}%
\bibitem [{\citenamefont {Ni}\ \emph {et~al.}(2016)\citenamefont {Ni}, \citenamefont {Jiang},\ and\ \citenamefont {Bambi}}]{Ni:2016uik}%
  \BibitemOpen
  \bibfield  {author} {\bibinfo {author} {\bibfnamefont {Y.}~\bibnamefont {Ni}}, \bibinfo {author} {\bibfnamefont {J.}~\bibnamefont {Jiang}}, \ and\ \bibinfo {author} {\bibfnamefont {C.}~\bibnamefont {Bambi}},\ }\href {\doibase 10.1088/1475-7516/2016/09/014} {\bibfield  {journal} {\bibinfo  {journal} {JCAP}\ }\textbf {\bibinfo {volume} {09}},\ \bibinfo {pages} {014} (\bibinfo {year} {2016})},\ \Eprint {http://arxiv.org/abs/1607.04893} {arXiv:1607.04893 [gr-qc]} \BibitemShut {NoStop}%
\bibitem [{\citenamefont {Konoplya}\ \emph {et~al.}(2018{\natexlab{b}})\citenamefont {Konoplya}, \citenamefont {Stuchl\'\i{}k},\ and\ \citenamefont {Zhidenko}}]{Konoplya:2018arm}%
  \BibitemOpen
  \bibfield  {author} {\bibinfo {author} {\bibfnamefont {R.~A.}\ \bibnamefont {Konoplya}}, \bibinfo {author} {\bibfnamefont {Z.}~\bibnamefont {Stuchl\'\i{}k}}, \ and\ \bibinfo {author} {\bibfnamefont {A.}~\bibnamefont {Zhidenko}},\ }\href {\doibase 10.1103/PhysRevD.97.084044} {\bibfield  {journal} {\bibinfo  {journal} {Phys. Rev. D}\ }\textbf {\bibinfo {volume} {97}},\ \bibinfo {pages} {084044} (\bibinfo {year} {2018}{\natexlab{b}})},\ \Eprint {http://arxiv.org/abs/1801.07195} {arXiv:1801.07195 [gr-qc]} \BibitemShut {NoStop}%
\bibitem [{\citenamefont {Konoplya}(2023{\natexlab{c}})}]{Konoplya:2022iyn}%
  \BibitemOpen
  \bibfield  {author} {\bibinfo {author} {\bibfnamefont {R.~A.}\ \bibnamefont {Konoplya}},\ }\href {\doibase 10.1103/PhysRevD.107.064039} {\bibfield  {journal} {\bibinfo  {journal} {Phys. Rev. D}\ }\textbf {\bibinfo {volume} {107}},\ \bibinfo {pages} {064039} (\bibinfo {year} {2023}{\natexlab{c}})},\ \Eprint {http://arxiv.org/abs/2210.14506} {arXiv:2210.14506 [gr-qc]} \BibitemShut {NoStop}%
\bibitem [{\citenamefont {Konoplya}\ and\ \citenamefont {Zhidenko}(2022{\natexlab{d}})}]{Konoplya:2022tvv}%
  \BibitemOpen
  \bibfield  {author} {\bibinfo {author} {\bibfnamefont {R.~A.}\ \bibnamefont {Konoplya}}\ and\ \bibinfo {author} {\bibfnamefont {A.}~\bibnamefont {Zhidenko}},\ }\href {\doibase 10.1103/PhysRevD.105.104032} {\bibfield  {journal} {\bibinfo  {journal} {Phys. Rev. D}\ }\textbf {\bibinfo {volume} {105}},\ \bibinfo {pages} {104032} (\bibinfo {year} {2022}{\natexlab{d}})},\ \Eprint {http://arxiv.org/abs/2201.12897} {arXiv:2201.12897 [gr-qc]} \BibitemShut {NoStop}%
\bibitem [{\citenamefont {Konoplya}\ and\ \citenamefont {Zinhailo}(2019)}]{Konoplya:2019ppy}%
  \BibitemOpen
  \bibfield  {author} {\bibinfo {author} {\bibfnamefont {R.~A.}\ \bibnamefont {Konoplya}}\ and\ \bibinfo {author} {\bibfnamefont {A.~F.}\ \bibnamefont {Zinhailo}},\ }\href {\doibase 10.1103/PhysRevD.99.104060} {\bibfield  {journal} {\bibinfo  {journal} {Phys. Rev. D}\ }\textbf {\bibinfo {volume} {99}},\ \bibinfo {pages} {104060} (\bibinfo {year} {2019})},\ \Eprint {http://arxiv.org/abs/1904.05341} {arXiv:1904.05341 [gr-qc]} \BibitemShut {NoStop}%
\bibitem [{\citenamefont {Konoplya}\ and\ \citenamefont {Zhidenko}(2019)}]{Konoplya:2019goy}%
  \BibitemOpen
  \bibfield  {author} {\bibinfo {author} {\bibfnamefont {R.~A.}\ \bibnamefont {Konoplya}}\ and\ \bibinfo {author} {\bibfnamefont {A.}~\bibnamefont {Zhidenko}},\ }\href {\doibase 10.1103/PhysRevD.100.044015} {\bibfield  {journal} {\bibinfo  {journal} {Phys. Rev. D}\ }\textbf {\bibinfo {volume} {100}},\ \bibinfo {pages} {044015} (\bibinfo {year} {2019})},\ \Eprint {http://arxiv.org/abs/1907.05551} {arXiv:1907.05551 [gr-qc]} \BibitemShut {NoStop}%
\bibitem [{\citenamefont {Ashtekar}\ \emph {et~al.}(2018)\citenamefont {Ashtekar}, \citenamefont {Olmedo},\ and\ \citenamefont {Singh}}]{Ashtekar:2018lag}%
  \BibitemOpen
  \bibfield  {author} {\bibinfo {author} {\bibfnamefont {A.}~\bibnamefont {Ashtekar}}, \bibinfo {author} {\bibfnamefont {J.}~\bibnamefont {Olmedo}}, \ and\ \bibinfo {author} {\bibfnamefont {P.}~\bibnamefont {Singh}},\ }\href {\doibase 10.1103/PhysRevLett.121.241301} {\bibfield  {journal} {\bibinfo  {journal} {Phys. Rev. Lett.}\ }\textbf {\bibinfo {volume} {121}},\ \bibinfo {pages} {241301} (\bibinfo {year} {2018})},\ \Eprint {http://arxiv.org/abs/1806.00648} {arXiv:1806.00648 [gr-qc]} \BibitemShut {NoStop}%
\bibitem [{\citenamefont {Bouhmadi-L\'opez}\ \emph {et~al.}(2020)\citenamefont {Bouhmadi-L\'opez}, \citenamefont {Brahma}, \citenamefont {Chen}, \citenamefont {Chen},\ and\ \citenamefont {Yeom}}]{Bouhmadi-Lopez:2020oia}%
  \BibitemOpen
  \bibfield  {author} {\bibinfo {author} {\bibfnamefont {M.}~\bibnamefont {Bouhmadi-L\'opez}}, \bibinfo {author} {\bibfnamefont {S.}~\bibnamefont {Brahma}}, \bibinfo {author} {\bibfnamefont {C.-Y.}\ \bibnamefont {Chen}}, \bibinfo {author} {\bibfnamefont {P.}~\bibnamefont {Chen}}, \ and\ \bibinfo {author} {\bibfnamefont {D.-h.}\ \bibnamefont {Yeom}},\ }\href {\doibase 10.1088/1475-7516/2020/07/066} {\bibfield  {journal} {\bibinfo  {journal} {JCAP}\ }\textbf {\bibinfo {volume} {07}},\ \bibinfo {pages} {066} (\bibinfo {year} {2020})},\ \Eprint {http://arxiv.org/abs/2004.13061} {arXiv:2004.13061 [gr-qc]} \BibitemShut {NoStop}%
\bibitem [{\citenamefont {Hod}(2007)}]{Hod:2006jw}%
  \BibitemOpen
  \bibfield  {author} {\bibinfo {author} {\bibfnamefont {S.}~\bibnamefont {Hod}},\ }\href {\doibase 10.1103/PhysRevD.75.064013} {\bibfield  {journal} {\bibinfo  {journal} {Phys. Rev. D}\ }\textbf {\bibinfo {volume} {75}},\ \bibinfo {pages} {064013} (\bibinfo {year} {2007})},\ \Eprint {http://arxiv.org/abs/gr-qc/0611004} {arXiv:gr-qc/0611004} \BibitemShut {NoStop}%
\bibitem [{\citenamefont {Lopez-Ortega}(2006)}]{Lopez-Ortega:2006aal}%
  \BibitemOpen
  \bibfield  {author} {\bibinfo {author} {\bibfnamefont {A.}~\bibnamefont {Lopez-Ortega}},\ }\href {\doibase 10.1007/s10714-006-0335-9} {\bibfield  {journal} {\bibinfo  {journal} {Gen. Rel. Grav.}\ }\textbf {\bibinfo {volume} {38}},\ \bibinfo {pages} {1565} (\bibinfo {year} {2006})},\ \Eprint {http://arxiv.org/abs/gr-qc/0605027} {arXiv:gr-qc/0605027} \BibitemShut {NoStop}%
\bibitem [{\citenamefont {Lopez-Ortega}(2012)}]{Lopez-Ortega:2012xvr}%
  \BibitemOpen
  \bibfield  {author} {\bibinfo {author} {\bibfnamefont {A.}~\bibnamefont {Lopez-Ortega}},\ }\href {\doibase 10.1007/s10714-012-1398-4} {\bibfield  {journal} {\bibinfo  {journal} {Gen. Rel. Grav.}\ }\textbf {\bibinfo {volume} {44}},\ \bibinfo {pages} {2387} (\bibinfo {year} {2012})},\ \Eprint {http://arxiv.org/abs/1207.6791} {arXiv:1207.6791 [gr-qc]} \BibitemShut {NoStop}%
\bibitem [{\citenamefont {Du}\ \emph {et~al.}(2004)\citenamefont {Du}, \citenamefont {Wang},\ and\ \citenamefont {Su}}]{Du:2004jt}%
  \BibitemOpen
  \bibfield  {author} {\bibinfo {author} {\bibfnamefont {D.-P.}\ \bibnamefont {Du}}, \bibinfo {author} {\bibfnamefont {B.}~\bibnamefont {Wang}}, \ and\ \bibinfo {author} {\bibfnamefont {R.-K.}\ \bibnamefont {Su}},\ }\href {\doibase 10.1103/PhysRevD.70.064024} {\bibfield  {journal} {\bibinfo  {journal} {Phys. Rev. D}\ }\textbf {\bibinfo {volume} {70}},\ \bibinfo {pages} {064024} (\bibinfo {year} {2004})},\ \Eprint {http://arxiv.org/abs/hep-th/0404047} {arXiv:hep-th/0404047} \BibitemShut {NoStop}%
\bibitem [{\citenamefont {Burgess}\ and\ \citenamefont {Lutken}(1985)}]{Burgess:1984ti}%
  \BibitemOpen
  \bibfield  {author} {\bibinfo {author} {\bibfnamefont {C.~P.}\ \bibnamefont {Burgess}}\ and\ \bibinfo {author} {\bibfnamefont {C.~A.}\ \bibnamefont {Lutken}},\ }\href {\doibase 10.1016/0370-2693(85)91415-7} {\bibfield  {journal} {\bibinfo  {journal} {Phys. Lett. B}\ }\textbf {\bibinfo {volume} {153}},\ \bibinfo {pages} {137} (\bibinfo {year} {1985})}\BibitemShut {NoStop}%
\bibitem [{\citenamefont {Son}\ and\ \citenamefont {Starinets}(2007)}]{Son:2007vk}%
  \BibitemOpen
  \bibfield  {author} {\bibinfo {author} {\bibfnamefont {D.~T.}\ \bibnamefont {Son}}\ and\ \bibinfo {author} {\bibfnamefont {A.~O.}\ \bibnamefont {Starinets}},\ }\href {\doibase 10.1146/annurev.nucl.57.090506.123120} {\bibfield  {journal} {\bibinfo  {journal} {Ann. Rev. Nucl. Part. Sci.}\ }\textbf {\bibinfo {volume} {57}},\ \bibinfo {pages} {95} (\bibinfo {year} {2007})},\ \Eprint {http://arxiv.org/abs/0704.0240} {arXiv:0704.0240 [hep-th]} \BibitemShut {NoStop}%
\bibitem [{\citenamefont {Konoplya}(2002{\natexlab{b}})}]{Konoplya:2002zu}%
  \BibitemOpen
  \bibfield  {author} {\bibinfo {author} {\bibfnamefont {R.~A.}\ \bibnamefont {Konoplya}},\ }\href {\doibase 10.1103/PhysRevD.66.044009} {\bibfield  {journal} {\bibinfo  {journal} {Phys. Rev. D}\ }\textbf {\bibinfo {volume} {66}},\ \bibinfo {pages} {044009} (\bibinfo {year} {2002}{\natexlab{b}})},\ \Eprint {http://arxiv.org/abs/hep-th/0205142} {arXiv:hep-th/0205142 [hep-th]} \BibitemShut {NoStop}%
\bibitem [{\citenamefont {Hod}(1998)}]{Hod:1998vk}%
  \BibitemOpen
  \bibfield  {author} {\bibinfo {author} {\bibfnamefont {S.}~\bibnamefont {Hod}},\ }\href {\doibase 10.1103/PhysRevLett.81.4293} {\bibfield  {journal} {\bibinfo  {journal} {Phys. Rev. Lett.}\ }\textbf {\bibinfo {volume} {81}},\ \bibinfo {pages} {4293} (\bibinfo {year} {1998})},\ \Eprint {http://arxiv.org/abs/gr-qc/9812002} {arXiv:gr-qc/9812002} \BibitemShut {NoStop}%
\bibitem [{\citenamefont {Motl}\ and\ \citenamefont {Neitzke}(2003)}]{Motl:2003cd}%
  \BibitemOpen
  \bibfield  {author} {\bibinfo {author} {\bibfnamefont {L.}~\bibnamefont {Motl}}\ and\ \bibinfo {author} {\bibfnamefont {A.}~\bibnamefont {Neitzke}},\ }\href {\doibase 10.4310/ATMP.2003.v7.n2.a4} {\bibfield  {journal} {\bibinfo  {journal} {Adv. Theor. Math. Phys.}\ }\textbf {\bibinfo {volume} {7}},\ \bibinfo {pages} {307} (\bibinfo {year} {2003})},\ \Eprint {http://arxiv.org/abs/hep-th/0301173} {arXiv:hep-th/0301173} \BibitemShut {NoStop}%
\bibitem [{\citenamefont {Motl}(2003)}]{Motl:2002hd}%
  \BibitemOpen
  \bibfield  {author} {\bibinfo {author} {\bibfnamefont {L.}~\bibnamefont {Motl}},\ }\href {\doibase 10.4310/ATMP.2002.v6.n6.a3} {\bibfield  {journal} {\bibinfo  {journal} {Adv. Theor. Math. Phys.}\ }\textbf {\bibinfo {volume} {6}},\ \bibinfo {pages} {1135} (\bibinfo {year} {2003})},\ \Eprint {http://arxiv.org/abs/gr-qc/0212096} {arXiv:gr-qc/0212096} \BibitemShut {NoStop}%
\bibitem [{\citenamefont {Cardoso}\ \emph {et~al.}(2004)\citenamefont {Cardoso}, \citenamefont {Natario},\ and\ \citenamefont {Schiappa}}]{Cardoso:2004up}%
  \BibitemOpen
  \bibfield  {author} {\bibinfo {author} {\bibfnamefont {V.}~\bibnamefont {Cardoso}}, \bibinfo {author} {\bibfnamefont {J.}~\bibnamefont {Natario}}, \ and\ \bibinfo {author} {\bibfnamefont {R.}~\bibnamefont {Schiappa}},\ }\href {\doibase 10.1063/1.1812828} {\bibfield  {journal} {\bibinfo  {journal} {J. Math. Phys.}\ }\textbf {\bibinfo {volume} {45}},\ \bibinfo {pages} {4698} (\bibinfo {year} {2004})},\ \Eprint {http://arxiv.org/abs/hep-th/0403132} {arXiv:hep-th/0403132} \BibitemShut {NoStop}%
\end{thebibliography}%
\end{document}